\documentclass[fleqn,usenatbib]{mnras}

\usepackage{newtxtext,newtxmath}

\usepackage{graphicx}
\usepackage{dcolumn}
\usepackage{slashbox}

\usepackage[usenames,dvipsnames]{xcolor}
\hypersetup{colorlinks=true,citecolor=blue,linkcolor=OliveGreen}

\usepackage{soul}

\usepackage{subfigure}
\usepackage{wasysym}
\usepackage{verbatim}
\usepackage{color}
\usepackage{mathtools}
\usepackage{hyperref}
\usepackage{soul}
\usepackage{pbox}
\usepackage[export]{adjustbox}

\usepackage{slashed}
\usepackage{multirow}
\usepackage{booktabs}  
\usepackage{natbib}
\usepackage{float}
\usepackage{afterpage}
\usepackage{dcolumn}
\usepackage{bm}
\usepackage{lipsum}
\usepackage{setspace}
\usepackage{etoolbox}

\newcommand{\Abacus}[1]{\textsc{#1}}

\def\beq{\begin{equation}}
\def\eeq{\end{equation}}
\def\L{{\bf L}}

\def\mnras{Mon. Not. R. Astron. Soc}
\def\physrep{Physics Reports}

\usepackage[T1]{fontenc}
\usepackage{ae,aecompl}

\usepackage{graphicx}	
\usepackage{amsmath}	

\usepackage{lineno}

\title[\textsc{CompaSO}]{\textsc{CompaSO}: A new halo finder for competitive assignment to spherical overdensities}

\author[B. Hadzhiyska et al.]{
Boryana Hadzhiyska,$^{1}$\thanks{E-mail: boryana.hadzhiyska@cfa.harvard.edu}
Daniel Eisenstein,$^{1}$
Sownak Bose,$^{1,2}$
\newauthor
\ Lehman H. Garrison$^{3}$,
and Nina Maksimova$^{1}$
\\
$^{1}$Harvard-Smithsonian Center for Astrophysics, 60 Garden St., Cambridge, MA 02138, USA\\
$^{2}$Institute for Computational Cosmology, Department of Physics, Durham University, Durham DH1 3LE, UK\\
$^{3}$Center for Computational Astrophysics, Flatiron Institute, 162 Fifth Avenue, New York, NY 10010, USA\\
}
\date{Accepted XXX. Received YYY; in original form ZZZ}

\pubyear{2019}

\begin{document}
\label{firstpage}
\pagerange{\pageref{firstpage}--\pageref{lastpage}}
\maketitle

\begin{abstract}
We describe a new method (\textsc{CompaSO}) for identifying groups of particles in cosmological $N$-body simulations. \textsc{CompaSO} builds upon existing spherical overdensity (SO) algorithms by taking into consideration the tidal radius around a smaller halo before competitively assigning halo membership to the particles. In this way, the \textsc{CompaSO} finder allows for more effective deblending of haloes in close proximity as well as the formation of new haloes on the outskirts of larger ones. This halo-finding algorithm is used in the \textsc{AbacusSummit} suite of $N$-body simulations, designed to meet the cosmological simulation requirements of the Dark Energy Spectroscopic Instrument (DESI) survey. \textsc{CompaSO} is developed as a highly efficient on-the-fly group finder, which is crucial for enabling good load-balancing between the GPU and CPU and the creation of high-resolution merger trees. In this paper, we describe the halo-finding procedure and its particular implementation in \Abacus{Abacus}, accompanying it with a qualitative analysis of the finder. {We test the robustness of the \textsc{CompaSO} catalogues before and after applying the cleaning method described in an accompanying paper and demonstrate its effectiveness by comparing it with other validation techniques.}  We then visualise the haloes and their density profiles, finding that they are well fit by the NFW formalism. Finally, we compare other properties such as radius-mass relationships and two-point correlation functions with that of another widely used halo finder, \textsc{ROCKSTAR}. 
\end{abstract}




\section{Introduction}
\label{sec:intro}
One of the central goals of cosmological $N$-body simulations 
is to be able to associate clumps of
simulation particles
with real-world objects, be they galaxies, dark matter haloes, 
or clusters, in a way which provides a framework for testing
cosmological models \citep{1990MNRAS.242P..59P,1994ApJ...436..467G,2002ApJ...573....7E,2003ApJS..145....1B,2004NewA....9..111D}.
For this reason, simulation data are generally packaged into  
physical objects that have their counterparts in observations.
Examples of simulation objects are large
virialised structures, known as dark matter haloes
and halo clusters, which are often assumed to be (biased) tracers 
of the observable galaxies and galaxy clusters. This is motivated
by the idea that galaxies are known to form in the
potential wells of haloes, and therefore, the evolution of galaxies
might be closely related to that of their halo parents. 
Thus, the potential of numerical cosmology hinges
upon our ability to identify collapsed haloes in a cosmological 
simulation by assigning sets of simulation particles to virialised
dark matter structures (also known as groups, or haloes). However,
oftentimes, the objects identified as groups
have fuzzy edges which may overlap with other such groups, so no 
perfect algorithmic definition of a halo exists. As a consequence,
the particular model through which we
assign halo membership
determines the properties we extract from these objects. There are 
many existing methods up to date for dividing a set of particles
into groups, and all of them come with their advantages and drawbacks,
depending on the applications for which they are employed.

There are a number of popular group-finding algorithms such as
friends-of-friends \citep[FoF,][]{1985ApJ...292..371D,1987ApJ...319..575B},
DENMAX \citep{1991ComPh...5..164B,1994ApJ...436..467G},
AMIGA Halo Finder \citep[\textsc{AMF},][]{2009ApJS..182..608K},
Bound Density Maxima \citep[\textsc{BDM},][]{1999ApJ...516..530K},
HOP \citep{1998ApJ...498..137E},
SUBFIND \citep{Springel:2000qu},
spherical overdensity \citep[SO,][]{1992ApJ...399..405W,1994MNRAS.271..676L,1996ApJS..103...41B}
and \textsc{ROCKSTAR} \citep{2013ApJ...762..109B}.
Most relevant for this work are FoF, SO, and \textsc{ROCKSTAR}.
In the FoF algorithm \citep{1985ApJ...292..371D,1998A&A...333..779A}, 
two particles belong to the same group if their separation is less
than a characteristic value known as
the linking length, $l_{\rm FoF}$. This
characteristic length is usually about $0.2 \ l_{\rm mean}$, 
where $l_{\rm mean}$ is the mean particle separation.
The resulting group of linked particles is considered
a virialised halo. The advantages of the FoF method is 
that it is coordinate-free and that it has only one
parameter. In addition, the outer boundaries of the FoF
haloes tend to roughly correspond to a density contour
related to the inverse cube of the linking length.
The typical choice of $l_{\rm FoF}=0.2 \ l_{\rm mean}$
corresponds roughly to an overdensity contour of 
178 times the background density using the percolation
theory results of \citet{2011ApJS..195....4M}.
However, sometimes groups identified by this method 
appear as two or more clumps, linked by a small thread of particles.
Computing the halo properties of such dumbbell-like 
structures, e.g. the centre-of-mass, virial radius, 
rotation, peculiar velocity, and shape parameters,
is a challenge, as they are completely blended. 
In addition, the FoF method is not well-suited for
finding subhaloes embedded in a dense hosting halo.

Another popular halo-finding method is the spherical overdensity method,
or SO \citep{1992ApJ...399..405W,1994MNRAS.271..676L}. It adopts a criterion
for a mean overdensity threshold in order to detect virialised haloes. 
In a simple implementation of the SO method, 
the particle with the largest local density is selected and marked as the
center of a halo. The distance to all other 
particles is then computed, and the radius of
the sphere around the halo center is increased 
 until the enclosed density satisfies the virialisation criterion. 
Particles inside the sphere are assumed to be members of the spherical
halo. The search for a new halo center usually stops past a chosen minimum
threshold density so as to avoid $\mathcal{O}(N^2)$ distance
calculations between all of the remaining lower-density particles 
in the parent group. However, this choice must be made carefully, as otherwise,
one might miss viable SO haloes.

Similarly to FoF, the SO method fails to resolve subhaloes in
highly clustered regions. It also makes a number of other assumptions.  For
example, the search begins from the particle with the highest
density, but that particle may not be at the center which would maximise the 
mass. An over-conservative choice for the density threshold criterion 
often leads to spherically truncated haloes. Moreover,
the SO method does not consider any competition between centers for the
membership of a particle. In many cases, it is possible that a particle would have
had a higher overdensity if it belonged to another halo that had
a smaller maximum density at its center. For this reason, an implementation
of the SO method allowing for more competitive choices
might produce memberships similar to tidal radii and thus more physical.  

\textsc{Rockstar} is a temporal, phase-space finder that
uses information both about the phase space distribution of the
particles and also about their temporal evolution \citep{2013ApJ...762..109B}. 
Phase-space halo finders take into consideration information about the relative motion
of two haloes, which makes the process of finding tidal remnants and determining halo boundaries
substantially more effective. Having temporal information
also helps maximise the consistency of halo properties 
across time, rather than just within a single snapshot.
For this reason \textsc{Rockstar} is considered to be highly accurate in
determining particle-halo membership. 

In this paper, we describe a new halo-finding method, dubbed \textsc{CompaSO},
which builds on the SO algorithm but differs significantly in its implementation.
Most notably, the \textsc{CompaSO} halo finder takes into consideration
the tidal radius around a smaller halo in order to assign halo 
membership to the particles, instead of simply truncating
the haloes at the overdensity threshold as typically done by SO methods. That is necessary to address, as
a traditional weak point of position-space halo finders
has been deblending haloes during major mergers. In most
existing halo finders, the assignment of particles to haloes 
becomes essentially random in the overlap region. Exception
to that rule are phase-space
halo finders such as \textsc{ROCKSTAR} \citep{2013ApJ...762..109B}.
Another long-standing problem is the identification of
haloes close to the centers of larger haloes. To alleviate
this issue, our finder allows for the formation of new haloes
\textit{within} the density threshold radius, i.e.
on the outskirts of forming haloes.
In addition, we demand that the particle  proposed to become halo center (nucleus)
has the highest density among all of its immediate neighbors.
In this way, we ensure that the \textsc{CompaSO} haloes have centers
corresponding to ``true'' density peaks.  A point of contention in 
recent halo finder discussions has been the idea of overlapping
halo boundaries. While the process of drawing clear cuts between 
is somewhat arbitrary and artistic by nature particularly for merging
objects, our method does impose stark boundaries between haloes
resulting from our competitive assignment of particles. This decision
is taken so as to ensure mass conservation within the halo catalogue,
which is a central requirement for our particular application --
namely, the creation of halo occupation distribution (HOD) mock catalogues.

To estimate the statistical and physical properties of the haloes
obtained with the competitively assigning SO algorithm (\textsc{CompaSO}),
we run a variety of comparison tests with \textsc{ROCKSTAR}.
Because of its robust approach to halo finding, 
we consider the \textsc{ROCKSTAR} halo catalogues as the norm
in our comparative evaluations against \textsc{CompaSO}.

This paper is organised as follows. In Section \ref{sec:meth},
we introduce the \textsc{CompaSO} algorithm (Section 
\ref{sec:meth.algo}) and mention particular
details of its implementation and of the optimization techniques
employed (Section \ref{sec:meth.opt}). {In Section \ref{sec:clean}, we summarise the cleaning technique adopted to weed out unphysical haloes from the \textsc{CompaSO} catalogues and compare it with other commonly used cleaning techniques.} In Section \ref{sec:anal},
we describe various characteristics of the \textsc{CompaSO} haloes
such as halo density and mass profiles, mass-radius relationships, 
halo center definitions, and auto- and cross-correlation 
functions, comparing it with the \textsc{ROCKSTAR} catalogues wherever 
feasible. We further outline additional consistency checks we have done
including particle visualizations and merger tree associations
to test the persistence of haloes through time. Finally, in Section
\ref{sec:disc}, we summarise the features of our
method and point out its most useful areas of application 
for large-scale cosmology.

\section{Methodology}
\label{sec:meth}
In this section, we introduce the \textsc{CompaSO} halo-finding algorithm,
which has been employed to create halo catalogues for the \textsc{AbacusSummit} suite of simulations.
We first describe the procedure for competitively assigning particles to haloes
and then go into specifics about the particular implementation and optimization
strategies adopted.

\subsection{\textsc{AbacusSummit}}
\label{sec:meth.summit}
This halo finding algorithm was 
developed in anticipation of
the \textsc{AbacusSummit} suite of high-performance
cosmological $N$-body simulations,
designed to meet the Cosmological Simulation
Requirements of the Dark Energy Spectroscopic
Instrument (DESI) survey and run on the
Summit supercomputer at the Oak Ridge
Leadership Computing Facility.
The simulations are run with \Abacus{Abacus} \citep{2019MNRAS.485.3370G},
a high-accuracy cosmological $N$-body simulation
code, which is optimised for Graphics Processing Unit (GPU) architectures and 
for large-volume, moderately clustered 
simulations. \Abacus{Abacus} is extremely fast, performing 
70 million particle updates per second on 
each node of the Summit supercomputer, and 
also extremely accurate, with typical force 
accuracy below $10^{-5}$. 
The near-field computations run on a GPU architecture, whereas
the far-field computations run on CPUs.  

The production of halo catalogs was a core requirement of \textsc{AbacusSummit}, 
and the sheer size of the data set requires that this be done as on-the-fly,
as part of the simulation code itself.  This is particularly true given the
desire to output halos at a few dozen redshifts to support the creation of
merger trees.
We therefore sought to augment beyond the FOF algorithm with an SO-based
algorithm.  However, the conditions of \textsc{AbacusSummit} are quite demanding:
the Abacus code is evolving the simulation very quickly because of the powerful
GPUs on each Summit node, and even though group finding is happening only on
a few percent of time steps, we need to have the group finding be comparably
fast, lest we find that an undue amount of total allocation on a GPU-based 
supercomputer is being consumed with CPU-bound work.  High speed was a priority,
driving a number of algorithmic choices and optimizations; we ended up achieving a rate of about 30 million particles/second/node
\citep{Maksimova+2021}.

It also is a requirement of the larger simulation code and in particular the parallelization
method that the group-finding algorithm rely on a strict segmentation of the 
simulation volume that could be defined uniquely regardless of data further away.
We also had to assure
that we could use many cores efficiently, as we were using Summit with 84 threads.
This drove us to do the initial segmentation with the FoF algorithm, and then 
proceed with CompaSO with an embarrassingly parallel application of threads to 
the list of FoF halos.

\subsection{The algorithm}
\label{sec:meth.algo}
In this section, we describe our new hybrid 
algorithm \textsc{CompaSO}.
There are three levels of group finding
performed on the particle sets of the \textsc{AbacusSummit}
simulations: Level 0 (L0) which
uses a modified FoF algorithm (see following paragraph), 
Level 1 (L1)  which adopts the competitive assignment to 
spherical overdensities algorithm \textsc{CompaSO}, and 
Level 2 (L2) which also uses \textsc{CompaSO}, but sets 
the density threshold to a much higher value.
These halo finding steps are performed in a nested
fashion, starting with the L0 (modified FoF) group finding. 
Thus, the main halo catalogue output from the simulation boxes
contains the L1 haloes, the L0 groups are large ``fluffy'' sets
of particles that usually contain several L1 haloes, and the
L2 subhaloes correspond to the substructure within the L1
haloes, but they do not have their own catalogue entries.

Short descriptions of the relevant quantities for the
\textsc{CompaSO} algorithm are displayed in Table
\ref{tab:names} along with the corresponding notation in 
the \Abacus{Abacus} code. In Appendix \ref{app:choices},
we discuss our particular choices for the
halo-finding parameters used to create the \textsc{AbacusSummit} suite of simulations.

\subsubsection{Pre-processing of the particles}

Prior to the FoF group identification,
an estimate of the ``local density,'' $\Delta$, is computed 
for each particle using the weighting kernel 
\begin{equation}
    W(r; b_{\rm kernel}) = 1-r^2/b_{\rm kernel}^2,    
\end{equation} 
where we have set $b_{\rm kernel}$ to be 
0.4 of the mean interparticle spacing, $l_{\rm mean}$. 
Obtaining a measure of the local density makes it 
easier to find substructure as well as 
identify the core of a halo. One of the advantages of choosing
this kernel as opposed to a top-hat kernel is that it avoids a discontinuity at $r = b_{\rm kernel}$, thus
reducing the noise from small perturbations 
in the particle positions. We note that since we need
the squared distances for computing the near-field forces,
we can essentially obtain the local density for ``free''.

We then group the particles
into L0 haloes, using a modified FoF algorithm with 
linking length $l_{\rm FoF}$, set to 0.25 of the mean
interparticle spacing, $l_{\rm mean}$, but only for particles
with high enough values of the local density, $\Delta>\Delta_{{\rm L0, min}}$, where $\Delta_{{\rm L0, min}} = 60$. 
We note that $l_{\rm FoF}=0.25 \ l_{\rm mean}$ normally would percolate at a 
noticeably lower density, $\Delta \approx 41$ \citep{2011ApJS..195....4M},
so the kernel density limit ($\Delta>60$)
imposes a physical smoothing scale. 
The reason we choose this density cut
is that the bounds of the L0 halo 
set are determined by the kernel density estimate, 
which has lower variance than the nearest 
neighbor method of FoF and imposes a physical
smoothing scale (see Appendix \ref{app:smoothing} for discussion of the choice of smoothing scale for the weighting kernel). {We discuss the choice of linking length, $l_{\rm FoF}$, smoothing scale, $b_{\rm kernel}$, and density threshold, $\Delta_{{\rm L0, min}}$, in Appendix \ref{app:choices} and find that their effect on halo properties is negligible.}

\subsubsection{The \textsc{CompaSO} method}
Within each L0 halo, we construct L1 and L2 haloes by the
\textsc{CompaSO} algorithm as follows:
\begin{itemize}
    \item[1.]
    We select the particle with highest kernel density ($\Delta$) in the
    L0 group to be the first halo nucleus. We then search outward
    to find the innermost radius at which the enclosed density dips
    below the L1 threshold density, $\Delta_{\rm L1}$.
    This L1 radius ($R_{\rm L1}$) determines the absolute outer 
    bound on the eventual member particles. 
    Particles interior to that 
    radius are tentatively assigned to that group.
    For large haloes, the local density at the
    boundary will be considerably less than the overdensity
    interior to the boundary (roughly 3 times lower,
    assuming that the halo profile resembles that of a
    singular isothermal sphere), so it is
    possible that some of the large haloes have satellite
    haloes orbiting on their outskirts (but still within the density
    threshold). For this reason, while we
    mark the particles interior to
    $R_{\rm L1, elig}$ as 
    ``ineligible'' to be future halo nuclei, the rest of 
    the particles \textit{are} ``eligible'' to become new nuclei. We set $R_{\rm L1, elig}/R_{\rm L1} = 0.8$.\\
    The choice of 80\% was made after
    testing several different
    cases between 70\% and 100\% and finding that smaller
    values lead to wedge-like gouges out of the large haloes, 
    while larger values are not as efficient at deblending
    haloes in close proximity ({see Appendix~\ref{app:choices} for more details}).

  \item[2.] We then search the remaining ``eligible''
    particles to find the one with next highest
    kernel density that also meets a
    minimum local density criterion. This latter condition is that
    a particle must be denser than \textit{all} other particles
    (``eligible'' or not) within a radius of $b_{\rm kernel}$ of 
    the interparticle spacing, where we have chosen $b_{\rm kernel}/l_{\rm mean} = 0.4$. If the condition
    is satisfied, we initiate a new nucleus and 
    again search outwards for the
    L1 radius $R_{\rm L1}$, using \textit{all} L0 particles.

  \item[3.]
    A particle
    is attributed to the new group if it is previously unassigned
    or if it is estimated to have an enclosed density
    with respect to the new group that is at least $c_{\rm Roche}$ times larger than
    that of the enclosed density with respect to its currently assigned group, where we have selected $c_{\rm Roche} = 2$.
    In this way, we allow for the particles to be divided into haloes
    by a principle similar to that of the tidal radius. For
    two rigid spherical isothermal spheres
    truncated at the separation of their centers, the tidal radius
    lies at the point between them for which the ratio of
    enclosed densities is $c_{\rm Roche}$. The value of $c_{\rm Roche}$ ranges from 1, 
    if the two spheres have equal masses, to 2, if one sphere is much 
    lighter \citep{1987gady.book.....B}. For the 
    \textsc{CompaSO} algorithm, we set the value of $c_{\rm Roche}$ to 2.

  \item[4.]
    The search for new centers of nucleation continues until
    we reach the minimum density threshold. This threshold 
    is set by estimating what central density would
    be generated by a singular isothermal sphere consisting of
    $N_{\rm new, min} = 35$ particles within a radius encompassing 200 times the 
    mean background density, $M_{\rm 200m}$.
\end{itemize}

Finally, within each L1 halo, 
we repeat the competitive SO algorithm steps to find the L2
haloes (subhaloes) within the enclosed radius $R_{\rm L2}$.\\
For the \textsc{AbacusSummit} suite, we only store
the masses of the 5 largest subhaloes, as it may
help to mark cases of over-merged L1 haloes. The
main purpose of the L2 haloes is to use the center-of-mass
of the largest L2 subhalo to define a
center for the output of the L1 statistics.

\begin{figure}
\centering  
\includegraphics[width=0.48\textwidth]{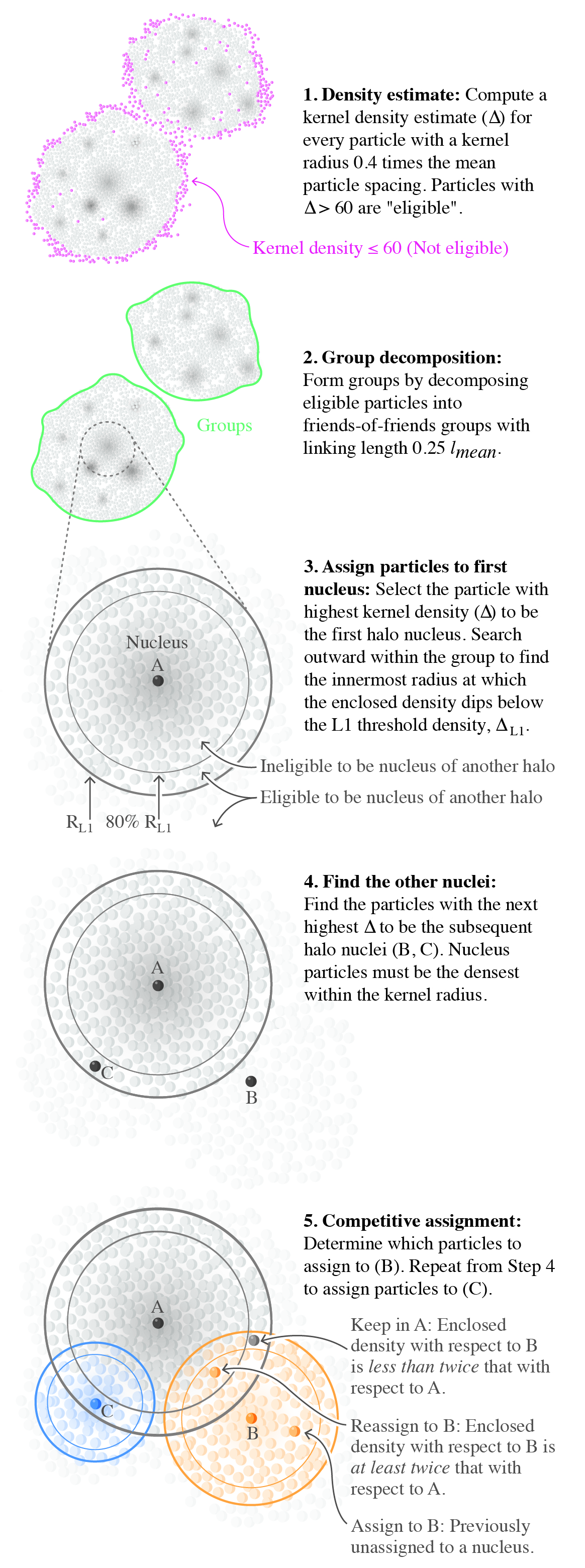}
\caption{Visualization of the \textsc{CompaSO} algorithm detailed in Section~\ref{sec:meth.algo}.}
\label{fig:algo}
\end{figure}

In Fig.~\ref{fig:algo}, we present a visual description of the
halo-finding procedure, summarizing the steps taken to identify L0 groups and search for the L1 \textsc{CompaSO} haloes within them. We do not perform any unbinding of the particles, as is sometimes
done through estimates of the gravitational potential and
resulting particle energy. {This is because, in addition to the computational expense associated with the procedure,
in dynamically evolving situations,
the energy of a single particle is not conserved and the
binding energy is not a guarantee of long-term membership in a halo. For a detailed discussion and tests of the effect of unbinding, see Section~\ref{sec:clean.unbind}.}
For the \textsc{AbacusSummit} simulations, we output properties
for all L1 haloes with more than $N_{\rm L1,min} = 35$ particles.

\subsection{Optimization and implementation}
\label{sec:meth.opt}
In the following, we describe particular features
of the implementation of our algorithm, which
have helped us to significantly speed up the halo
finding process.

\subsubsection{Density scaling with redshift}
For Einstein deSitter cosmology, the density threshold
for L1 and L2 haloes are $\Delta_{\rm L1}=200$
and $\Delta_{\rm L2}=800$, respectively. As the 
cosmology departs from Einstein deSitter, 
we scale upward the density
threshold for L1, $\Delta_{\rm L1}$ in order to keep in
agreement with spherical collapse
estimates for low-density universes.
The FoF linking length, $l_{\rm FoF}$, is scaled as
the inverse cube root of that change, while
The kernel density smoothing scale, $b_{\rm kernel}$, remains unchanged.
We make use of the redshift scaling of the fitting
function provided by \citet{1998ApJ...495...80B}, which
defines the density with respect to the critical
density as:
\begin{equation}
  \Delta_{\rm base}(z) = 18 \pi^2 + 82 x - 39 x^2,
\end{equation}
where  $x = \Omega_m(z) - 1$ and $\Omega_m(z)$ is
the matter energy density at redshift $z$ and
$\Delta_{\rm L1}(z) = (200/18 \pi^2)\Delta_{\rm base}(z)$
is the corresponding density threshold $\Delta_{\rm L1}(z)$ 
relative to the critical density at a given redshift $z$
\footnote{To obtain the value of $\Delta_{\rm L1}(z)$ (\texttt{SODensityL1(z)}) relative
to the mean density, one needs to divide by $\Omega_m(z)$.}.
Equivalently, in the case of L2 haloes, it is determined
by the relation $\Delta_{\rm L2} = (800/18 \pi^2)\Delta_{\rm base}$. {Throughout the paper, we will refer to overdensities defined using the \citet{1998ApJ...495...80B} scaling relation as ``virial.'' The default mass definition of the \textsc{ROCKSTAR} algorithm also adopts the \citet{1998ApJ...495...80B} fitting formula.}

\subsubsection{Density-radius relation}
For our estimates of the enclosed density of a particle
with respect to an existing nuclei, we do not compute
its exact value, but rather use a fitted model of 
the relation between the density and radius. This
speeds up the code significantly for two main reasons:
(a) it avoids having to do a 
complete sort on the distances (which we never do for
finding the L1 radius), and consequently,
(b) it avoids having to permute those enclosed
densities back into the original particle order,
which would incur disordered memory access.

Our fitted model is as simple as assuming a flat rotation
curve based on the L1 radius, so the enclosed density is
\begin{equation}
  \Delta_{\rm encl.}(r) = \Delta_{\rm L1} \frac{R_{\rm L1}^2}{r^2}.
\end{equation}
We emphasize that this relation is only being used to deblend
zones between haloes. The actual halo boundary for any unblended
region is still taken to be the L1 enclosed density radius
(based on all particles, not just members).

\subsubsection{Distances computation}
The algorithm is sped up by storing the position and original
particle index in a compact quad of floats, aligned to 16 bytes to
aid in the vectorization \citep[see][for details on the implementation]{Maksimova+2021,Garrison+2021b}.
We store the index as an integer; when interpreted bitwise as a float, 
it becomes a tiny number.  We then compute the square distances in the
4-dimensional space, as this is faster with the vector instructions.

Note that throughout the \textsc{CompaSO} algorithm, we work with 
square distances (rather than linear ones), so as to avoid taking many square roots.  Furthermore, 
we track the inverse enclosed density because we can compute it 
directly without having to take its reciprocal. 
Searching for the enclosed density threshold is logically equivalent
to sorting the distance list and searching outwards until one finds
the element for which the index in the sorted array exceeds the cubed 
distance (assuming the particle mass is unity, the index number is equivalent
to the enclosed mass).

\subsubsection{Threshold radius search using cells}
The most straightforward implementation of the algorithm in
Section \ref{sec:meth.algo} involves sorting the whole list and looking
for the threshold radius $R_{\rm L1/L2}$, 
but usually we do not need to go far from the nucleus
to find this radius. Therefore, sorting the entire list
is unnecessary effort, as it is a 
$\mathcal{O}(N\log{}N)$ process, while doing a single partition
is only $\mathcal{O}(N)$. Instead, one can 
do a partitioning on
a square radius and determine that the mass interior
to that radius does not reach the density threshold,
repeating that process until the threshold radius is found.

This is precisely what we use to optimise the
inside-out radius search. Internally, \Abacus{Abacus}
has already
ordered the particles in terms of their 
cell membership, i.e. their bounding boxes\footnote{\Abacus{Abacus} 
relies heavily on these cells to compute the 
forces and accelerations of the particles \citep{Maksimova+2021,Garrison+2021b}.}. 
We can thus use the the cell boundaries positions to compute the
minimum and maximum bounds for each cell, and we also
know the mass in each group.
Working outward in steps of $\sqrt{3}/3 L$ (radial bins), where
$L$ is the cell size, yields at most 3
crossings (or two bisection passes) per cell.
Note that this needs to be done only for the
cells that satisfy the condition for
threshold radius search. We determine this
in the following way: if we place the
mass contained within the \textit{interior} 
edge of the radial bin under consideration at the 
\textit{exterior} edge of that radial
bin and find that it satisfies the density
threshold, then we know that the searched-for radius
cannot be inside that radial bin. 
If the mass does not satisfy the 
threshold, then we have to search within it. 
Note that in the case where the search fails,
we simply need to keep looking outwards.
The innermost radial bin will always satisfy the
condition for threshold searching.

\subsubsection{Comparative sweep and collection of particles into group order}
In our algorithm, we need to have a way of denoting whether a given particle 
is eligible to become a halo center (nucleus) and also what its current halo
affiliation is (i.e. which existing halo it has been assigned to most recently). 
To do so, we create a single array that combines the group identification number and active
flag into one integer. The absolute value of the integer indicates the index of the
halo nucleus it is assigned to and negative integer means that the particle is 
ineligible to become a halo nucleus. This saves computation time since we
reduce the number of variables which need to be swapped and sorted.

We finally sweep through the particles to segment 
them (by partitioning them based on their halo affiliation
index) into groups. A simple optimization we have 
applied is to first
partition for the first halo, as it is most likely to
be the largest. Then before going to the second,  we
partition to sweep all of the unassigned particles, so 
that we do not have to pass through them again. This is
because in most cases the largest halo particles and 
the unassigned particles are the two biggest sets
and getting those out of the way before we segment the
other haloes helps save computation time.

\begin{table*}
\begin{center}
\begin{tabular}{c c c c } 
 \hline\hline
 Code notation & Paper notation & Value & Description \\ [0.5ex]
 \hline
\texttt{BoxSize} & $L_{\rm box}$ & 1 Gpc$/h$ & Size of the simulation box \\
\texttt{NP} & $N_{\rm part}$ & $3456^3$ & Number of particles in the box \\
\texttt{ParticleMassHMsun} & $M_{\rm part}$ & $2.109 \times 10^9 \ M_\odot/h$ & Mass of each dark matter particle \\
\texttt{Omega\_M} & $\Omega_m$ & 0.315 & Energy density of the total matter in the Universe at $z = 0$ \\
\texttt{FoFLinkingLength} & $l_{\rm FoF}$ & 0.25 & Linking length for the modified FoF \\
\texttt{MinL1HaloNP} & $N_{\rm L1,min}$ & 35 & Minimum number of particles in the L1 halo for including in halo catalogue \\
\texttt{L0DensityThreshold} & $\Delta_{{\rm L0, min}}$ & 77.5 & Minimum kernel density for a particle to be part of an L0 group (scaled to 60) \\
\texttt{SODensityL1} & $\Delta_{\rm L1}$ & 258.2 & L1 halo density threshold at $z = 0.5$ (scaled to 200) \\
\texttt{SODensityL2} & $\Delta_{\rm L2}$ & 1033.0 & L2 halo density threshold at $z = 0.5$ (scaled to 800) \\
\texttt{DensityKernelRad} & $b_{\rm kernel}$ & 0.4 $l_{\rm mean}$ & \pbox{20cm}{\vspace{0.1cm}The scale over which a new nucleus must have highest \\ kernel density $\Delta$ in units of the mean interparticle distance, $l_{\rm mean}$/\\
The smoothing scale of the weighting kernel $W(r;b_{\rm kernel})$ \\ used for computing the density of each particle \vspace{0.1cm}} \\
\texttt{SO\_RocheCoeff} & $c_{\rm Roche}$ & 2 & \pbox{20cm}{\vspace{0.1cm}Factor for attributing particle to new nucleus \\ motivated by the tidal radius condition\vspace{0.1cm}} \\
\texttt{SO\_NPForMinDensity} & $N_{\rm new, min}$ & 35 & Kernel density criterion for ceasing to look for new nuclei \\
\texttt{SO\_alpha\_eligible} & $R_{\rm L1, elig}$ & 80\% of $R_{\rm L1}$ & \pbox{20cm}{\vspace{0.1cm}Radius for determining the eligibility boundary of new \\ nuclei in units of the L1 threshold radius $R_{\rm L1}$\vspace{0.1cm}} \\
\texttt{ParticleSubsampleB} & Subsample \texttt{B} & 7\% & Percentage of particles output in Subsample \texttt{B} \\ 
\hline
\texttt{N} & $N$ & --- & Number of particles in the L1 halo \\
\texttt{r\_XX\_L2com} & $r_{\rm XX}$ & --- & \pbox{20cm}{\vspace{0.1cm}Radius within which XX\% of the L1 halo particles \\ are contained wrt the particle center of the first L2 halo\vspace{0.1cm}} \\ 
\texttt{r\_50\_L2com} & $r_{\rm halfmass}$ & --- & \pbox{20cm}{\vspace{0.1cm}Radius within which 50\% of the L1 halo particles \\ are contained wrt the particle center of the first L2 halo\vspace{0.1cm}} \\ 
\texttt{r\_98\_L2com} & $r_{\rm halo}$, $r_{98}$ & --- & \pbox{20cm}{\vspace{0.1cm}Radius within which 98\% of the L1 halo particles \\ are contained wrt the particle center of the first L2 halo\vspace{0.1cm}} \\ 
\texttt{SO\_central\_particle} & L1 particle & --- & Particle center of the L1 halo \\
\texttt{SO\_L2max\_central\_particle} & L2max particle & --- & Particle center of the first L2 halo \\
\texttt{x\_com} & L1 com & --- & Center-of-mass of the L1 halo \\
\texttt{x\_L2com} & L2max com & --- & Center-of-mass of the first L2 halo \\
\texttt{vcirc\_max\_L2com} & $V_{\rm max}$ & --- & The maximum circular velocity any particle in the L1 halo attains \\
\texttt{rvcirc\_max\_L2com} & $r_{\rm v,max}$ & --- & The radius at which the maximum circular velocity is reached \\
\hline \hline
\end{tabular}
\end{center}
\caption{Names of the variables as they appear in the \Abacus{Abacus} code and in this paper, accompanied by short descriptions. The corresponding values are also provided for the box used in this analysis, \texttt{AbacusSummit\_highbase\_c000\_ph100}. The first set of variables can be found in the \texttt{Header} files of the individual simulations, whereas the second set can be found in the field names of the halo catalogues.}
\label{tab:names}
\end{table*}

\section{Catalogue cleaning}
\label{sec:clean}

{In this section, we explore the robustness of the \textsc{CompaSO} halo catalogues. As is the case for all configuration-space finders, \textsc{CompaSO} has to deblend haloes in close proximity to each other without using velocity information of the particles. In some cases, this results in the algorithm defining unphysical objects as haloes (as illustrated in Fig.~\ref{fig:split}), which should instead be merged into nearby haloes or altogether discarded. Some halo finders try to address this problem by taking additional steps: e.g., performing particle unbinding or using merger tree information. In the case of \textsc{CompaSO}, we opt to perform cleaning of the halo catalogues in post-processing, as detailed in \citet{Bose+2021} and summarised below. In this section, we test how robust the cleaned catalogues are, comparing and complementing them with additional consistency checks. In particular, we test the persistence of haloes through time and the presence of particles with excessively high velocities, as measures of their ``trustworthiness''. We show that while the cleaning does not fix all issues, it certainly gets rid of the majority of problematic haloes. Additional cleaning of the catalogues can be performed using already existing halo statistics.}

\subsection{Cleaning method}
\label{sec:clean.meth}
The choice of whether an object is identified as a halo by any halo-finding
algorithm can be somewhat arbitrary. In the case of SO-based methods,
the halo boundary is set starkly by the SO threshold density, while for FoF-based
finders, it is strongly dependent on the linking length parameter. This
choice becomes even more challenging in dense regions of the simulation
as well as during halo merging and splashback events.
A frequent interaction between a satellite halo orbiting around a larger companion
is the expulsion of the satellite after several time steps. 
This typically results in a significant reduction of the mass of the
expelled object, as it is stripped of its outermost layer. The inner dense core
of the halo, which is also expected to be the location that the baryons would
occupy in a realistic full-physics setting, would be less prone to tidal stripping. 
However, an empirical population model such as the
halo occupation distribution (HOD) model applied after the satellite gets expelled
would underweight it and assign a galaxy to it based on its ``present-day'' mass.
This could clearly result in non-physical outcomes, where e.g. prior to a two-halo
interaction, the HOD prescription predicts two galaxies, but a few snapshots
later, when the smaller halo gets expelled by the larger one and stripped of its
outer shell, the HOD prescription attributes a single galaxy to the two-halo system.

Since one of the main goals of the \textsc{AbacusSummit} project is to provide the DESI
collaboration with a suite of simulations for creating mock catalogues via empirical
models, our focus is on adapting the final products to suit those needs
by ameliorating some of the aforementioned issues. To this end,
we provide our users with a ``cleaned'' version of the \textsc{CompaSO} catalogues.
The procedure for cleaning the catalogues is detailed in \citet{Bose+2021}.
and relies on utilizing merger-tree information about each halo.
There are two types of merger-tree outputs we have combined in an
attempt to weed out unhealthy haloes: a flag that marks ``potential splits''
and the ratio between the peak halo mass and the present-day halo mass.
The first flag checks the consistency of the halo
by tracing its main progenitors in previous steps and
tracking what percentage of the particles are shared between
the main progenitor and the present-day halo. If the number
is unphysical, i.e. the present-day halo received most of its
particles from a much larger halo in the previous time step
and therefore cannot be its main descendant, we mark this halo
as a ``potential split'' ({\tt IsPotentialSplit} flag in the merger tree catalogues). 
The second way in which we diagnose these halo-finding pathologies 
is by declaring unphysical all haloes for which the peak mass exceeds the
present day mass by more than a factor of $\kappa = 2$ \citep[see][for details]{Bose+2021}.
We merge haloes that have failed either of the above criteria into other nearby haloes 
from which they have presumably split off and record the updated halo masses
(the haloes that are merged donate their entire mass to the object they
are merged into). We show halo mass and correlation statistics of the cleaned 
\textsc{CompaSO} catalogues in Section \ref{sec:anal.hmf} and Section 
\ref{sec:anal.corr}. In very rare cases ($\sim0.1\%$), the \texttt{MainProgenitor} of a halo marked for cleaning may not be identified correctly, so we search for its true main progenitor in the complete \texttt{Progenitors} list and merge it onto that object.

\subsection{Persistence of haloes through time}
\label{sec:clean.persist}

\begin{figure*}
\centering
\includegraphics[width=.48\textwidth]{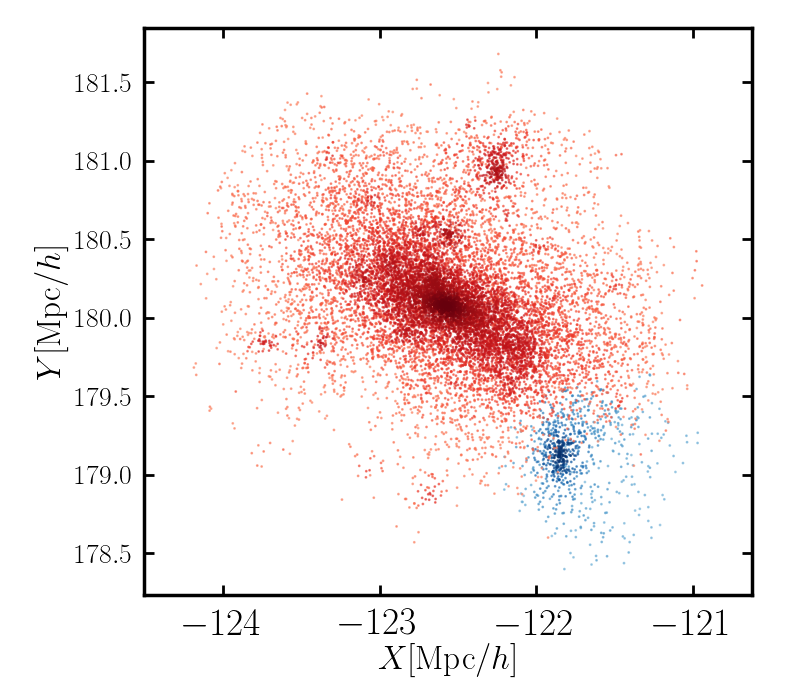}
\includegraphics[width=.48\textwidth]{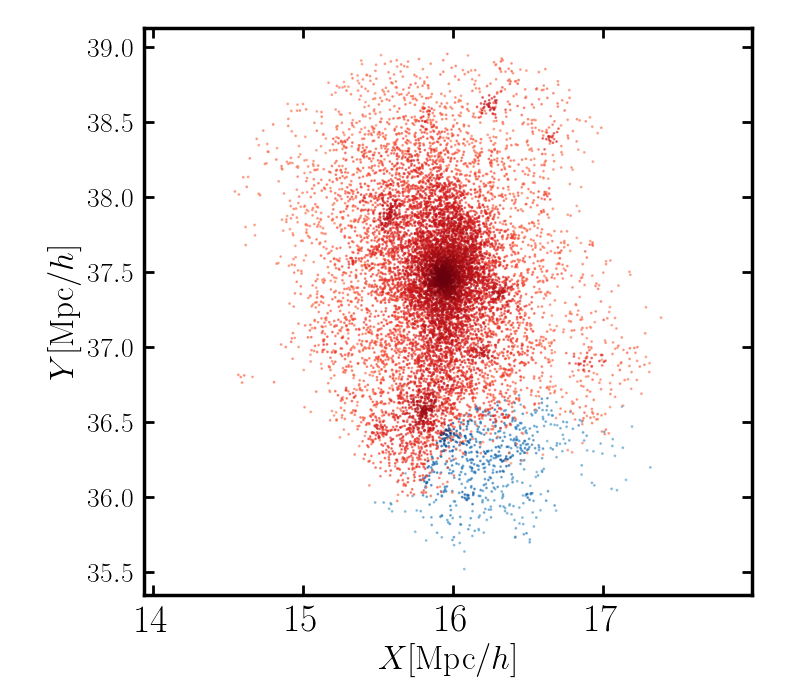}
\caption{Scatter plots of haloes
identified as ``potential
splits'' (\textit{blue dots}) and their central 
(largest) companions in the vicinity (\textit{red 
dots}). \textit{The left panel} shows an example
in which the ``potentially split'' halo does have
a dense core, distinct from that of its companion.
In this case, the smaller halo has been orbiting
in close proximity to the larger one for several
snapshots (during which the algorithm has identified
the two as a singular object) after which it has been expelled.
\textit{The right panel} shows a case of
a non-physical ``potential split'' for which
the small halo is lacking a dense center
of particles. A ``potential split'' is defined as
an object which in the previous time step, was considered 
part of a larger halo, but in the last time step,
has been split by the \textsc{CompaSO} algorithm. 
Objects lacking a well-defined core can be 
isolated by using diagnostics combining 
the merger tree flag {\texttt IsPotentialSplit}
(see statistics for how often these occur
in Table \ref{tab:mtree_stats})
and the ratio of the radius containing all the
particles to $r_{{\rm v,max}}$ (see Fig. \ref{fig:split}). 
The scatter plots are using 3\% of the halo  particles.}
\label{fig:scatter}
\end{figure*}

A powerful test of the ``reality'' of the small haloes
found on the outskirts of big ones
is their persistence between redshift slices. 
In other words, we can potentially track satellite
haloes at $z = 0.5$ and check if they were present
at earlier times.
For these smaller haloes, the effects which
contribute to their disruption and inconsistency
throughout the merger history are both physical as
well as algorithmic. Smaller haloes are less likely to survive
flybys past larger haloes because their cores are less dense and
thus more easily destructible. 
Another frequent interaction between a satellite and a central halo
is the expulsion of the satellite by the larger halo,
which typically significantly reduces the mass of the
expelled object, stripping it of its outermost layer. 
In addition, sometimes dark-matter substructure on the
outskirts of large haloes may have been identified
as an autonomous halo by the \textsc{CompaSO} algorithm at
a given snapshot, but this object might not have a distinct
merger history. In this section, we examine how often
we find such occurrences by using statistics output
by the accompanying merger tree associations\footnote{These can also be found alongside the other products of the \textsc{AbacusSummit} suite at https://abacussummit.readthedocs.io/en/latest/data-access.html} \citep{Bose+2021}.

In Fig. \ref{fig:scatter}, we demonstrate two cases of haloes
that have been marked as ``potential splits'' (see Section 
\ref{sec:clean.meth} for a definition). On the
left panel, we show a case of a non-physical algorithmic split while
on the right panel, the smaller halo has a
healthy core and has only spent a small portion of its history
as part of its larger companion (but otherwise has a distinct history). In Table \ref{tab:mtree_stats},
we show the percentage of ``potential splits'' for different
mass ranges in the third column. We see that their numbers
are very small in particular on the high-mass end ($\log(M)>13.7$), where they are less than 0.004\%.
The overall percentage is around 12\% and it
is highly dominated by the smallest haloes
($\log(M)\sim 10^{11}$). The reason for this relatively larger
percentage is that for less massive objects flying through dense
regions, the boundary is harder to draw, as they have less dense
cores and are more difficult to distinguish from the background.
This is less and less frequent at higher halo masses.

\begin{figure}
\centering  
\includegraphics[width=0.5\textwidth]{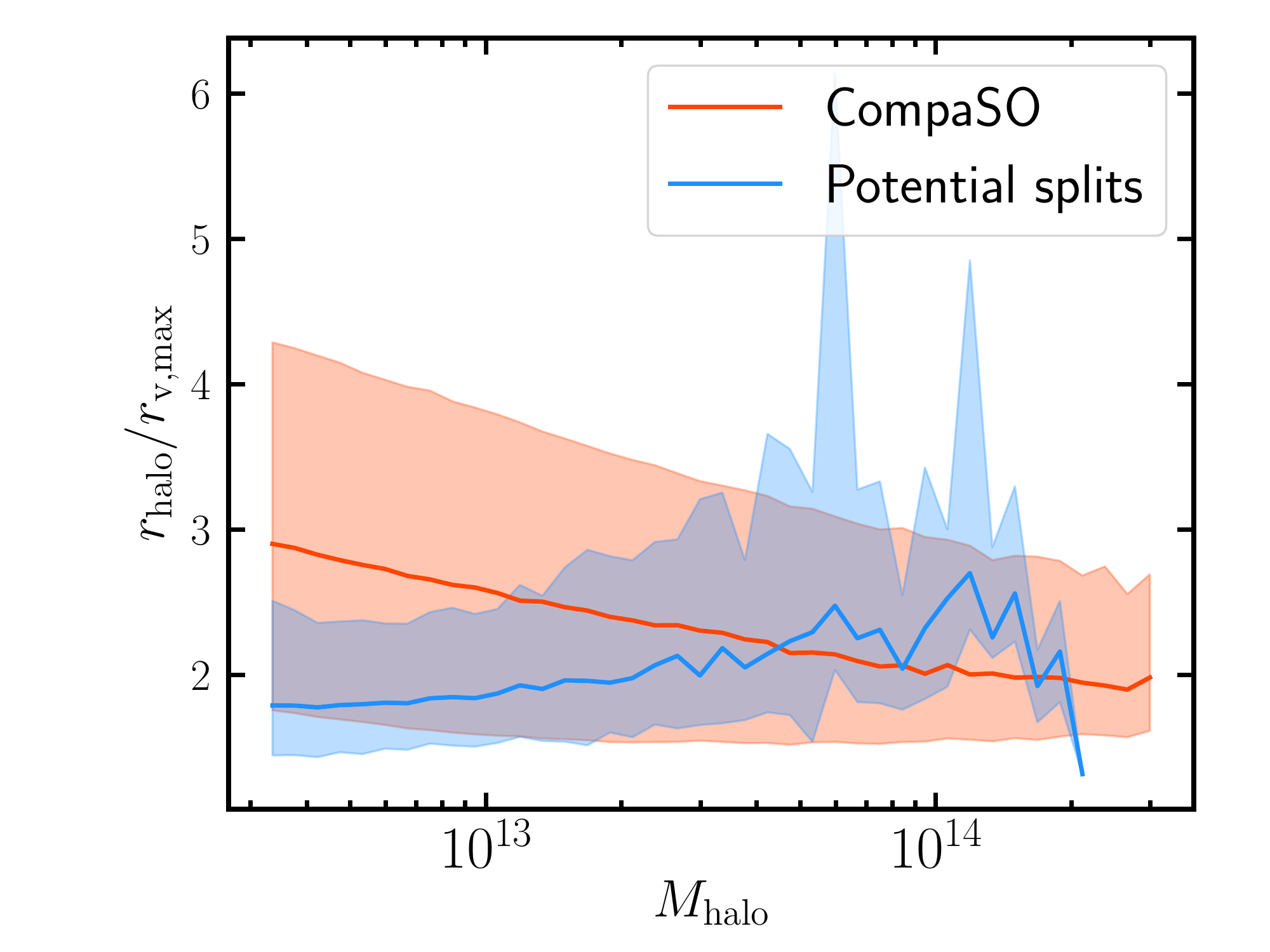}
\caption{Ratio of the radius containing all
halo particles to the radius where the maximum circular
velocity is attained, $r_{{\rm v,max}}$. We see that the
haloes flagged as ``potential splits'' have a distinct
$r_{\rm halo}/r_{{\rm v,max}}$ distribution (\textit{blue
shaded curve}) compared with the typical \textsc{CompaSO}
population (\textit{red shaded curve}). This indicates
that looking at ratios of the inner and outer halo radii
can help us diagnose problematic cases, since for these
haloes their centers are typically not correctly identified.}
\label{fig:split}
\end{figure}

\begin{table*}
\begin{center}
\begin{tabular}{c c| c c c } 
 \hline\hline
 $\log(M)$ range & Number of haloes & Main Prog. matching at $z = 0.875$ & Main Prog. matching at $z = 1.625$ & ``Potential splits'' \\ [0.5ex] 
 \hline
$11.5 \pm 0.1$ & 5096084 & 94.8\% & 73.1\% & 10.5\% \\ 
$12.5 \pm 0.1$ & 610240 & 96.7\% & 91.4\% & 7.86\% \\ 
$13.5 \pm 0.1$ & 46585 & 96.9\% & 89.9\% & 5.11\% \\ 
$14.5 \pm 0.3$ & 2823 & 96.3\% & 85.3\% & 0.42\% \\ 
all haloes & 24721876 & 89.7\% & 64.2\% & 12.8\% \\ 
 \hline
 \hline
\end{tabular}
\end{center}
\caption{Merger trees statistics for the haloes at $z = 0.5$ 
using the \textsc{AbacusSummit} merger trees associations 
\citep{Bose+2021}. The first two columns of the table show
the mass range selected as well as the number of 
\textsc{CompaSO} haloes in it at $z = 0.5$. The next two 
columns show what percentage of these haloes' main 
progenitors inferred through the ``fine'' tree associations
agree with those inferred using the ``coarse'' tree ones
at $z = 0.875$ and $z = 1.625$, respectively. The last
column displays what percentage of the haloes are identified
as ``potential splits'' by the merger tree algorithm in
each of the mass sets.}\label{tab:mtree_stats}
\end{table*}

Another reliable diagnostic is to examine the
persistence of haloes across longer periods of time. The
simulation products include merger 
histories based upon stringing together consecutive
snapshots and thus obtaining a full merger tree, which we
also dub as ``fine.''
Additionally, one can also utilise the ``coarse'' trees
built over non-adjacent time steps, spanning longer
time epochs. To illustrate, the next two steps of the ``coarse''
merger trees for haloes at redshift of $z = 0.5$ are at
$z = 0.875$ and $z = 1.625$. One of the ``realness'' criteria for
the haloes at $z = 0.5$ we propose is to compare their progenitors
at earlier times and note how often their ``coarse''
and ``fine'' histories point to the same progenitor.
We do this at time steps $z = 0.875$ and $z = 1.625$
and display the results in the first and second columns of
Table \ref{tab:mtree_stats}. At $z = 0.875$, the progenitors
of the haloes at $z = 0.5$ match at $\gtrsim 95\%$ at the
mass scales typically of interest for
creating mock catalogues (i.e. $\log(M) \sim 11.5$ and higher)
and close to 90\% for the entire halo 
population. haloes for which we find a match
can be certified as ``healthy,'' though we should
not be discarding those for which no match,
as there could be multiple reasons for this
(e.g., a halo has orbited inside a larger
companion for a significant portion of its
life and as a result has lost a substantial
amount of mass, which has caused the step-by-step
merger tree to fail at tracing its history).
At $z = 1.625$, the percentages are noticeably lower. For the high-mass objects, they are close to 90\%. However, we
note that on the low-mass end, oftentimes the haloes
have not yet been formed at these earlier redshifts,
so the merger tree ``fine''-to-``coarse'' 
comparison might not be an accurate test of persistence.

There are potentially other ways to spot and diagnose
unphysical haloes. As mentioned above, some objects are
identified by the \textsc{CompaSO} algorithm as
distinct haloes while in reality they may lack a dense core
and might simply be carved out of a nearby companion due to
the SO boundary. For these objects, the ratio between 
radial quantities such as $r_{\rm halo}$ and $r_{\rm v,max}$
will most likely be atypical compared with the rest of the
population at that mass scale. In Fig. \ref{fig:split},
we demonstrate what the distribution looks like for haloes flagged
as ``potential splits'' by the merger tree algorithm
and the rest of the population. We can notice that the ratios
are particularly different for the lower halo masses, which
is also where we expect to find most of the unphysical algorithmic
choices. On the higher mass end, the ``potential
splits'' are more consistent with the general population. 
As discussed, some of the objects
labelled ``potential splits'' may in fact be ``real'' haloes
expelled after orbiting a companion halo (particularly for
haloes with denser cores and more particles), so a more robust
way of ``pruning'' the merger tree would involve
flagging objects with odd ratios of their radial quantities
which have also been identified as ``splits.''
 
\subsection{{Presence of high-velocity particles}}
\label{sec:clean.unbind}

{
A problematic mode for many halo finders is the deblending of close pairs of haloes with a large mass ratio, where particles get assigned to one halo, even though they may not be bound to that halo. Neglecting the presence of these particles may bias the inferred properties of the haloes such as their spins and velocity dispersions. A way to identify such failed cases is by locating particles with high infall velocities. Some configuration-space-based finders address this issue by performing energy-based unbinding, but in the case of \textsc{CompaSO}, we opt not to perform particle unbinding because of both the speed requirement of \Abacus{Abacus} and the problems associated with it, some of which we mention below. Instead, we adopt alternative strategies for cleaning the halo catalogues (see Section~\ref{sec:clean.meth}). In this section, we analyse the cleaned catalogues in terms of their effectiveness in removing haloes with a large presence of high-velocity (``unbound'') particles. We verify that indeed after the cleaning, very few problematic haloes remain in the catalogues. In addition, we find alternative statistics that help us weed out unphysical haloes that were missed in the cleaning.}

{
Energy-based unbinding is typically done to discard particles that are tentatively assigned to a particular (sub)structure, but do not appear gravitationally bound to it. For subhaloes, removing unbound particles is of particular importance, since their particle lists are often contaminated with particles from the host halo due to the relatively lower density of the subhalo. This can influence significantly the inferred subhalo properties such as its mass and dispersion velocity. Similarly, interacting haloes often acquire stray particles whose velocities are too high to keep them bound to their host. Standard unbinding treats every halo individually and removes all particles whose kinetic energies exceed their potential energy.
}

{
However, there are several well-known issues associated with energy-based unbinding. The presence of particles with high velocities inside the boundaries of the halo is often a consequence of close fly-bys with other haloes. Therefore, the high-velocity particles should instead be assigned to a neighbour (which is also the spirit of the competitive assignment algorithm for \textsc{CompaSO} and the cleaning procedure). In standard unbinding, these particles are usually returned to the next level up the halo hierarchy, but when applying unbinding to halo hosts, the unbound are altogether deprived of halo membership rather than merged onto a close neighbour. The unbinding procedure assumes that the halo is isolated, and any effects that arise from outside, e.g. particles on the edges, interactions, tidal effects, are neglected. In particular, even particles in isolated haloes whose orbits take them outside of the arbitrarily defined ``halo radius'' have their ability to return underestimated because the mass contribution from the particles outside is neglected (we address this below). In addition, in the case of interacting haloes (e.g, two haloes orbiting around each other, or falling into each other), calculating the potential of each particle with respect to a single halo would not be reflective of the gravitational forces experienced by the particle. 
Enlarging the scope and considering the global potential of the simulation, on the other hand, is also not the correct quantity to use, as it is large-scale dominated and does not answer the question of whether the particle is bound to a particular halo.  
The simplification of ignoring interactions is also motivated by the enormous computational expense that would arise from taking the entirety of structures in the computational domain into account for each structure anew. A more precise approach involves boosting the gravitational potential to make it a locally meaningful quantity, which can be used to define a binding criterion that incorporates the effect of tidal fields \citep{2021arXiv210713008S}.
}

{
We test the effect of applying energy-based unbinding to the \textsc{CompaSO} catalogues of a small test simulation with box size $L_{\rm box} = 296 \ {\rm Mpc}/h$ and particle mass $M_{\rm part} = 2.1 \times 10^9 \ M_\odot/h$, containing 1.3 million haloes. We also apply the `cleaning' procedure from Section~\ref{sec:clean.meth} and \citet{Bose+2021}, which identifies 1.6\%, or 21000, of the haloes in the simulation as `unphysical'. We perform the process of unbinding iteratively for each halo in its centre of momentum frame of its core, i.e. subtracting $\mathbf{x}_{\rm L2com}$ and $\mathbf{v}_{\rm L2com}$ from the position and velocity of each particle, $\mathbf{x}$ and $\mathbf{v}$. All comoving positions are converted into proper coordinates, and the Hubble flow term is added to the particle velocities, $H(z) (x-x_{\rm L2com})$. We compute the spline potential adopted by \Abacus{Abacus} \citep{Garrison+2021b} at each iteration, discarding the particles with kinetic energies higher than their potential energies. This process continues until no more particles are found to be unbound.
}

\begin{figure}
\centering  
\includegraphics[width=.5\textwidth]{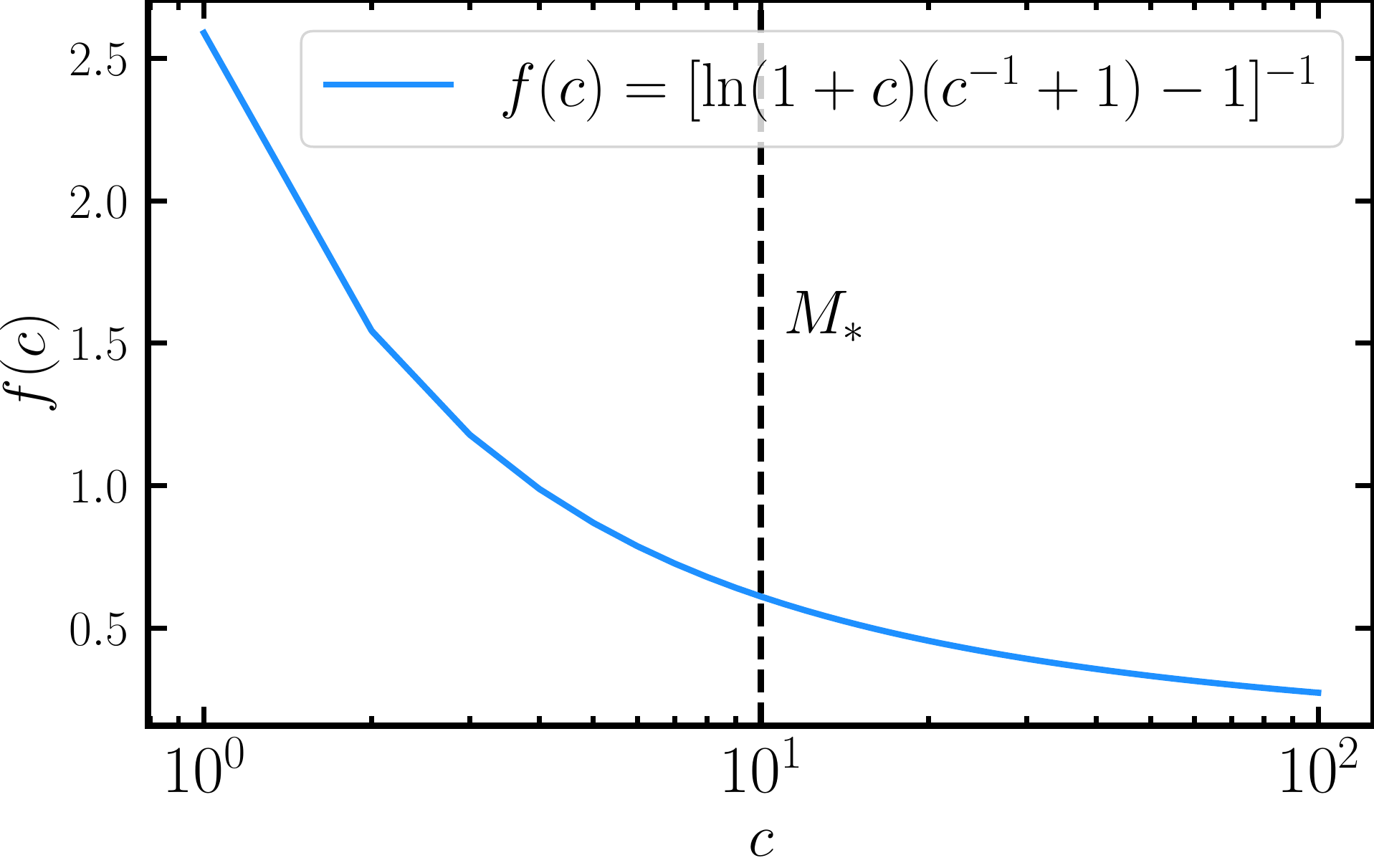}
\caption{Factor $f(c)$ determining the magnitude of the extra potential term as a function of concentration, as derived in Eq.~\ref{eq:f_c}. Less concentrated haloes receive a larger contribution to their potentials from particles lying outside $r_{\rm vir}$ compared with more highly concentrated haloes. The black vertical dashed line shows roughly the concentration of Mily-Way-like haloes. We note that halo concentration and halo mass are weakly anti-correlated.}
\label{fig:f_c}
\end{figure}

{
In addition, we propose a simple, yet well-motivated modification to the standard unbinding algorithm, which ameliorates the issue of the arbitrary drawing of halo boundaries. Let us imagine an isolated halo whose density profile follows an NFW distribution. Typically, we define the halo boundaries, $r_{\rm vir}$, by requesting that the enclosed density is $X \times \rho_{\rm crit}$. In the case of \textsc{CompaSO}, the `virial mass' is defined using the \citet{1998ApJ...495...80B} scaling relation (see Section~\ref{sec:meth.opt}). Nevertheless, the particles outside $r_{\rm vir}$ still contribute to deepening the potential well of the halo despite not strictly belonging to it. Not accounting for this outside contribution would lead us to falsely discard bound particles that, as a result of the deepening of the potential, are in thermal equilibrium despite their higher velocities. This extra contribution to the potential, $\phi_{\rm extra}$, assuming a spherical halo, is computed below for an NFW profile. 
}
\begin{equation}
   \phi_{\rm extra} = \int_{r_{\rm vir}}^\infty \frac{4 \pi G \rho_{\rm NFW}(r) r^2}{r} dr = -\frac{G M_{\rm vir}}{r_{\rm vir}} \frac{c/(c+1)}{A_{\rm NFW}} ,
\end{equation}
where $A_{\rm NFW} \equiv \ln{(1+c)} - c/(1+c)$, $x \equiv r/r_{\rm vir}$, and $c$ is the halo concentration. Simplifying this, we get:
\begin{equation}
    \phi_{\rm extra} = - \frac{G M_{\rm vir}}{r_{\rm vir}} [\ln({1+c})(c^{-1} + 1) - 1]^{-1} \equiv - \frac{G M_{\rm vir}}{r_{\rm vir}} f(c) 
\label{eq:f_c}
\end{equation}
{
and show the function $f(c)$ in Fig.~\ref{fig:f_c}. We note that in the case of elliptical haloes, the assumption of spherical shapes tends to underestimate the contribution of the outside potential and its ability to stabilise the particles on the edges of the halo. Nevertheless, to approximately account for this effect, we add the term $\phi_{\rm extra}$ to the potential energy of each halo when removing unbound particles. As demonstrated in Fig.~\ref{fig:density}, the truncated NFW profile is a good approximation to the \textsc{CompaSO} halo profiles. We thus obtain the halo concentration $c$ by fitting an NFW profile to every halo. For haloes for which a fit cannot be obtained (mostly small haloes, making up less than 3\%), we approximate the concentration as $r_{100}/r_{10}$.
}

\begin{figure}
\centering
\includegraphics[width=.48\textwidth]{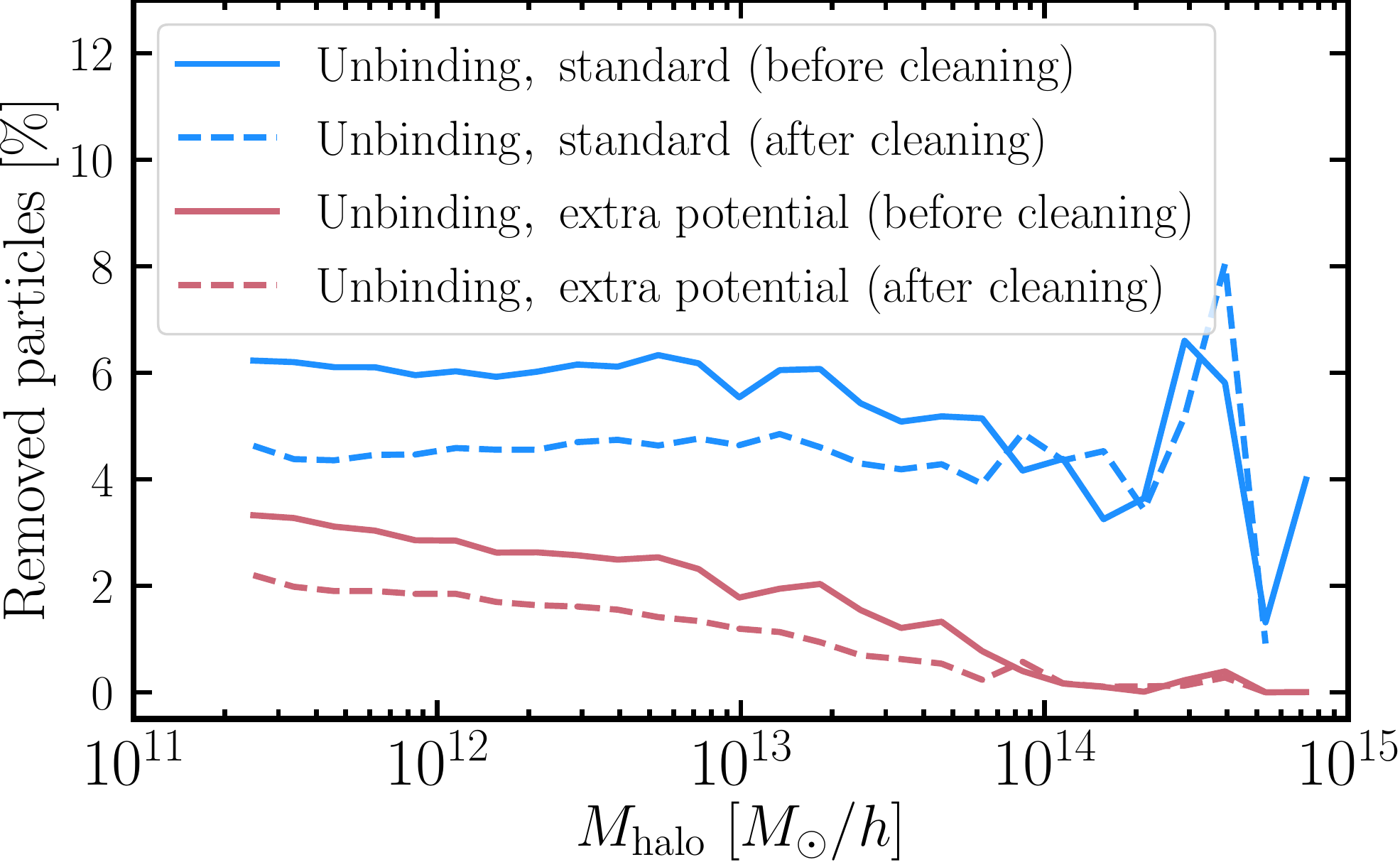}
\includegraphics[width=.48\textwidth]{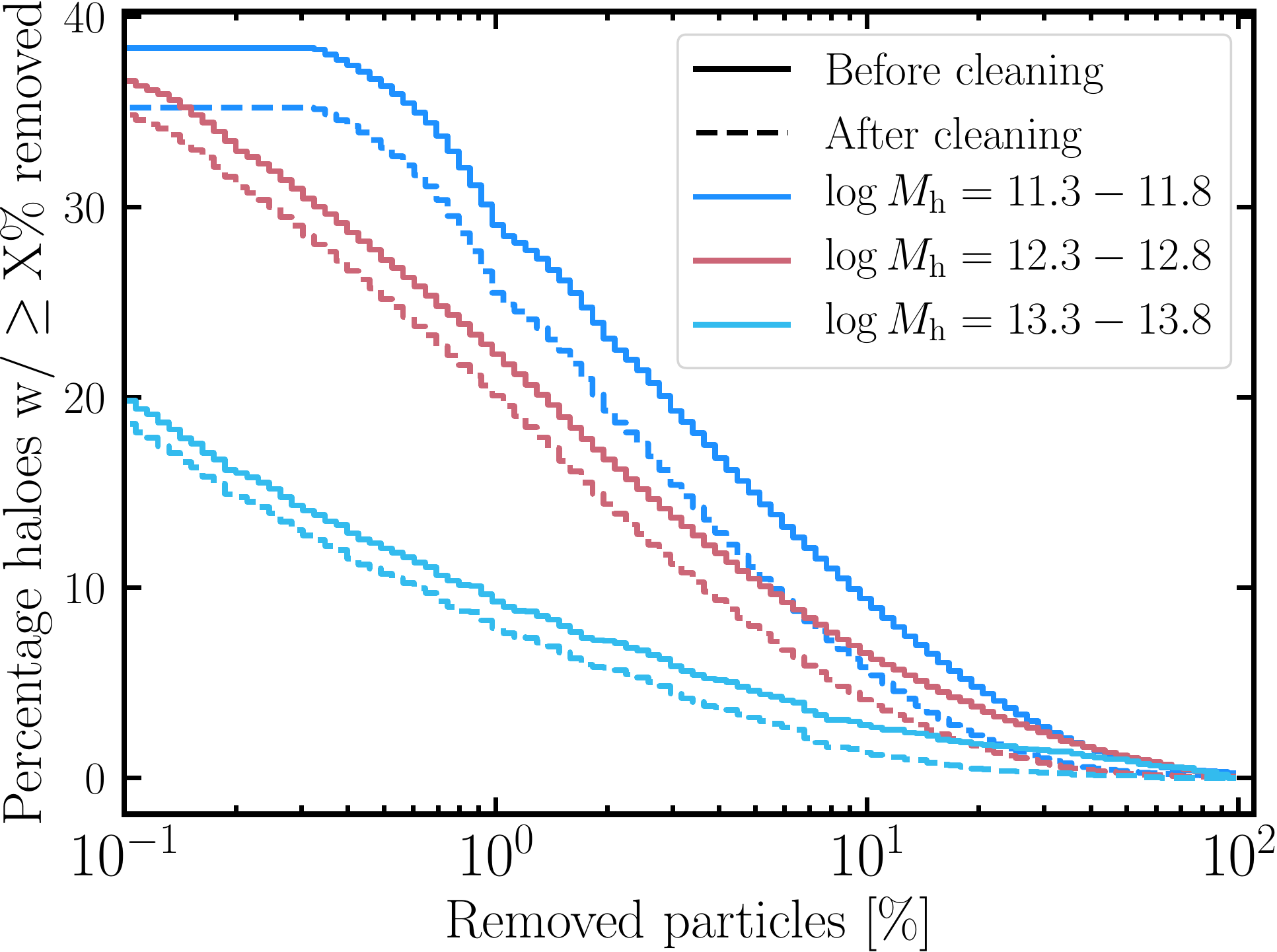}
\includegraphics[width=.48\textwidth]{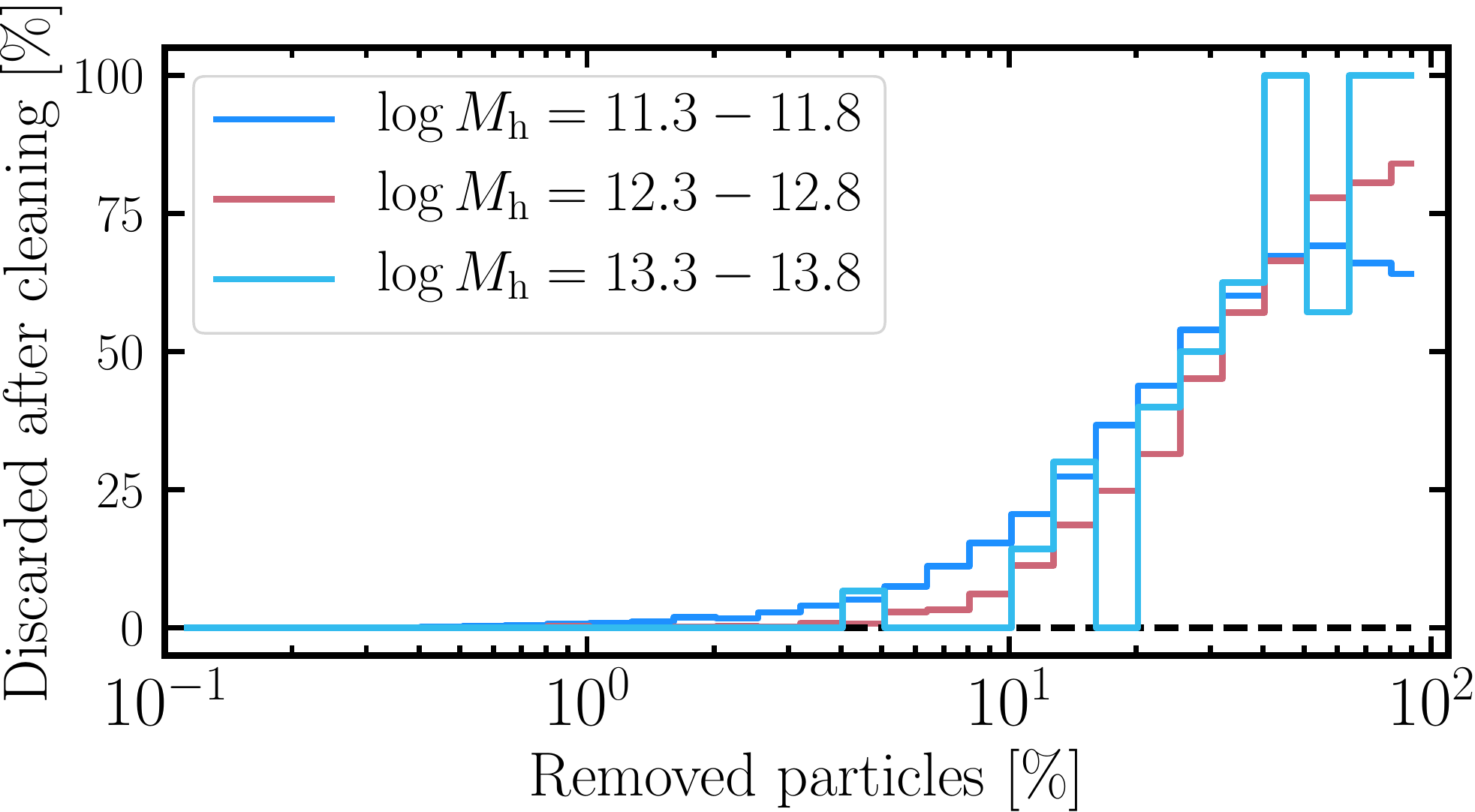}
\caption{\textit{Top panel:} Average percentage of removed particles through unbinding as a function of halo mass. In blue, we show the result when performing the standard cleaning procedure (see Section~\ref{sec:clean.meth} and \citet{Bose+2021}), whereas in red, we show the outcome when adding an extra term to the potential energy, $\phi_{\rm extra}$ (see Eq.~\ref{eq:f_c}). The percentage of removed particles is significantly reduced (\textit{top panel}) when including $\phi_{\rm extra}$ -- by a factor of $\sim$ 2 for smaller haloes and $\sim$ 10 for higher-mass haloes. Similarly, when we remove the discarded haloes (dashed lines), the average percentage of removed haloes drops further. The objects that are cleaned away make up 1.6\% of all haloes in the simulation. \textit{Middle panel:} Cumulative percentage of haloes as a function of removed particles before and after cleaning for three different mass bins: $\log(M_{\rm halo}) = 11.3 - 11.8$, $\log(M_{\rm halo}) = 12.3 - 12.8$, and $\log(M_{\rm halo}) = 13.3 - 13.8$. We see that 38\%, 36\% and 30\% of the before-cleaning haloes for these mass bins have more than 0.1\% of their particles removed, while 10\%, 7\% and 3\$, respectively, have more than 10\% removed. The cleaning reduces these numbers by several per cent.
\textit{Bottom panel:} Percentage of retained haloes after cleaning as a function of removed particles for three different mass bins. It is encouraging to see that the cleaning weeds out an increasingly larger number of haloes (close to 100\% for the highest mass bin shown), as we consider haloes with an increasingly large number of ``unbound'' particles.}
\label{fig:removed}
\end{figure}

{
In the top panel of Fig.~\ref{fig:removed}, we show the averaged percentage of removed particles as a function of halo mass when we apply the standard unbinding procedure and the $\phi_{\rm extra}$-augmented procedure (see Eq.~\ref{eq:f_c}). We see that the standard procedure removes 6.5\% of the particles in smaller haloes and 4\% of the particles in higher-mass haloes. These number are significantly reduced when including the $\phi_{\rm extra}$ term, to 3.5\% and 0\%. Additionally, we discard the haloes considered `unphysical' by the cleaning procedure of Section~\ref{sec:clean.meth} and \citet{Bose+2021}, noting that the average percentage of removed particles is brought down by nearly 1-2\% for $M_{\rm halo} < 10^{14} \ M_\odot/h$. This implies that there is an overlap between haloes with a large percentage of removed particles and the haloes that were cleaned away. To test this, in the middle panel, we show the cumulative percentage of haloes with more than X\% ``unbound particles'' for three mass bins and find that the majority of haloes have less than 0.1\% removed particles, with higher mass bins performing better. Overall, as also illustrated in the bottom  of Fig.~\ref{fig:removed}, the cleaning makes a big difference, removing very effectively haloes that have more than 10\% ``unbound particles''. For the higher masses, the cleaning is even more efficient and removes up to 100\% of the haloes with a large unbound particle population. If the cleaning procedure from Section~\ref{sec:clean.meth} is regarded as reliable, then this finding suggests that the unbinding procedure augmented with an extra potential term yields a more robust halo catalogue compared with the standard unbinding procedure.
}

\begin{figure}
\centering  
\includegraphics[width=.48\textwidth]{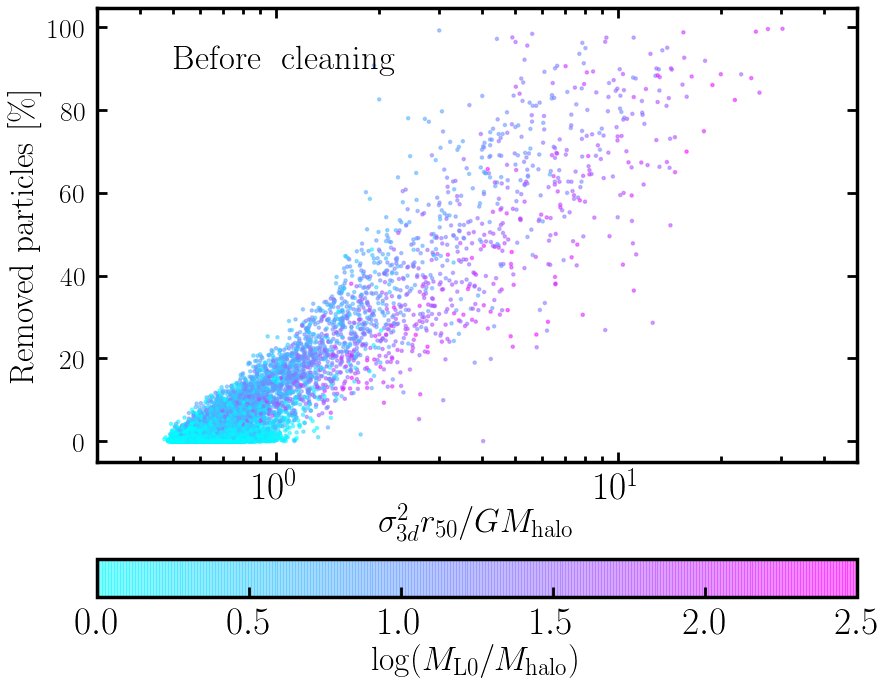}
\includegraphics[width=.48\textwidth]{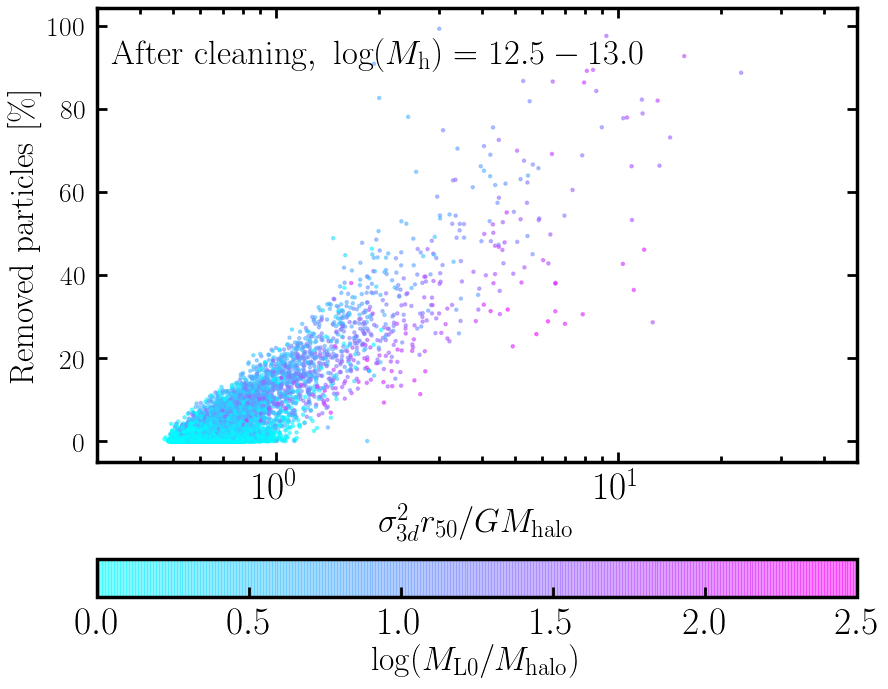}
\includegraphics[width=.48\textwidth]{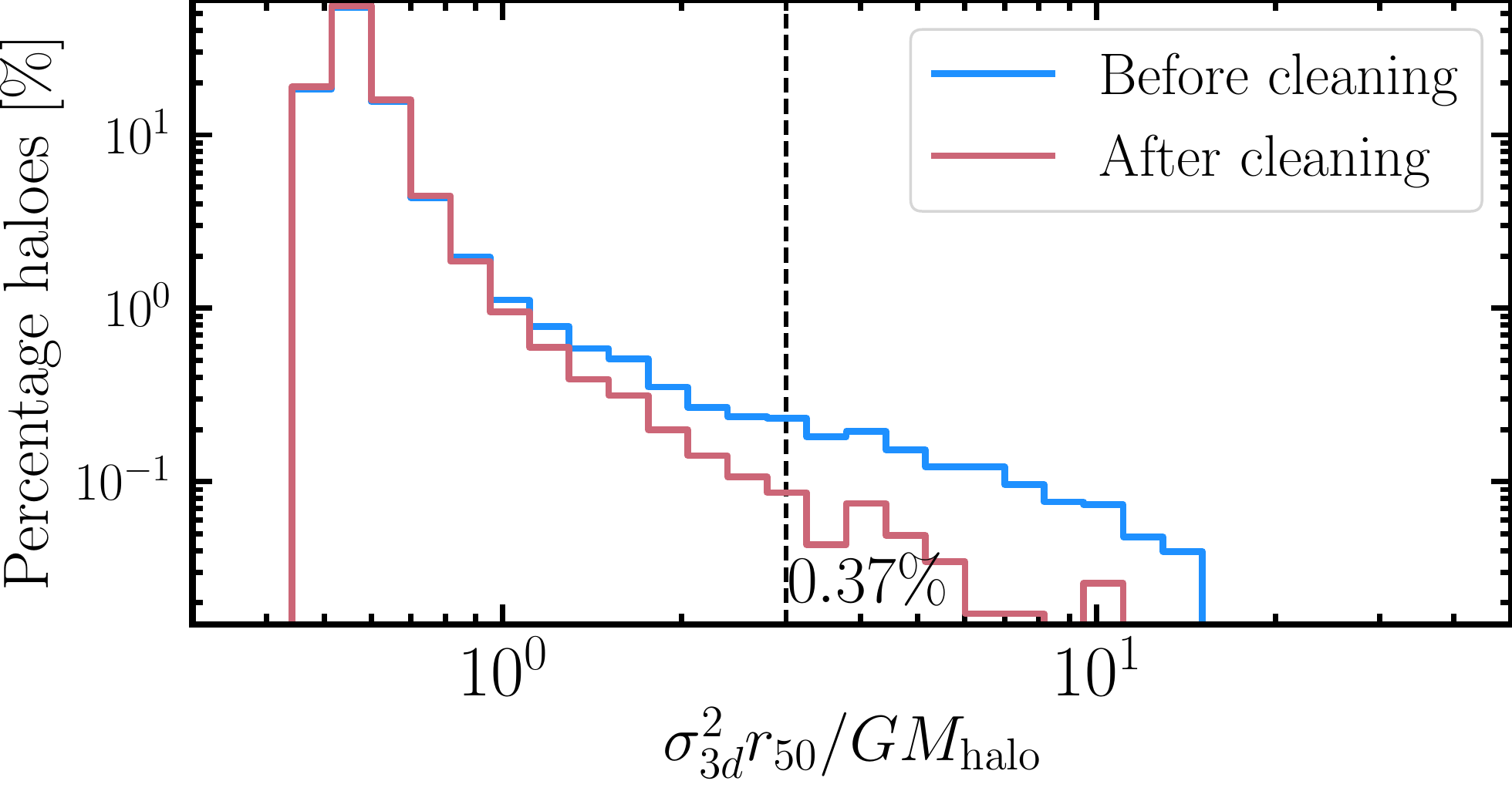}
\caption{Coloured scatter plot of the percentage of removed particles due to unbinding as a function of the virialisation parameter, defined as $\sigma^2_{3d} r_{50}/G M_{\rm halo}$. The colour denotes the ratio of the mass of the ``L0'' parent group and the halo mass, $\log(M_{\rm L0}/M_{\rm halo})$, with purple corresponding to a high ratio and blue corresponding to a low ratio. The plot shows all haloes in the range $\log M_{\rm halo} = 12.5 - 13.0$, measured in units of $M_\odot/h$, before applying the cleaning procedure (\textit{top panel}) and after (\textit{middle panel}). We see a strong positive correlation between the virialisation ratio and the percentage of removed particles, suggesting that a viable mechanism for cleaning the halo catalogues in addition to the default \textsc{CompaSO} cleaning may involve discarding haloes with a high virialisation parameter ($\gtrsim$ 2-3). Furthermore, haloes with a high ratio $\log(M_{\rm L0}/M_{\rm halo})$ tend to lose a higher percentage of their particles to unbinding, indicating that the most problematic mode of \textsc{CompaSO} is the assignment of particles to haloes in a cluster. In \textit{the bottom panel}, we show the percentage of haloes as a function of virial ratio before and after cleaning. Haloes to the right of the dashed black line account for 0.37\% of all haloes after cleaning in the mass bin. This demonstrates that the cleaning takes care of the majority of bad actors and leftover objects can be discarded by combining output statistics such as the virialisation parameter and the cluster mass ratio.}
\label{fig:vir_rem}
\end{figure}

{
In addition, we test the conjecture that unphysical objects (e.g., haloes with a bimodal profile or haloes lacking a nucleus) can be identified through other internal halo properties that are already being output. In Fig.~\ref{fig:vir_rem}, we consider two such quantities. The first one is the virial ratio $\sigma_{3d}^2 r_{50}/G M_{\rm vir}$, where $\sigma_{3d}$ is the three-dimensional dispersion velocity. Fig.~\ref{fig:vir_rem} indicates that there is a very strong correlation between haloes with a large percentage of unbound particles and a large virial ratio. One can therefore imagine augmenting the cleaning procedure with a requirement that the halo virial ratio does not exceed a certain threshold such as 2 or 3, which as seen in the figure for the bin $\log(M_{\rm halo}) = 12.5-13.0$, gets rid of the majority of offenders. In fact, after performing the cleaning, only 1.14\% of the haloes with mass $\log(M_{\rm halo}) = 11.5-12.0$, 0.63\% of the haloes with mass $\log(M_{\rm halo}) = 12.0-12.5$, 0.37\% of the haloes with mass $\log(M_{\rm halo}) = 12.5-13.0$, 0.19\% of the haloes with mass $\log(M_{\rm halo}) = 13.0-13.5$, 0.038\% of the haloes with mass $\log(M_{\rm halo}) = 13.5-14.0$, and none above $\log(M_{\rm halo}) = 14.0$, are retained that have virial ratio above 3. All masses are reported in units of $M_\odot/h$.
}

{ 
We conjecture that this is due to issues that \textsc{CompaSO} has with deblending haloes in close proximity and in particular, with determining the boundaries of small haloes near larger ones. To test this, we have colour-coded the points in Fig.~\ref{fig:vir_rem} by the logarithmic ratio of the ``L0'' parent group mass and the halo mass, $\log(M_{\rm L0}/M_{\rm halo})$, with small haloes living in large halo clusters having large $\log(M_{\rm L0}/M_{\rm halo})$ ratio. Fig.~\ref{fig:vir_rem} shows that the vast majority of haloes with high percentage of removed particles (and thus high virial ratio), have high ratio $\log(M_{\rm L0}/M_{\rm halo}) > 1.5$. A sensible and easy solution to this issue is thus to discard these small haloes rather than try to remerge them, as their contributions to the masses of large nearby structures are negligible. Additionally, we perform visual inspection of hundreds of these objects and find that the majority of the leftover haloes have well-defined nuclei with extended envelopes that likely contain some number of particles that are gravitationally bound to nearby massive structures.
}

{
We conclude that the cleaning mechanism (from Section~\ref{sec:clean.meth} and \citet{Bose+2021}) takes care of the vast majority of unphysical objects and that remaining wrongdoers can be identified and discarded (or remerged into the main progenitor branch) through their higher virial and/or mass ratios during postprocessing.
}

\section{Halo analysis}
\label{sec:anal}
In this section, we analyse the properties of the
haloes derived through the \textsc{CompaSO} algorithm 
using the \texttt{AbacusSummit\_highbase\_c000\_ph100} 
simulation\footnote{Further
details about these simulations can be found at 
\url{https://abacussummit.readthedocs.io/en/latest/data-products.html}}.
This simulation contains
$N_{\rm part} = 3456^3$ dark matter particles in a box of size 
$L_{\rm box} = 1$ Gpc$/h$, which corresponds to a
particle mass resolution of $M_{\rm part} = 2.109 \times 10^9 \ M_\odot/h$. 
For the subsequent investigation of halo properties,
we have chosen to focus on the redshift slice at
$z = 0.5$ and have also obtained an alternative
halo catalogue in post-processing via the
phase-space halo finder \textsc{ROCKSTAR}. We compare the two
halo finders in terms of their halo mass functions,
auto- and cross-correlation functions, and radius-mass
relationships. We further study the density profiles of
the \textsc{CompaSO} haloes and alternative definitions
of their halo centers. In the mass
range $M=10^{12}-10^{15} M_\odot/h$, the \textsc{ROCKSTAR}
halo catalogue contains a total of haloes of 4,137,230,
while \textsc{CompaSO} has 4,016,661. 
Throughout this section, the masses of objects are
presented in units of $M_\odot/h$.

Short descriptions of the relevant quantities for the
halo analysis are displayed in Table \ref{tab:names} along
with the corresponding notation for the halo catalogue fields in
the \textsc{AbacusSummit} output products.

\subsection{Halo mass function}
\label{sec:anal.hmf}
The halo mass function is central in cosmology, as it
describes how many haloes of a given mass exist at a 
given redshift. Obtaining this information
is important since 
dark matter haloes play an essential role in
modeling galaxies and galaxy clusters 
and thus, in studying large-scale structure. 
The halo mass function allows us
to understand the statistics of primordial matter
inhomogeneities and is also used to compute 
the effects of nonlinear structure on 
observations through, for instance, the
Sunyaev-Zeldovich effect and lensing. Another
feature of the halo mass function is that
it can be expressed as a universal function 
that relates the mass of haloes to the variance of
the mass fluctuations (e.g. \citep{1974ApJ...187..425P,1999MNRAS.308..119S,2001MNRAS.321..372J,2001ApJ...550L.129W,2005Natur.435..629S,2006ApJ...646..881W}.

For our comparative analysis of halo algorithms,
we compute the halo mass function at redshift
$z = 0.5$ using both the \textsc{ROCKSTAR} and \textsc{CompaSO}
algorithms as
\begin{equation}
dN_{\rm halo}(M, z) \equiv N_{\rm halo}(M, z) \, d \log M , 
\end{equation}
where we have defined $dN_{\rm halo}$ as the number of haloes
in the mass range $d \log M$.

\begin{figure}
\centering  
\includegraphics[width=.48\textwidth]{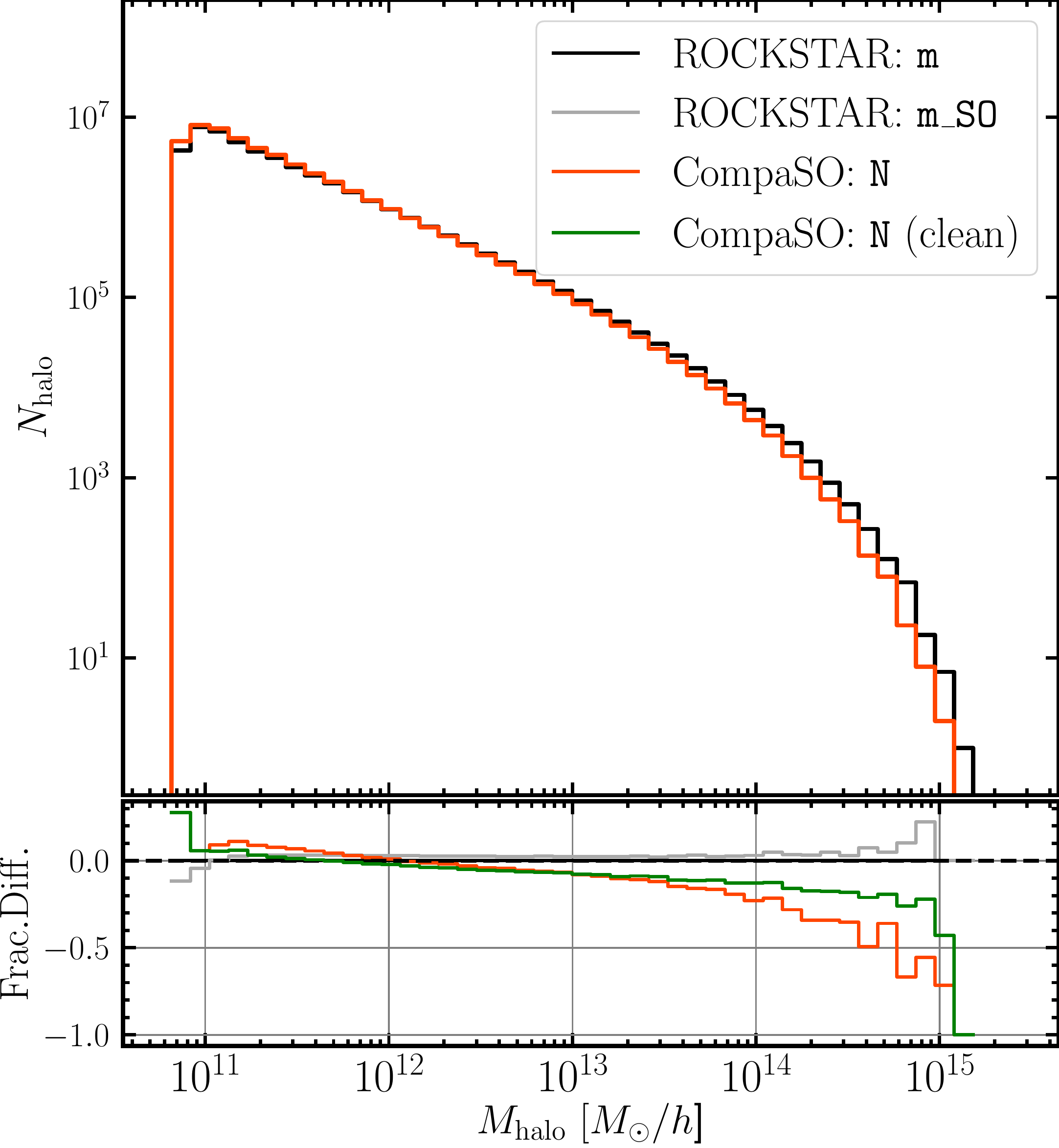}
\includegraphics[width=.48\textwidth]{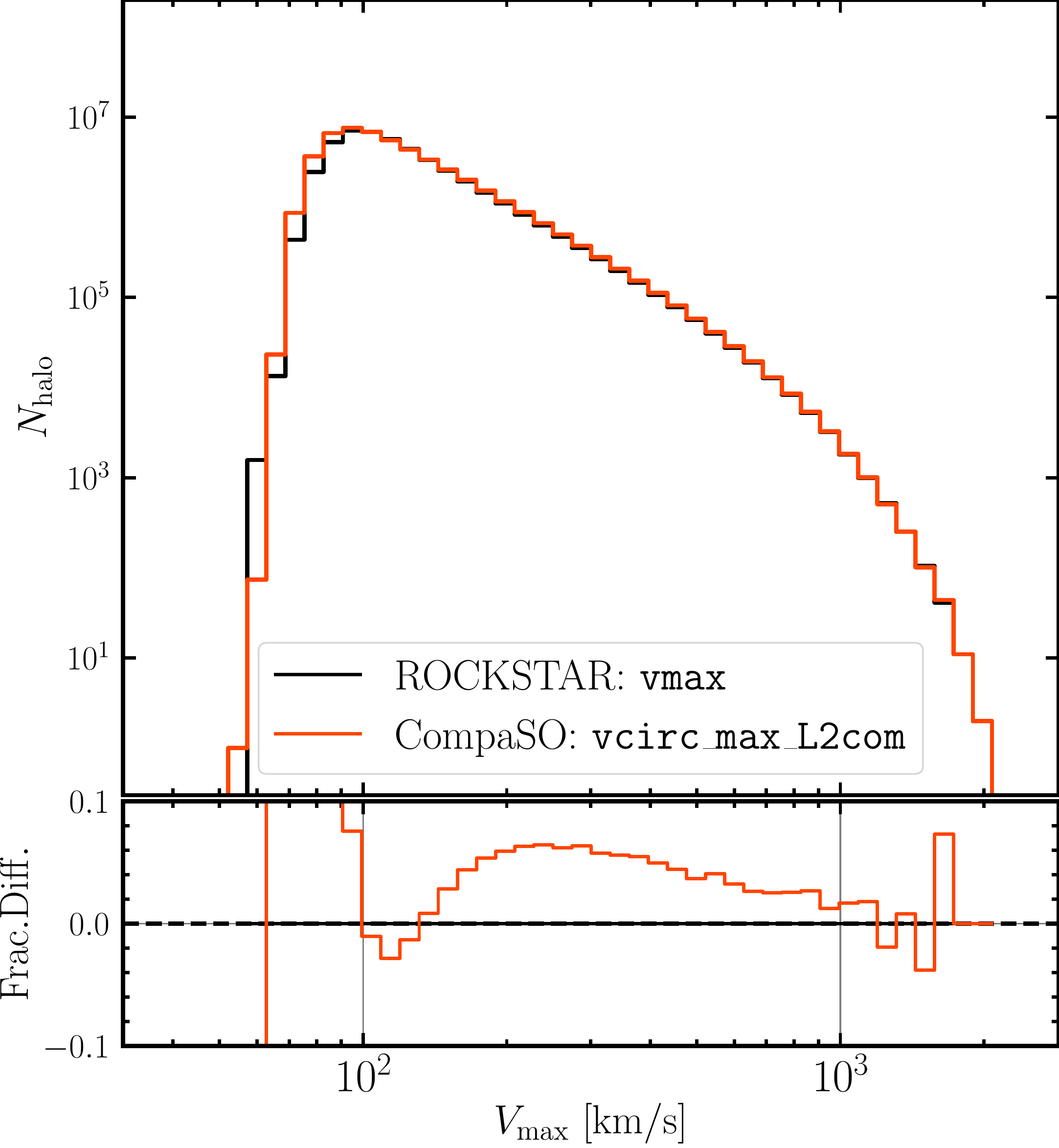}
\caption{Halo mass functions of the \textsc{ROCKSTAR} and \textsc{CompaSO} groups. The \textit{upper panel} shows the traditional halo mass function computed as the number of haloes as a function of halo mass for both finders. The \textit{black curve} uses the `virial' (default) mass definition of the \textsc{ROCKSTAR} finder, whereas the \textit{gray} curve uses the `strict SO' definition for the same choice of virial density. The \textit{red curve} uses the \textsc{CompaSO} halo field \texttt{N}, corresponding to the total number of particles, multiplied by the particle mass, while the \textit{green curve} uses cleaned haloes \citep[see][for details]{Bose+2021}. The lower segment shows the fractional differences between each curve and \textsc{ROCKSTAR}. \textsc{CompaSO} haloes exhibit a small excess ($\lesssim 10\%$) of low-mass haloes compared with \textsc{ROCKSTAR} and a deficiency of high-mass objects. The cleaned \textsc{CompaSO} catalogues are in a better agreement with the \textsc{ROCKSTAR} halo mass function, exhibiting only half the discrepancy observed compared with the raw catalogues. The \textit{lower panel} shows the number of haloes as a function of the maximum circular velocity. We see that the two halo functions are in a considerably better agreement -- within $\sim$5\% on all scales, suggesting that the two algorithms find the same objects for the most part.}
\label{fig:hmf}
\end{figure}

The halo mass
functions of \textsc{CompaSO} and \textsc{ROCKSTAR} are
presented in Fig. \ref{fig:hmf} adopting virial mass in the top panel and maximum circular velocity in the bottom.
The lower segment of each panel shows the fractional difference between the two curves with respect to the black \textsc{ROCKSTAR} curve.
We note that we have imposed a mass cut of $M_{\rm min} = N_{\rm L1,min} \times 
M_{\rm part} = 7.4 \ 10^{10} \ M_\odot/h$, where $N_{\rm L1,min} = 35$, on the \textsc{ROCKSTAR} haloes, since we output CompaSO haloes of only roughly that mass and above. We note that the difference in the choice of minimum mass accounts for the majority of the observed discrepancy of the halo mass function for the lowest-mass haloes ($\log(M_{\rm halo}) < 11$).

In the upper panel, the masses of the \textsc{ROCKSTAR} haloes are obtained by using the default `virial' mass, $M_{\rm vir}$, {which similarly to \textsc{CompaSO}},
adopts the fitting function of \citet{1998ApJ...495...80B},
while the raw \textsc{CompaSO} masses (in red) are defined as the
total number of particles in a halo (\texttt{N}, see Table \ref{tab:names}) 
multiplied by the particle mass. We also show in green the cleaned \textsc{CompaSO} catalogues (see Section \ref{sec:clean.meth}).
We can see that the \textsc{CompaSO} haloes exhibit a small excess ($\lesssim 10\%$) of low-mass haloes near $10^{11} \ M_\odot/h$ compared with \textsc{ROCKSTAR}.
On the high-mass
end ($\sim 10^{15} \ 
M_\odot/h$), the \textsc{CompaSO} catalogue displays a ($\sim 50\%$) deficiency of high-mass objects near $10^{15} \ M_\odot/h$ compared with \textsc{ROCKSTAR}. These differences are nearly two times smaller when using the cleaned \textsc{CompaSO} catalogues \citep[see][for a discussion]{Bose+2021}. The remaining difference is most likely
the result of the choice of density threshold 
($\Delta_{\rm L1}$) and the competitive assignment we impose on
the \textsc{CompaSO} haloes as described in Section \ref{sec:meth.algo}.
In the mass range where the vast majority of haloes
reside ($\sim 10^{11} - 10^{14} \ M_\odot/h$), the
agreement is good (within 25\%) for our particular application
of creating mock catalogues via empirical models such as the
HOD model, although there is a
slight excess of smaller haloes in the cleaned \textsc{CompaSO} catalogue.
This finding, together with the relative paucity of large
haloes, may be an indication of a tendency of 
\textsc{CompaSO} to oversplit large DM structures into
central haloes surrounded by multiple satellites in
their outskirts. We examine this further in Section
\ref{sec:anal.corr} and Section
\ref{sec:clean.persist}.
We also show the resulting fractional difference when using the `strict SO' \textsc{ROCKSTAR} mass definition (see \textsc{ROCKSTAR} \href{https://bitbucket.org/gfcstanford/rockstar/src/master/README.md}{documentation} for more details) for the same virial density choice, demonstrating that our curves of the
halo mass function are strongly dependent
on the mass definition adopted. 

In the lower panel, we show the number of haloes as a function of another popular choice for halo mass proxy, the maximum circular velocity $V_{\rm max}$. We can now see that the two halo functions are in a considerably better agreement -- roughly within 5\% of each other on all $V_{\rm max}$ scales. This finding suggests that the two halo algorithms find the same objects for the most part and that the observed differences in the upper panel can be largely attributed to halo boundary choices and mass definitions. 

\subsection{Density profile}
\label{sec:anal.prof}
\begin{figure*}
\centering  
\includegraphics[width=1\textwidth]{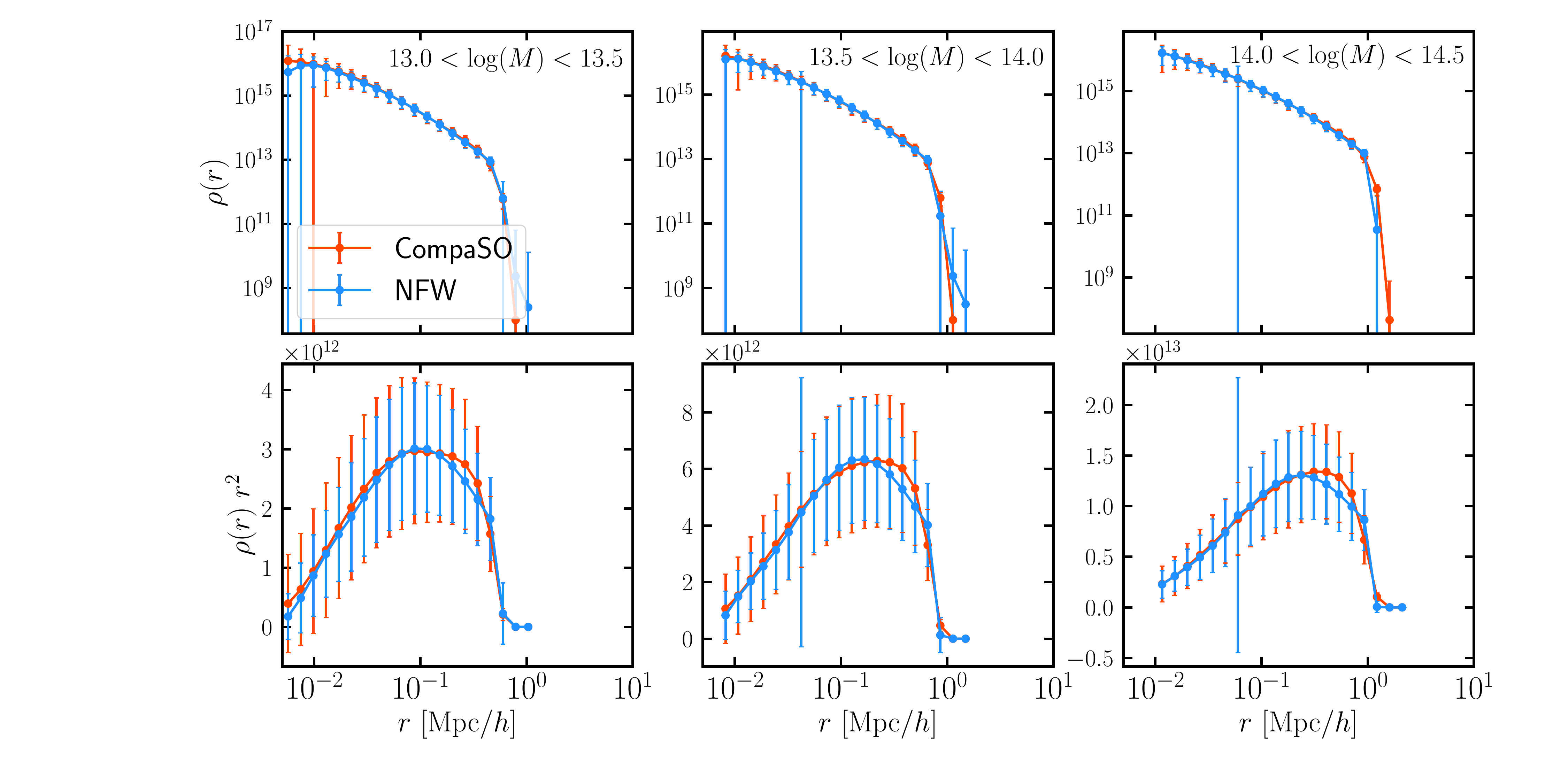}
\caption{Average density profiles in three mass bins: 13 < $\log M$ < 13.5 (\textit{left panels}), 13.5 < $\log M$ < 14 (\textit{middle panels}), and 14 < $\log M$ < 14.5 (\textit{right panels}), along with their respective averaged NFW fits. The upper panels shows the density profile, $\rho(r)$ in logarithmic space, while the lower panels show $\rho(r) r^2$, which is helpful for illustrating the mass differential ($dM/dr$) as well as the ``knee'' of the density profile, which is directly related to the concentration. The NFW-fitted concentration values for the three mass bins shown are $c_{\rm NFW} = [7.7, 7.6, 5.2]$, respectively.} 
\label{fig:density}
\end{figure*}

According to \citet{1996ApJ...462..563N},
the density profiles of dark matter haloes in high-resolution
cosmological simulations for a wide range
of masses and for different initial power spectra of density fluctuations
are well fitted by the formula
\begin{equation}    
    \rho_{\rm NFW}(r) = \frac{\rho_{\rm s}}{(r/r_{\rm s})\,(1+r/r_{\rm s})^{2}}
\end{equation}
with $\rho_{\rm s}$ being the characteristic
density and $r_{\rm s}$ the scale radius defined as
\begin{equation}     
    r_{\rm s} = \frac{r_{\rm vir}}{c} \ ,
\end{equation}
where $r_{\rm vir}$ is the virial radius usually defined as the distance from the
center of the halo to the outer boundary within which
the mean density is $\sim 200$ times the
present-day mean or critical density. 

The concentration parameter introduced above
describes the
shape of the density profile, and in particular
where the ``knee'' of the curve is located.
From cosmological
$N$-body simulations, extended 
Press-Schechter theory, and the spherical
infall model, \citep{2000MNRAS.311..423L,2001MNRAS.321..559B}, we know 
that $c$ depends on the mass of the
object and the form of the initial power
spectrum of the density fluctuation.
A well-known result is that higher mass objects
have a lower concentration parameter and vice versa \citep[e.g.][]{2014MNRAS.441.3359D}.

To study the particle distributions of the \textsc{CompaSO} haloes,
we have computed the spherically-averaged density
profiles of all haloes
by binning the halo mass in equally spaced bins in
log-space, between the virial radius and 
$\log(r/r_{\rm vir}) = -2$.
We are using 20 bins, which correspond to
$\Delta \log(r) = -0.1$ and are sufficient to produce
robust results. We compute the density profiles of
haloes in the following 3 mass bins:
$13 < \log M < 13.5$, $13.5 < \log M < 14$ and
$14 < \log M < 14.5$ together with fits using the NFW
profile. In this form, the NFW profile has two free 
parameters, $\rho_s$ and $c$, which are both
adjusted through a least-squares minimization 
between the binned $\rho(r)$ and the NFW profile,
\begin{equation}
\sigma^2 = \sum_{i=1}^{N_{\rm bins}} [\log(\rho(r_i))-
\log(\rho_{\rm NFW} (r_i,[\rho_s,c]))]^2,
\end{equation}
where $r_i$ are the radial bins of the density
profile, and $\rho_s$ and $c$ are the two
parameters we are fitting for.
Note that here we assign an equal weight to each bin.

The averaged density profiles and NFW fits of the haloes 
in the three mass bins are presented
in Fig. \ref{fig:density}.
The top panels are displaying $\rho(r)$ as
a function of distance with both axes in logarithmic space,
while the bottom ones have $\rho(r) r^2$ as a function of
distance. The latter are useful for seeing where
the ``knee'' of the profiles is, while the former
exemplify the typical power law expected from the NFW
formalism and $N$-body simulations using alternative halo
finders. We can see that the averaged NFW
fits provide reasonably good fits to the halo profiles
within the expected standard deviations across all haloes in
each mass bin. The NFW-fitted concentration values for the
three mass bins are $c_{\rm NFW} = [7.7, 7.6, 5.2]$, 
respectively, in accordance with the general expectation
that concentration should decrease with mass. 
There are a few radial bins for which
the error bars are substantially larger. Those are due
to individual haloes whose density profiles follow an
atypical form and thus, no good NFW fit has been found
for them.

\subsection{Two-point halo correlation functions}
\label{sec:anal.corr}

The two-point statistic is one of
the most important probes of clustering
in configuration space, and the main 
ingredient for the computation of
the higher point statistics.
The two-point correlation function, $\xi(r)$, is 
defined as the excess probability compared with a
Poisson process of finding two objects at a certain
spatial separation, $r$, \citep[e.g.][]{2002PhR...367....1B}
\begin{equation}
dP_{12}(r) = \bar n^2 (1 + \xi(r)) \, dV1 \ dV2 
\label{eq:corr}
\end{equation}
where $\bar n$ is the mean density. These objects in
our case are the centers of the dark matter haloes. 
Clustered objects are indicated by $\xi(r) > 0$,
whereas objects are anticorrelated when $\xi(r) < 0$.
Eq. \ref{eq:corr} can also be viewed
as the conditional probability of having an object
at $dV_1$ with a probability of $\bar n dV_1$ and
of finding another object at distance $r$ in $dV_2$.

The auto-correlation functions in this work are 
estimated from $N$-body simulations using the natural
estimator \citep{1980lssu.book.....P}
\begin{equation}
\hat \xi(r) = \frac{\langle DD(r)\rangle}{\langle RR(r) \rangle} - 1
\end{equation}
where $\langle DD(r)\rangle$ and 
$\langle RR(r)\rangle$ are the normalised number of all 
possible pairs within the data and the random set,
respectively. 

Here we examine the two-point correlation
functions of the haloes defined by the \textsc{ROCKSTAR} 
and \textsc{CompaSO} halo finders. We use two versions of
the \textsc{CompaSO} catalogue: a raw version, which is directly taken from the simulation outputs, and a cleaned version,
created for purposes of constructing mock catalogues and measuring
correlation functions \citep[for details about the cleaning procedure, see][]{Bose+2021}.
We do that by splitting the haloes into 5 bins:
$\log M = 11.5 \pm 0.1$, $\log M = 12.0 \pm 0.1$,
$\log M = 12.5 \pm 0.1$, $\log M = 13.5 \pm 0.1$, 
$\log M = 14.5 \pm 0.3$, respectively,
with the mass measured in $M_\odot/h$. We also ensure 
that the number of \textsc{ROCKSTAR}
objects is equal to the number of \textsc{CompaSO} objects
in each set. We do this by rank-ordering the haloes in
\textsc{ROCKSTAR} by mass and selecting all objects
in the range $\log(M_i) \pm [\Delta\log(M)]_i$, where $i$
is the current mass bin. We then similarly
rank-order the \textsc{CompaSO} objects by mass
and select the indices corresponding to the \textsc{ROCKSTAR}
choice, adopting an abundance matching procedure. 
In this way, we can better
normalise the correlation function on large scales
and observe physical differences resulting from the
halo definitions. 

\begin{figure*}
\centering  
\includegraphics[width=1.\textwidth]{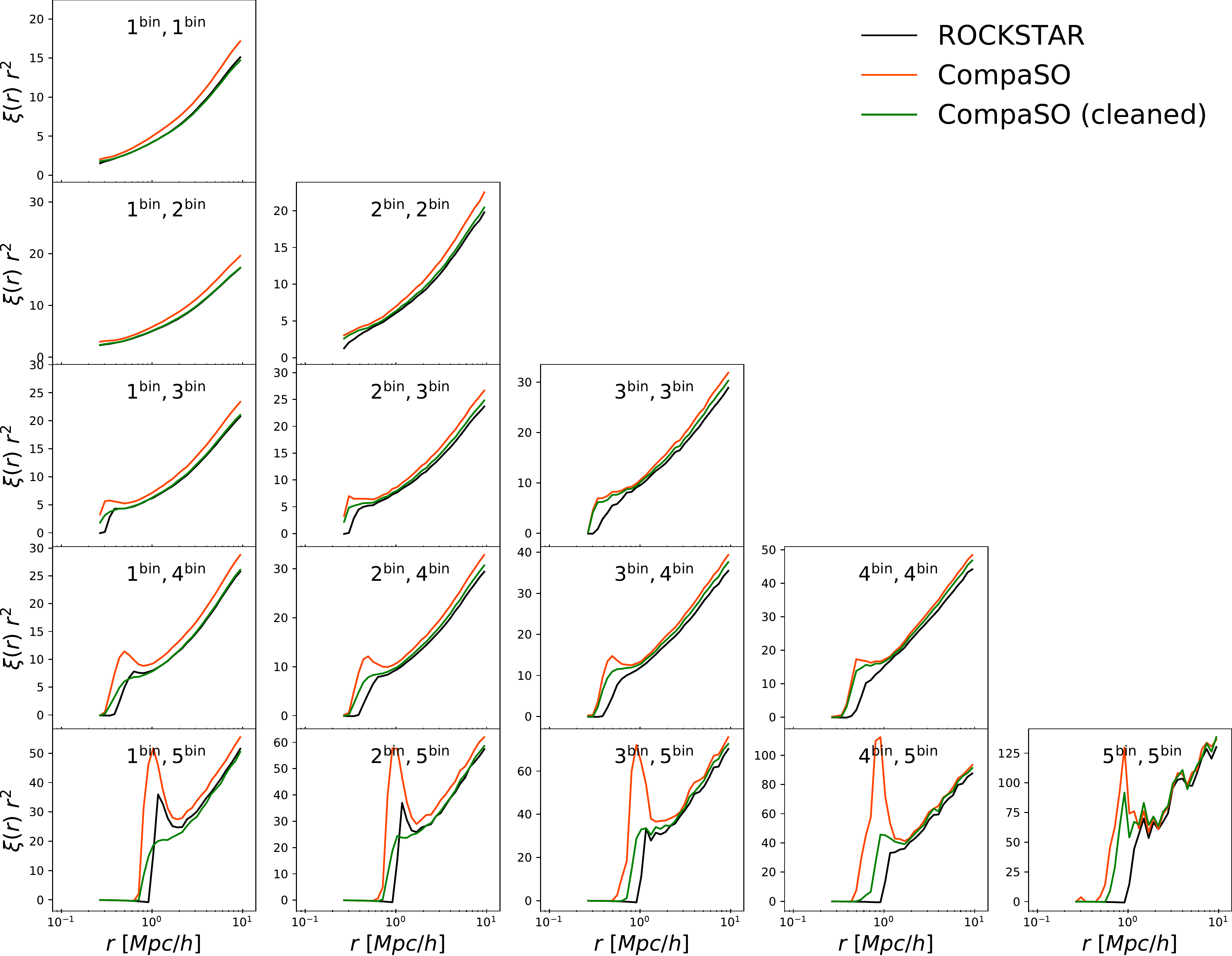}
\caption{Two-point correlation functions
of the \textsc{ROCKSTAR} (in \textit{black}), \textsc{CompaSO} (in \textit{red}), and cleaned 
\textsc{CompaSO} haloes (in \textit{green}). The haloes are split into the following
5 bins:
$1^{\rm st} \ {\rm bin}$: $\log M = 11.5 \pm 0.1$,
$2^{\rm nd} \ {\rm bin}$: $\log M = 12.0 \pm 0.1$, 
$3^{\rm rd} \ {\rm bin}$: $\log M = 12.5 \pm 0.1$,
$4^{\rm th} \ {\rm bin}$: $\log M = 13.5 \pm 0.1$, and
$5^{\rm th} \ {\rm bin}$: $\log M = 14.5 \pm 0.3$,
measured in $M_\odot/h$, and cross-correlations between
each pair are shown. We see that there is an excess
of \textsc{CompaSO} haloes around $\sim 1 \ {\rm Mpc}/h$
relative to \textsc{ROCKSTAR} and the cleaned \textsc{CompaSO} catalogue, which implies that
\textsc{CompaSO} is more likely to identify structures
located in the outskirts of other haloes as separate
haloes. We see a similar trend for the other two halo
samples (\textsc{ROCKSTAR} and cleaned \textsc{CompaSO}) of finding a surplus of haloes
around larger objects,
but it is less pronounced than the raw sample.}
\label{fig:corr}
\end{figure*}

In Fig.~\ref{fig:corr}, we study the cross-correlations between
each pair for the 5 mass bins defined above.
This test serves to show us whether there is a significant
number of haloes lurking on the outskirts
of other, typically larger haloes. The correlation 
functions are given in Fig. \ref{fig:corr} and 
suggest that the \textsc{CompaSO}
halo finder is more likely to identify small dark matter
structures and define them as distinct haloes on the boundaries
of other haloes compared with the \textsc{ROCKSTAR} and
cleaned \textsc{CompaSO} catalogues. We see this effect across all mass
ranges, but most prominently for the highest mass sets, where 
\textsc{CompaSO} identifies the largest amount of substructure
relative to \textsc{ROCKSTAR} (bottom row of the figure).
The cleaned catalogue appears to be more similar to the 
\textsc{ROCKSTAR} one in terms of its halo clustering. It 
seems to contain more examples of large haloes orbiting other large
haloes than \textsc{ROCKSTAR}, which suggests that large objects
are oversplit (bottom right corner) while on the other hand,
\textsc{ROCKSTAR} tends to identify more small haloes
near large haloes (bottom left corner).
Another feature of this plot is that the peaks of the correlation
functions seem to be shifting towards larger scales as
we go to higher-mass haloes (top to bottom), since these 
smaller haloes have correspondingly smaller radii.
On the other hand, the locations of the peaks do not seem to change
significantly for a given row. That is because the
scale of the peaks is set by the average radius of the 
heavier haloes in a given cross-correlation panel.

\subsection{Definitions of halo center}
\label{sec:anal.cent}

\begin{figure*}
\centering  
\includegraphics[width=1.\textwidth]{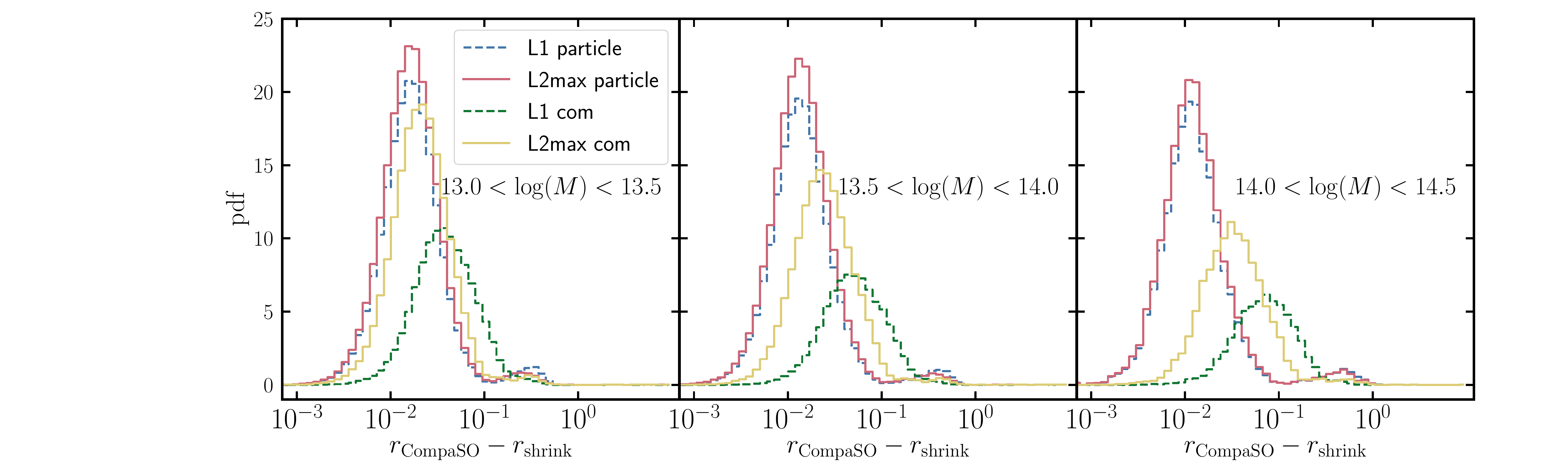}
\caption{Difference between the various halo 
centers output by \textsc{CompaSO}
and the ``shrinking-sphere'' halo center
defined according to the method of \citet{2003MNRAS.338...14P}.
The haloes are split into three different mass bins:
13 < $\log M$ < 13.5 (\textit{left panel}),
13.5 < $\log M$ < 14 (\textit{middle panel}), and
14 < $\log M$ < 14.5 (\textit{right panel}),
measured in $M_\odot/h$. We see that
the center of the first L2 halo, L2max,
(denoted as L2max particle in the figure) is
closest to the shrinking-radius
definition (which we take as a proxy for the
``true'' density peak), while the center of
mass of the L1 group (L1 com) exhibits
the largest amount of scatter and
is centered further away than the rest.
Both particle center definitions, however,
have a small locus of points located around
$\Delta_r \approx 0.5$ Mpc$/h$, which
is a consequence of the densest particle
not coinciding with the halo density peak.
For this reason, we recommend using the
halo properties computed with respect to \texttt{L2max\_com}.}
\label{fig:shrink}
\end{figure*}

Since much of our post-processing analysis relies on
halo properties computed with respect to the halo
centre, it is important to define it carefully. 
For each halo, we store the locations of four 
different halo center definitions: particle center 
of the L1 halo (i.e. the position of the halo nucleus particle), particle 
center of the first L2 halo (L2max),
center-of-mass of the L1 halo and center-of-mass
of the first L2 halo. L2max is the first Level 2 halo
to start forming within a given L1 halo and is typically
also the largest (since the halo nuclei are chosen
to be the particles with densest nuclei as described in
Section \ref{sec:meth.algo}).
We output all halo properties
with respect to both the L1 and L2max
centers-of-mass (L1 com and L2max com, respectively) and recommend
using the L2 com ones. 

Although this
seems like a sensible choice, it is important to 
compare it with other plausible options and
check whether this leads to a substantial
offset from the location of the ``true'' halo center.
We have therefore compared all of these centers
with the ``shrinking-sphere'' center \citep{2003MNRAS.338...14P}, 
which converges
towards the density maximum of the most massive 
substructure independently
of the halo-finding algorithm and is
thus a good proxy for the ``true'' center. 
In the ``shrinking sphere'' algorithm, 
one computes iteratively the center-of-mass of
all particles within a sphere of radius
\begin{equation}
r_i = (1 - 0.05)^i \ r_0, 
\end{equation}
where $i$ is the
iteration index and $r_0$ is the total 
radius of the \textsc{CompaSO} sphere,
rejecting all particles outside the sphere.
The iteration stops when the shrinking sphere 
contains 1\%
of the initial number of particles or
fewer than 50 subsample \texttt{B} particles in the 
halo\footnote{Subsample \texttt{B} consists of 7\% of 
all particles in the simulation, so 50 subsample
\texttt{B} particles corresponds to roughly 700 total
particles}.

We carried out this comparison in three
different mass bins: 
$13 < \log M < 13.5$, $13.5 < \log M < 14$, 
and $14 < \log M < 14.5$ with mass measured in $M_\odot/h$,
and are showing the result in Fig. \ref{fig:shrink}. 
We can see that all curves but the L1 com are sharply
peaking around $10^{-2}$ Mpc$/h$, which is
comparable to the softening length of the simulation
(7.2 kpc$/h$). The L1 com exihibits the largest amount
of scatter, while the L2max particle center is closest to the
``shrinking-sphere'' definition. One can also notice a
second much smaller locus of points located away from
the ``shrinking-sphere'' center. 

Those are likely the
consequence of three effects related to the geometry
of the \textsc{CompaSO} L1 group. The first one is that
if a halo is composed of two or more smaller haloes
of similar mass that are identified as one
by our halo finder, but the center-of-mass
is far from the center of the most massive substructure,
the shrinking sphere may converge to
another slightly less massive substructure.
The second effect may come from the fact that
in some cases the first
L2 subhalo does not end up being the largest (its center
particle has the highest local density) due
to the competitive aspect of our algorithm (e.g. another
L2 halo which is part of the same L1 group ends up
eating up a larger number of the particles). Finally,
sometimes the center of the first L2 subhalo
is the better choice for the group center, as it is identified
as the most dense particle \textit{after} the competitive
assignment takes place and the final particle line-up
for the L1 halo has been decided on.
This methodology could be used to flag and further examine
potentially problematic haloes whose mass
distributions deviate significantly from spherical symmetry.
Another observation from Fig. \ref{fig:shrink} is that
with increasing halo mass the L1 and L2max center-of-mass
definitions seem to exhibit larger scatter,
while the L1 and L2max particle centers do not get affected.
This finding makes sense as the center-of-mass for these
larger objects is derived by averaging the positions of
more particles. In addition, high-mass haloes are expected
to have less spherical shapes, which may further bias the
center-of-mass measurements.

These findings corroborate our recommendation
to use quantities computed with respect to the L2 center-of-mass
(L2 com), which is not only closer to the shrinking-radius center than 
L1 com, but at the same time free of the issue 
exhibited by the L1 and L2max particle
centers of occasionally finding a secondary
peak which does not coincide with the location
of the ``true'' density peak.

\subsection{Relationships between halo properties}
\label{sec:anal.rel}
\subsubsection{Radius-mass}
\begin{figure*}
\centering  
\includegraphics[width=.32\textwidth]{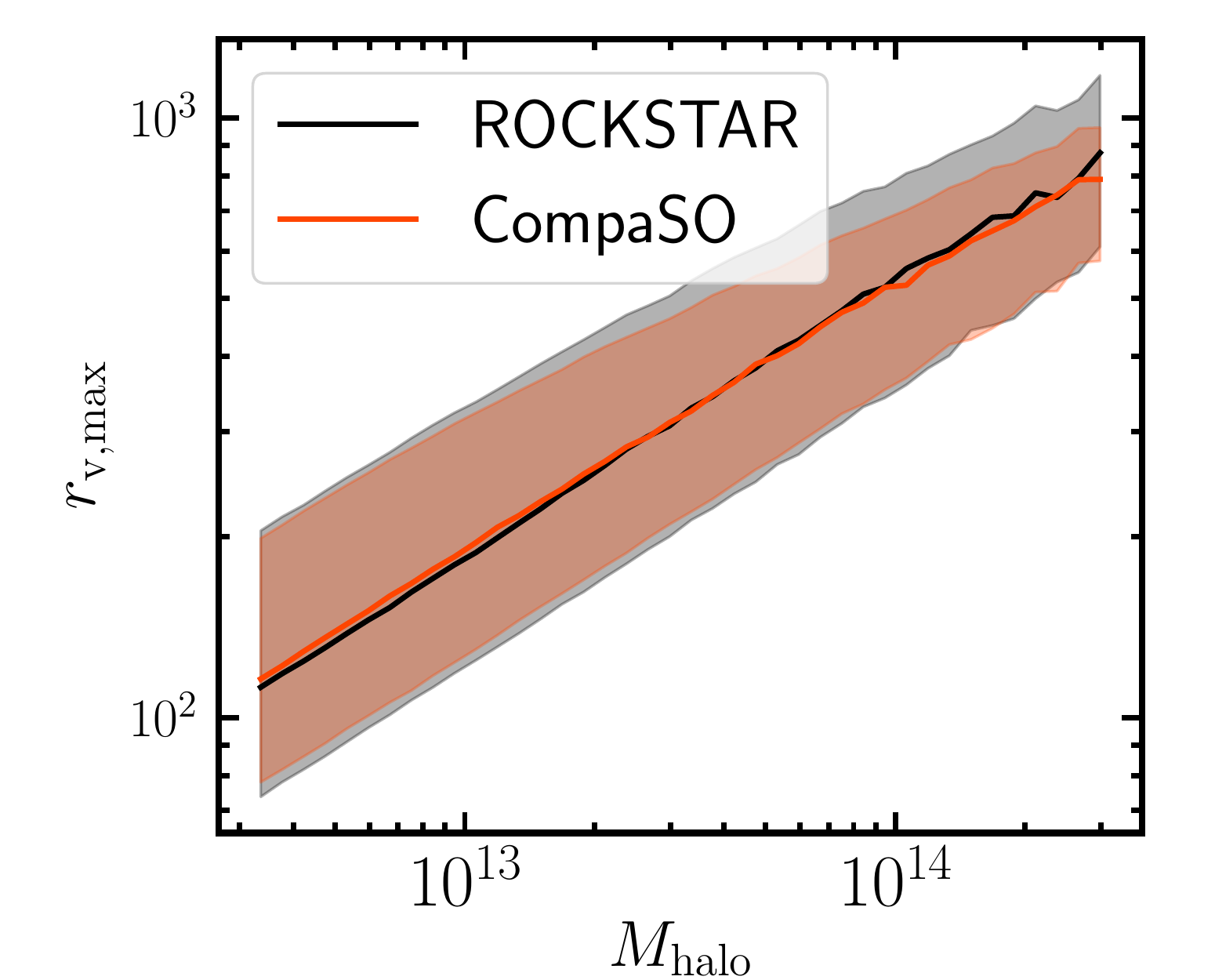}
\includegraphics[width=.32\textwidth]{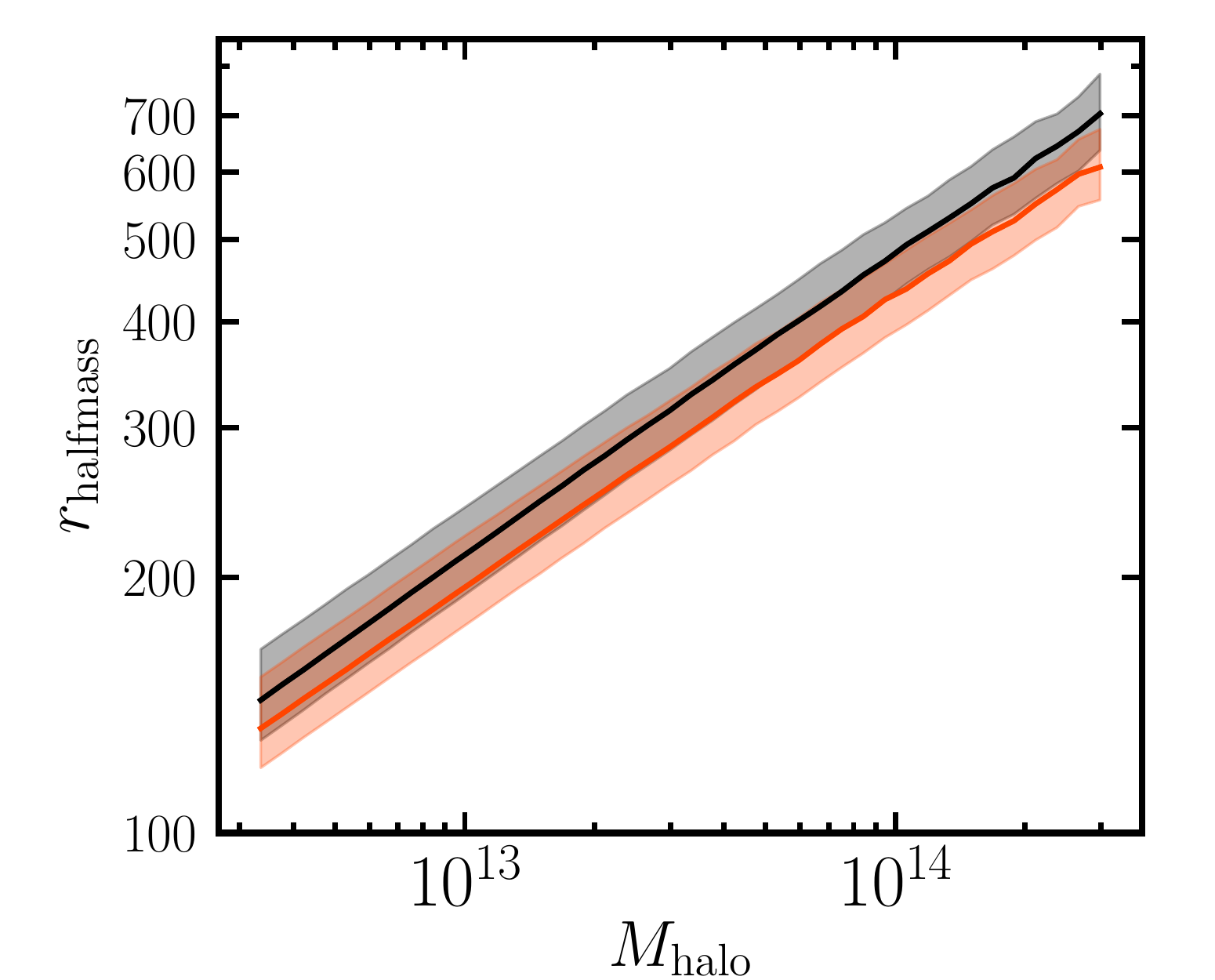}
\includegraphics[width=.32\textwidth]{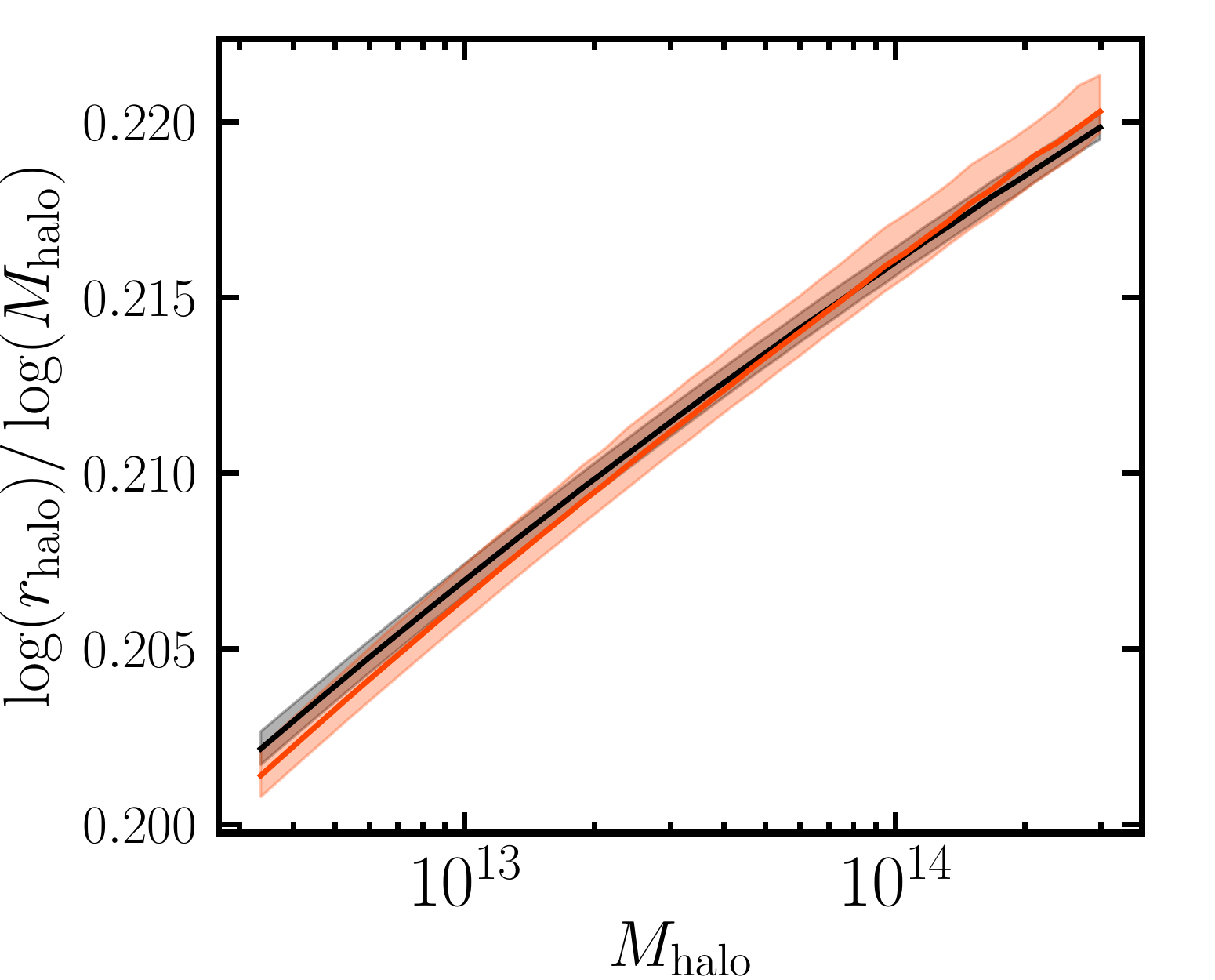}
\caption{Various measures of the halo radius as a
function of the total mass of the halo for the two
halo finders \textsc{CompaSO} and \textsc{ROCKSTAR}.
\textit{The left panel} illustrates the radius
at which the maximum circular velocity is
attained, \textit{the middle panel} 
shows the radius within which half of the halo
mass is contained, and \textit{the right panel} shows
the recommended parameter choices which 
yield a measure of
the virial halo radius for both halo finders.} 
\label{fig:rad_mass}
\end{figure*}

We analyse the distributions of 
halo radius and halo mass at a fixed redshift
of $z = 0.5$ for both the \textsc{CompaSO} and \textsc{ROCKSTAR}
halo catalogues. Fig. \ref{fig:rad_mass} shows examples
of these relations for three definitions of radius, 
namely, $r_{\rm v,max}$ (the radius at which the maximum
circular velocity is reached), $r_{\rm 50}$ (the radius
containing 50\% of the halo particles) and $r_{\rm vir}$
(the virial radius, approximated by $r_{98}$ in the
\textsc{CompaSO} catalogue).
Generally, the distributions of the three radii 
and mass are in reasonably good agreement between
both halo finders in terms of their median values
and scatters which exhibit a slightly longer tail towards
positive values. Particularly consistent between
the two halo catalogues are the measures of the maximum
circular velocity whose median values and scatter
appear to be in very good agreement. The largest
discrepancy is in the halfmass radius panel, which
shows that at fixed mass, the \textsc{CompaSO} haloes
have smaller values of $r_{\rm halfmass}$. This
is consistent with the finding that the \textsc{CompaSO}
algorithm tends to identify a larger number of small
haloes on the outskirts of large ones, which eat up
some of the substructure and leave dents in the
outer regions of the haloes, but preserve the innermost
particles.

\subsubsection{Maximum circular velocity-radius}
\begin{figure*}
\centering  
\includegraphics[width=.48\textwidth]{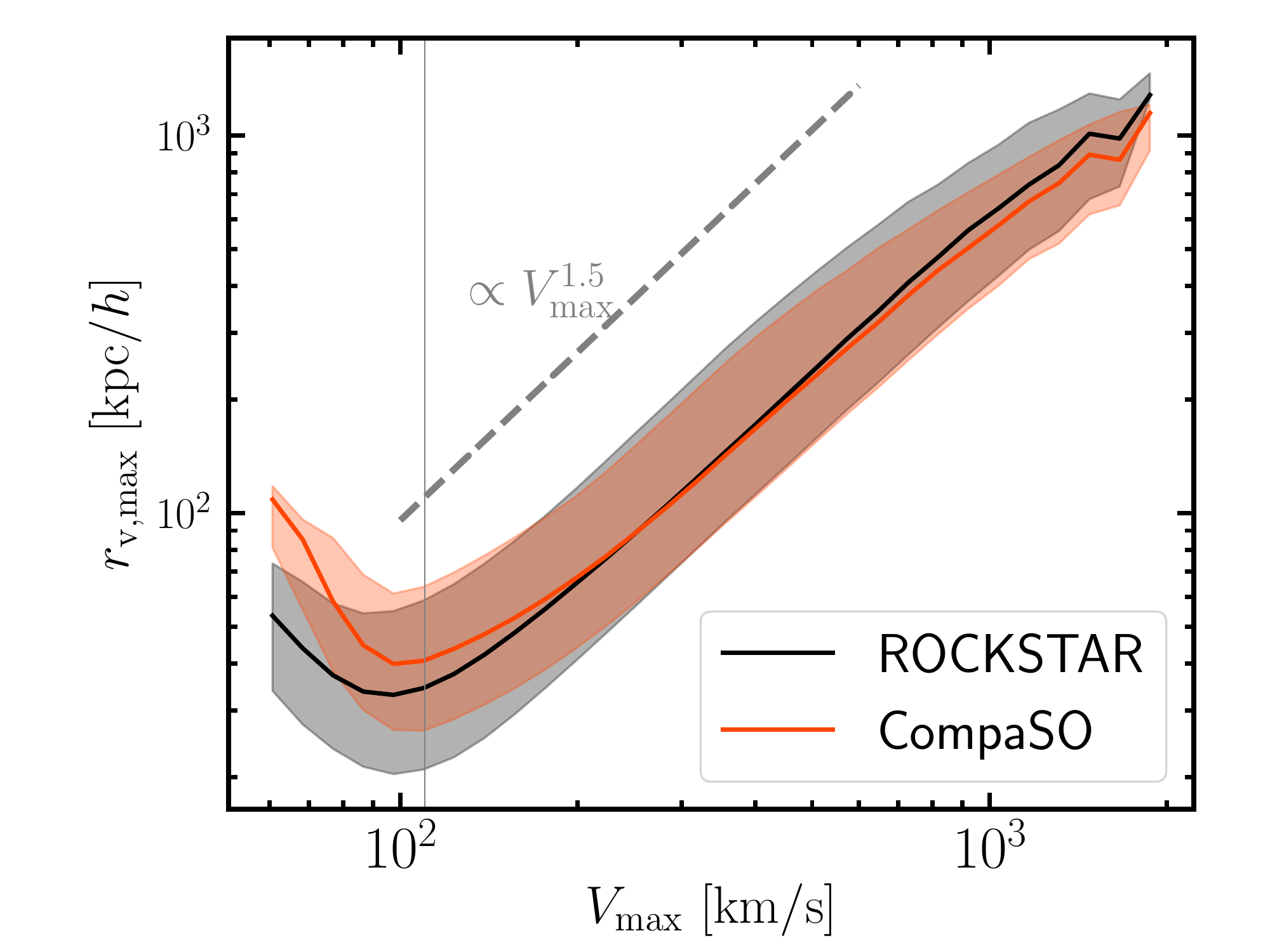}
\includegraphics[width=.48\textwidth]{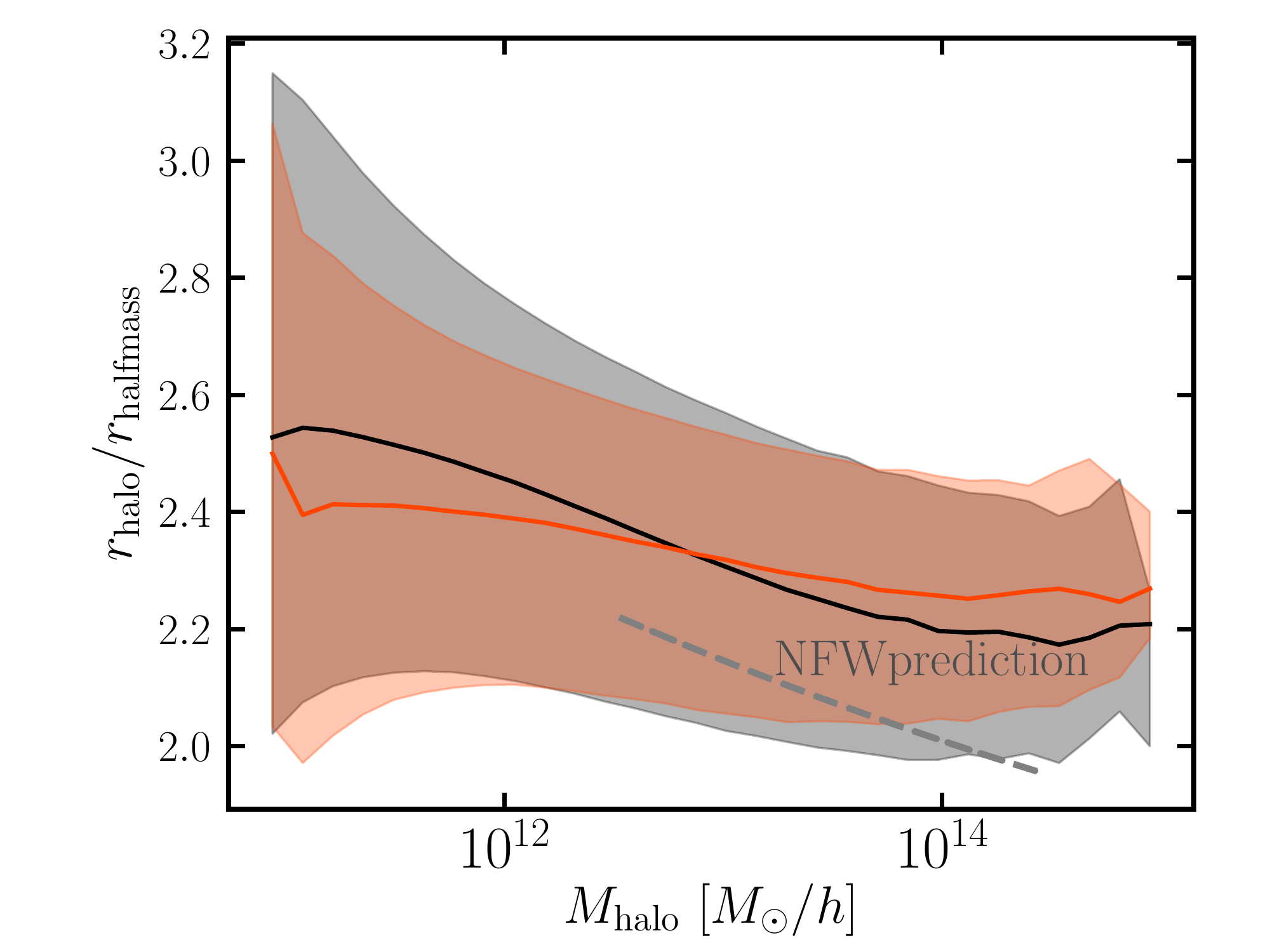}
\caption{\textit{Left panel:} relation between
$V_{\rm max}$ and the radius at which it is
attained. We see very good agreement between
the two halo finders, with the \textsc{CompaSO}
curve appearing a little flatter. The 
\textit{gray dotted} line shows the typical
shape found in other $N$-body simulations.
\textit{Right panel:} relation between the
concentration proxy, $r_{\rm halo}/r_{\rm halfmass}$, and the halo mass.
We see that the two curves compare well with the
mass-concentration relationship appearing
slightly flatter in the \textsc{CompaSO} case.
The \textit{gray dotted} line shows the NFW
prediction for this ratio.
The \textsc{CompaSO} quantity used as proxy
for $r_{\rm halo}$ here is $r_{98}$, the radius
containing 98\% of the particles, while the
\textsc{ROCKSTAR} $r_{\rm halo}$ proxy is the
default virial radius.} 
\label{fig:vmax_conc}
\end{figure*}

The left panel of Fig. \ref{fig:vmax_conc}
presents the maximum circular velocity versus
radius at which this velocity is attained
for the \textsc{ROCKSTAR} and \textsc{CompaSO} halo catalogues. 
The two catalogues are broadly in a good agreement,
although there are differences at 
$V_{\rm max} \gtrsim 1000$
km/s, corresponding to the regime of
low-mass clusters. The vertical line
shows roughly where 100-particle haloes are located, which
is the scale below which resolution effects kick in quite
strongly. The \textsc{CompaSO} objects typically have a smaller value near $V_{\rm max} \gtrsim 1000$ km/s and a higher value for
objects with $V_{\rm max} \sim 100$ km/s, which is
consistent with our observation from the halo
mass function (Fig. \ref{fig:hmf}). In general, however,
both halo catalogues show a 
scaling relation and $1\sigma$ scatter 
comparable to standard $\Lambda$CDM runs -- the gray
dashed curve shows the scaling relation found 
universally in $N$-body simulations.

\subsubsection{Mass-concentration}
In the right panel of Fig. \ref{fig:vmax_conc},
we visualise the mass-concentration relation in
\textsc{CompaSO} and \textsc{ROCKSTAR}. As a proxy for concentration, we
use the ratio $c = r_{\rm halo}/r_{\rm halfmass}$, where 
$r_{\rm halfmass}$ is the radius enclosing 50\% of the
particles and $r_{\rm halo}$ in the case of \textsc{CompaSO}
is $r_{98}$ (the radius encompassing 98\% of
the particles in a halo), 
whereas for \textsc{ROCKSTAR} we use the default
virial radius output as $r_{\rm halo}$. 
The two halo finders exhibit a very similar
scatter, with the contours of \textsc{CompaSO} 
being a little tighter 
than \textsc{ROCKSTAR}, particularly at low halo mass. In
addition, the low-mass \textsc{CompaSO} haloes appear to be less 
concentrated than the \textsc{ROCKSTAR} ones. That finding
is very sensitive to the concentration proxy used
to obtain the \textsc{CompaSO} concentrations (e.g., 
$c = r_{75}/r_{33}$ yields higher values),
but for consistency, we are keeping the definitions
across the two catalogues analogous. 
The results might differ depending on the halo
center definition used to compute these
quantities (see Fig. \ref{fig:shrink}).
The observed relationship between mass and concentration
is similar to what is found by other
$N$-body simulation codes and group finders.

\section{Discussion}
\label{sec:disc}
At heart, our new halo-finding method (\textsc{CompaSO})
can be thought of as an extension of the SO
algorithm, and yet, its particular implementation 
differs quite significantly.
Most notably, instead of simply truncating
the haloes at the overdensity threshold as done in the
standard SO algorithms, our method takes into account
the tidal radii around all nearby haloes before assigning halo 
membership to a given particle. By adopting this approach,
we aim to alleviate a
traditional weakness of position-space halo finders -- namely, their ability to
deblend haloes. Phase-space
halo finders such as \textsc{ROCKSTAR} \citep{2013ApJ...762..109B} 
typically circumvent this issue by
using 6-dimensional information about the
particle distributions. For this reason, we take
particular interest in the comparison between \textsc{ROCKSTAR}
and \textsc{CompaSO}. Furthermore, many of the traditionally used
halo finders struggle to identify
haloes close to the centers of larger haloes due to the
very high density of these regions. To mitigate 
this issue, our finder allows new haloes to form
\textit{within} the density threshold radius, i.e.
on the outskirts of other (typically larger) haloes. {In addition, we merge unphysical haloes to close companions via the cleaning procedure described in \citet{Bose+2021} and summarised in Section~\ref{sec:clean}. We demonstrate that the cleaning successfully removes a large number of the ill-defined haloes and leads to more ``sensible'' answers for some diagnostic statistics (e.g., two-point correlations). We consider other ways of identifying such objects including particle unbinding, virialisation and radial ratio tests. We find that the cleaning alone does a sufficiently good job in removing the vast majority of wrongdoers. For some particular applications, we show that combining additional halo statistics such as the virialisation ratio may help remove the remaining outliers.}

In this work, we have described in detail the \textsc{CompaSO}
halo finding method which is the
default halo algorithm in the \textsc{AbacusSummit} suite of $N$-body
simulations. We have made several
parameter choices discussed in the text and extensively
tested in smaller simulation boxes, such as the choice
of the inner eligibility radius (set at $R_{\rm L1, elig} = 80\% R_{\rm L1}$) and the Level 1 (L1) density threshold, $\Delta_{L1}$.
We have also listed the various optimization
and implementation techniques we employed in order to
fit into the Summit allocation constraints. More details
on the performance and optimization strategies are provided
in the main \textsc{AbacusSummit} papers \citep{Maksimova+2021,Garrison+2021b}.

To study the properties of the \textsc{CompaSO} haloes,
we have provided a variety of comparison tests
with the temporal phase-space halo-finding method 
\textsc{ROCKSTAR}, which is considered highly accurate
albeit computationally expensive. We have first
compared the halo mass functions derived from the
\textsc{CompaSO} and \textsc{ROCKSTAR} catalogues using two different
mass proxies: the virial mass of the halo and the maximum circular
velocity. We have shown (see Fig.~\ref{fig:hmf}) that in the virial
mass case, the two mass functions exhibit a slight tilt with respect
to each other and that the largest differences come from the very
low-mass ($\sim 10\%$, $\log(M_{\rm halo}) = 11$)
and the very high-mass regime ($\sim 50\%$, $\log(M_{\rm halo}) = 15$). 
Repeating this comparison for the maximum circular velocity $V_{\rm
max}$, we have observed that the discrepancy between the two finders is 
much smaller, at $\sim 5\%$ across the $V_{\rm max}$ range. This suggests 
that \textsc{CompaSO} and \textsc{ROCKSTAR} are finding largely the same 
objects in the simulation and that the difference can be attributed 
predominantly to the mass definitions and the drawing of halo boundaries. 
We have also demonstrated that our clean catalogues reconcile some of the 
tension with \textsc{ROCKSTAR}. This finding also suggests that when 
constructing HOD catalogues, adopting the cleaned catalogues is likely to 
give us a galaxy sample that is in a better agreement with 
\textsc{ROCKSTAR}. We have also shown that the \textsc{CompaSO} halo
profiles agree reasonably
well with the predictions of the NFW formalism (see Fig.~\ref{fig:density}).

The general trend
corroborated by the comparison of their auto-
and cross-correlation functions is that the \textsc{CompaSO} algorithm
finds a larger number of smaller haloes on the outskirts
of the largest haloes and more readily deblends these large objects.
In Fig.~\ref{fig:corr} we also show that the cleaned \textsc{CompaSO}
catalogues result in clustering that appears to be much more similar
to that of the \textsc{ROCKSTAR} haloes.
To follow up on this finding, we test the persistence of the
\textsc{CompaSO} haloes by studying their past merger histories. 
A powerful diagnostic we explore is the comparison of
whether a merger tree
built snapshot-by-snapshot (dubbed as ``fine'') yields the same
progenitors for a given halo as obtained by a merger tree
built across significantly longer time jumps (dubbed as ``coarse'').
For the haloes at $z = 0.5$, we obtain a high percentage 
of matches at the first ``coarse'' time step 
($z=0.875$,  more than 97\% for $\log(M) > 10^{12.5}$) and a relatively
high one at the second ($z=1.625$, close to 90\%). We find
that in the higher mass regime, $\log(M) > 10^{12.5}$, the
percentage of haloes flagged as ``potential splits'' is negligible
($\ll 0.1\%$).
In terms of the best halo center choice, our recommendation 
(motivated by Fig.~\ref{fig:shrink}) is
to use quantities computed with respect to the L2 center-of-mass
(L2 com), which is both closer to the shrinking-radius center than the 
L1 com, and at the same time does not exhibit the issue that
the L1 and L2max particle centers have of occasionally not coinciding
with the location of the ``true'' density peak.

Overall, we have presented a highly efficient halo-finding
method which provides a recipe for deblending haloes
and determining particle membership on the outskirts of
large haloes at a negligible computational expense
compared with other accurate halo finders. While 
\textsc{CompaSO} does not currently output sufficient information
about its subhaloes (L2 haloes), combining the L1 haloes
with the merger tree outputs can allow us to build
reliable halo catalogues, which are central to
galaxy population approaches such as the HOD model. Creating highly realistic
galaxy mock catalogues will be one of the biggest challenges
for current cosmological efforts, and improving the
accuracy of the halo finding techniques 
can therefore be of crucial importance for the advancement
of observational large-scale structure cosmology.

\section*{Acknowledgements}
We would like to thank Alex Smith and Shaun Cole for their very helpful and illuminating comments, and Lucy Reading-Ikkanda for illustrating Figure~\ref{fig:algo}. We thank the referee for the informative discussion and helpful suggestions.

This work has been supported by NSF AST-1313285, DOE-SC0013718, and NASA ROSES grant 12-EUCLID12-0004.
DJE is supported in part as a Simons Foundation investigator. 
NAM was supported in part as a NSF Graduate Research Fellow.  
LHG is supported by the Center for Computational Astrophysics at the Flatiron Institute, which is supported by the Simons Foundation.  
SB is supported by Harvard University through the ITC Fellowship.

This research used resources of the Oak Ridge Leadership Computing Facility, which is a DOE Office of Science User Facility supported under Contract DE-AC05-00OR22725. 
Computation of the merger trees used resources of the National Energy Research Scientific Computing Center (NERSC), a U.S. Department of Energy Office of Science User Facility located at Lawrence Berkeley National Laboratory, operated under Contract No. DE-AC02-05CH11231.
The \textsc{AbacusSummit} simulations have been supported by OLCF projects AST135 and AST145, the latter through the Department of Energy ALCC program.

We would like to thank the OLCF and NERSC staff for their highly responsive and expert assistance, both scientific and administrative, during the course of this project.


\section*{Data Availability}

The simulation data is available as part of \Abacus{AbacusSummit} and is subject to the academic citations described at \url{https://abacussummit.readthedocs.io/en/latest/citation.html}.

Data access is available through OLCF's Constellation portal.  The persistent DOI describing the data release is \href{https://www.doi.org/10.13139/OLCF/1811689}{10.13139/OLCF/1811689}.  Instructions for accessing the data are given at \url{https://abacussummit.readthedocs.io/en/latest/data-access.html}.




\bibliographystyle{mnras}
\bibliography{refs} 

\begin{thebibliography}{}
\makeatletter
\relax
\def\mn@urlcharsother{\let\do\@makeother \do\$\do\&\do\#\do\^\do\_\do\%\do\~}
\def\mn@doi{\begingroup\mn@urlcharsother \@ifnextchar [ {\mn@doi@}
  {\mn@doi@[]}}
\def\mn@doi@[#1]#2{\def\@tempa{#1}\ifx\@tempa\@empty \href
  {http://dx.doi.org/#2} {doi:#2}\else \href {http://dx.doi.org/#2} {#1}\fi
  \endgroup}
\def\mn@eprint#1#2{\mn@eprint@#1:#2::\@nil}
\def\mn@eprint@arXiv#1{\href {http://arxiv.org/abs/#1} {{\tt arXiv:#1}}}
\def\mn@eprint@dblp#1{\href {http://dblp.uni-trier.de/rec/bibtex/#1.xml}
  {dblp:#1}}
\def\mn@eprint@#1:#2:#3:#4\@nil{\def\@tempa {#1}\def\@tempb {#2}\def\@tempc
  {#3}\ifx \@tempc \@empty \let \@tempc \@tempb \let \@tempb \@tempa \fi \ifx
  \@tempb \@empty \def\@tempb {arXiv}\fi \@ifundefined
  {mn@eprint@\@tempb}{\@tempb:\@tempc}{\expandafter \expandafter \csname
  mn@eprint@\@tempb\endcsname \expandafter{\@tempc}}}

\bibitem[\protect\citeauthoryear{{Audit}, {Teyssier}  \& {Alimi}}{{Audit}
  et~al.}{1998}]{1998A&A...333..779A}
{Audit} E.,  {Teyssier} R.,   {Alimi} J.-M.,  1998, \aap, \href
  {https://ui.adsabs.harvard.edu/abs/1998A&A...333..779A} {333, 779}

\bibitem[\protect\citeauthoryear{{Barnes} \& {Efstathiou}}{{Barnes} \&
  {Efstathiou}}{1987}]{1987ApJ...319..575B}
{Barnes} J.,  {Efstathiou} G.,  1987, \mn@doi [\apj] {10.1086/165480}, \href
  {https://ui.adsabs.harvard.edu/abs/1987ApJ...319..575B} {319, 575}

\bibitem[\protect\citeauthoryear{{Behroozi}, {Wechsler}  \& {Wu}}{{Behroozi}
  et~al.}{2013}]{2013ApJ...762..109B}
{Behroozi} P.~S.,  {Wechsler} R.~H.,   {Wu} H.-Y.,  2013, \mn@doi [\apj]
  {10.1088/0004-637X/762/2/109}, \href
  {https://ui.adsabs.harvard.edu/abs/2013ApJ...762..109B} {762, 109}

\bibitem[\protect\citeauthoryear{{Bernardeau}, {Colombi}, {Gazta{\~n}aga}  \&
  {Scoccimarro}}{{Bernardeau} et~al.}{2002}]{2002PhR...367....1B}
{Bernardeau} F.,  {Colombi} S.,  {Gazta{\~n}aga} E.,   {Scoccimarro} R.,  2002,
  \mn@doi [\physrep] {10.1016/S0370-1573(02)00135-7}, \href
  {https://ui.adsabs.harvard.edu/abs/2002PhR...367....1B} {367, 1}

\bibitem[\protect\citeauthoryear{{Bertschinger} \& {Gelb}}{{Bertschinger} \&
  {Gelb}}{1991}]{1991ComPh...5..164B}
{Bertschinger} E.,  {Gelb} J.~M.,  1991, \mn@doi [Computers in Physics]
  {10.1063/1.4822978}, \href
  {https://ui.adsabs.harvard.edu/abs/1991ComPh...5..164B} {5, 164}

\bibitem[\protect\citeauthoryear{{Binney} \& {Tremaine}}{{Binney} \&
  {Tremaine}}{1987}]{1987gady.book.....B}
{Binney} J.,  {Tremaine} S.,  1987, {Galactic dynamics}

\bibitem[\protect\citeauthoryear{{Bode} \& {Ostriker}}{{Bode} \&
  {Ostriker}}{2003}]{2003ApJS..145....1B}
{Bode} P.,  {Ostriker} J.~P.,  2003, \mn@doi [\apjs] {10.1086/345538}, \href
  {https://ui.adsabs.harvard.edu/abs/2003ApJS..145....1B} {145, 1}

\bibitem[\protect\citeauthoryear{{Bond} \& {Myers}}{{Bond} \&
  {Myers}}{1996}]{1996ApJS..103...41B}
{Bond} J.~R.,  {Myers} S.~T.,  1996, \mn@doi [\apjs] {10.1086/192268}, \href
  {https://ui.adsabs.harvard.edu/abs/1996ApJS..103...41B} {103, 41}

\bibitem[\protect\citeauthoryear{{Bose}, {Eisenstein}, {Hadzhiyska},
  {Garrison}, {Yuan}  \& {Maksimova}}{{Bose} et~al.}{2021}]{Bose+2021}
{Bose} S.,  {Eisenstein} D.,  {Hadzhiyska} B.,  {Garrison} L.,  {Yuan} S.,
  {Maksimova} N.,  2021, submitted

\bibitem[\protect\citeauthoryear{{Bryan} \& {Norman}}{{Bryan} \&
  {Norman}}{1998}]{1998ApJ...495...80B}
{Bryan} G.~L.,  {Norman} M.~L.,  1998, \mn@doi [\apj] {10.1086/305262}, \href
  {https://ui.adsabs.harvard.edu/abs/1998ApJ...495...80B} {495, 80}

\bibitem[\protect\citeauthoryear{{Bullock}, {Kolatt}, {Sigad}, {Somerville},
  {Kravtsov}, {Klypin}, {Primack}  \& {Dekel}}{{Bullock}
  et~al.}{2001}]{2001MNRAS.321..559B}
{Bullock} J.~S.,  {Kolatt} T.~S.,  {Sigad} Y.,  {Somerville} R.~S.,  {Kravtsov}
  A.~V.,  {Klypin} A.~A.,  {Primack} J.~R.,   {Dekel} A.,  2001, \mn@doi
  [\mnras] {10.1046/j.1365-8711.2001.04068.x}, \href
  {https://ui.adsabs.harvard.edu/abs/2001MNRAS.321..559B} {321, 559}

\bibitem[\protect\citeauthoryear{{Davis}, {Efstathiou}, {Frenk}  \&
  {White}}{{Davis} et~al.}{1985}]{1985ApJ...292..371D}
{Davis} M.,  {Efstathiou} G.,  {Frenk} C.~S.,   {White} S.~D.~M.,  1985,
  \mn@doi [\apj] {10.1086/163168}, \href
  {http://adsabs.harvard.edu/abs/1985ApJ...292..371D} {292, 371}

\bibitem[\protect\citeauthoryear{{Dubinski}, {Kim}, {Park}  \&
  {Humble}}{{Dubinski} et~al.}{2004}]{2004NewA....9..111D}
{Dubinski} J.,  {Kim} J.,  {Park} C.,   {Humble} R.,  2004, \mn@doi [\na]
  {10.1016/j.newast.2003.08.002}, \href
  {https://ui.adsabs.harvard.edu/abs/2004NewA....9..111D} {9, 111}

\bibitem[\protect\citeauthoryear{{Dutton} \& {Macci{\`o}}}{{Dutton} \&
  {Macci{\`o}}}{2014}]{2014MNRAS.441.3359D}
{Dutton} A.~A.,  {Macci{\`o}} A.~V.,  2014, \mn@doi [\mnras]
  {10.1093/mnras/stu742}, \href
  {https://ui.adsabs.harvard.edu/abs/2014MNRAS.441.3359D} {441, 3359}

\bibitem[\protect\citeauthoryear{{Eisenstein} \& {Hut}}{{Eisenstein} \&
  {Hut}}{1998}]{1998ApJ...498..137E}
{Eisenstein} D.~J.,  {Hut} P.,  1998, \mn@doi [\apj] {10.1086/305535}, \href
  {https://ui.adsabs.harvard.edu/abs/1998ApJ...498..137E} {498, 137}

\bibitem[\protect\citeauthoryear{Epanechnikov}{Epanechnikov}{1969}]{Epanechnikov:1969:NPE}
Epanechnikov V.~A.,  1969, \mn@doi [] {https://doi.org/10.1137/1114019}, 14,
  153

\bibitem[\protect\citeauthoryear{{Evrard} et~al.,}{{Evrard}
  et~al.}{2002}]{2002ApJ...573....7E}
{Evrard} A.~E.,  et~al., 2002, \mn@doi [\apj] {10.1086/340551}, \href
  {https://ui.adsabs.harvard.edu/abs/2002ApJ...573....7E} {573, 7}

\bibitem[\protect\citeauthoryear{{Garrison}, {Eisenstein}  \&
  {Pinto}}{{Garrison} et~al.}{2019}]{2019MNRAS.485.3370G}
{Garrison} L.~H.,  {Eisenstein} D.~J.,   {Pinto} P.~A.,  2019, \mn@doi [\mnras]
  {10.1093/mnras/stz634}, \href
  {https://ui.adsabs.harvard.edu/abs/2019MNRAS.485.3370G} {485, 3370}

\bibitem[\protect\citeauthoryear{{Garrison}, {Eisenstein}, {Ferrer},
  {Maksimova}  \& {Pinto}}{{Garrison} et~al.}{2021}]{Garrison+2021b}
{Garrison} L.,  {Eisenstein} D.,  {Ferrer} D.,  {Maksimova} N.,   {Pinto} P.,
  2021, submitted

\bibitem[\protect\citeauthoryear{{Gelb} \& {Bertschinger}}{{Gelb} \&
  {Bertschinger}}{1994}]{1994ApJ...436..467G}
{Gelb} J.~M.,  {Bertschinger} E.,  1994, \mn@doi [\apj] {10.1086/174922}, \href
  {https://ui.adsabs.harvard.edu/abs/1994ApJ...436..467G} {436, 467}

\bibitem[\protect\citeauthoryear{Gradshteyn \& Ryzhik}{Gradshteyn \&
  Ryzhik}{2007}]{gradshteyn2007}
Gradshteyn I.~S.,  Ryzhik I.~M.,  2007, Table of integrals, series, and
  products, seventh edn.
Elsevier/Academic Press, Amsterdam

\bibitem[\protect\citeauthoryear{{Jenkins}, {Frenk}, {White}, {Colberg},
  {Cole}, {Evrard}, {Couchman}  \& {Yoshida}}{{Jenkins}
  et~al.}{2001}]{2001MNRAS.321..372J}
{Jenkins} A.,  {Frenk} C.~S.,  {White} S.~D.~M.,  {Colberg} J.~M.,  {Cole} S.,
  {Evrard} A.~E.,  {Couchman} H.~M.~P.,   {Yoshida} N.,  2001, \mn@doi [\mnras]
  {10.1046/j.1365-8711.2001.04029.x}, \href
  {https://ui.adsabs.harvard.edu/abs/2001MNRAS.321..372J} {321, 372}

\bibitem[\protect\citeauthoryear{{Klypin}, {Gottl{\"o}ber}, {Kravtsov}  \&
  {Khokhlov}}{{Klypin} et~al.}{1999}]{1999ApJ...516..530K}
{Klypin} A.,  {Gottl{\"o}ber} S.,  {Kravtsov} A.~V.,   {Khokhlov} A.~M.,  1999,
  \mn@doi [\apj] {10.1086/307122}, \href
  {https://ui.adsabs.harvard.edu/abs/1999ApJ...516..530K} {516, 530}

\bibitem[\protect\citeauthoryear{{Knebe} et~al.,}{{Knebe}
  et~al.}{2011}]{2011MNRAS.415.2293K}
{Knebe} A.,  et~al., 2011, \mn@doi [\mnras] {10.1111/j.1365-2966.2011.18858.x},
  \href {https://ui.adsabs.harvard.edu/abs/2011MNRAS.415.2293K} {415, 2293}

\bibitem[\protect\citeauthoryear{{Knollmann} \& {Knebe}}{{Knollmann} \&
  {Knebe}}{2009}]{2009ApJS..182..608K}
{Knollmann} S.~R.,  {Knebe} A.,  2009, \mn@doi [\apjs]
  {10.1088/0067-0049/182/2/608}, \href
  {https://ui.adsabs.harvard.edu/abs/2009ApJS..182..608K} {182, 608}

\bibitem[\protect\citeauthoryear{{Lacey} \& {Cole}}{{Lacey} \&
  {Cole}}{1994}]{1994MNRAS.271..676L}
{Lacey} C.,  {Cole} S.,  1994, \mn@doi [\mnras] {10.1093/mnras/271.3.676},
  \href {https://ui.adsabs.harvard.edu/abs/1994MNRAS.271..676L} {271, 676}

\bibitem[\protect\citeauthoryear{{{\L}okas}}{{{\L}okas}}{2000}]{2000MNRAS.311..423L}
{{\L}okas} E.~L.,  2000, \mn@doi [\mnras] {10.1046/j.1365-8711.2000.03082.x},
  \href {https://ui.adsabs.harvard.edu/abs/2000MNRAS.311..423L} {311, 423}

\bibitem[\protect\citeauthoryear{{Maksimova}, {Garrison}, {Eisenstein},
  {Hadzhiyska}, {Bose}  \& {Satterthwaite}}{{Maksimova}
  et~al.}{2021}]{Maksimova+2021}
{Maksimova} N.,  {Garrison} L.,  {Eisenstein} D.,  {Hadzhiyska} B.,  {Bose} S.,
    {Satterthwaite} T.,  2021, submitted

\bibitem[\protect\citeauthoryear{{More}, {Kravtsov}, {Dalal}  \&
  {Gottl{\"o}ber}}{{More} et~al.}{2011}]{2011ApJS..195....4M}
{More} S.,  {Kravtsov} A.~V.,  {Dalal} N.,   {Gottl{\"o}ber} S.,  2011, \mn@doi
  [\apjs] {10.1088/0067-0049/195/1/4}, \href
  {https://ui.adsabs.harvard.edu/abs/2011ApJS..195....4M} {195, 4}

\bibitem[\protect\citeauthoryear{{Navarro}, {Frenk}  \& {White}}{{Navarro}
  et~al.}{1996}]{1996ApJ...462..563N}
{Navarro} J.~F.,  {Frenk} C.~S.,   {White} S. D.~M.,  1996, \mn@doi [\apj]
  {10.1086/177173}, \href
  {https://ui.adsabs.harvard.edu/abs/1996ApJ...462..563N} {462, 563}

\bibitem[\protect\citeauthoryear{{Park}}{{Park}}{1990}]{1990MNRAS.242P..59P}
{Park} C.,  1990, \mn@doi [\mnras] {10.1093/mnras/242.1.59P}, \href
  {https://ui.adsabs.harvard.edu/abs/1990MNRAS.242P..59P} {242, 59P}

\bibitem[\protect\citeauthoryear{{Peebles}}{{Peebles}}{1980}]{1980lssu.book.....P}
{Peebles} P.~J.~E.,  1980, {The large-scale structure of the universe}

\bibitem[\protect\citeauthoryear{{Power}, {Navarro}, {Jenkins}, {Frenk},
  {White}, {Springel}, {Stadel}  \& {Quinn}}{{Power}
  et~al.}{2003}]{2003MNRAS.338...14P}
{Power} C.,  {Navarro} J.~F.,  {Jenkins} A.,  {Frenk} C.~S.,  {White} S.~D.~M.,
   {Springel} V.,  {Stadel} J.,   {Quinn} T.,  2003, \mn@doi [\mnras]
  {10.1046/j.1365-8711.2003.05925.x}, \href
  {https://ui.adsabs.harvard.edu/abs/2003MNRAS.338...14P} {338, 14}

\bibitem[\protect\citeauthoryear{{Press} \& {Schechter}}{{Press} \&
  {Schechter}}{1974}]{1974ApJ...187..425P}
{Press} W.~H.,  {Schechter} P.,  1974, \mn@doi [\apj] {10.1086/152650}, \href
  {https://ui.adsabs.harvard.edu/abs/1974ApJ...187..425P} {187, 425}

\bibitem[\protect\citeauthoryear{{Sheth} \& {Tormen}}{{Sheth} \&
  {Tormen}}{1999}]{1999MNRAS.308..119S}
{Sheth} R.~K.,  {Tormen} G.,  1999, \mn@doi [\mnras]
  {10.1046/j.1365-8711.1999.02692.x}, \href
  {https://ui.adsabs.harvard.edu/abs/1999MNRAS.308..119S} {308, 119}

\bibitem[\protect\citeauthoryear{Springel, White, Tormen  \&
  Kauffmann}{Springel et~al.}{2001}]{Springel:2000qu}
Springel V.,  White S. D.~M.,  Tormen G.,   Kauffmann G.,  2001, \mn@doi [Mon.
  Not. Roy. Astron. Soc.] {10.1046/j.1365-8711.2001.04912.x}, 328, 726

\bibitem[\protect\citeauthoryear{{Springel} et~al.,}{{Springel}
  et~al.}{2005}]{2005Natur.435..629S}
{Springel} V.,  et~al., 2005, \mn@doi [\nat] {10.1038/nature03597}, \href
  {https://ui.adsabs.harvard.edu/abs/2005Natur.435..629S} {435, 629}

\bibitem[\protect\citeauthoryear{{St{\"u}cker}, {Angulo}  \&
  {Busch}}{{St{\"u}cker} et~al.}{2021}]{2021arXiv210713008S}
{St{\"u}cker} J.,  {Angulo} R.~E.,   {Busch} P.,  2021, arXiv e-prints, \href
  {https://ui.adsabs.harvard.edu/abs/2021arXiv210713008S} {p. arXiv:2107.13008}

\bibitem[\protect\citeauthoryear{{Warren}, {Quinn}, {Salmon}  \&
  {Zurek}}{{Warren} et~al.}{1992}]{1992ApJ...399..405W}
{Warren} M.~S.,  {Quinn} P.~J.,  {Salmon} J.~K.,   {Zurek} W.~H.,  1992,
  \mn@doi [\apj] {10.1086/171937}, \href
  {https://ui.adsabs.harvard.edu/abs/1992ApJ...399..405W} {399, 405}

\bibitem[\protect\citeauthoryear{{Warren}, {Abazajian}, {Holz}  \&
  {Teodoro}}{{Warren} et~al.}{2006}]{2006ApJ...646..881W}
{Warren} M.~S.,  {Abazajian} K.,  {Holz} D.~E.,   {Teodoro} L.,  2006, \mn@doi
  [\apj] {10.1086/504962}, \href
  {https://ui.adsabs.harvard.edu/abs/2006ApJ...646..881W} {646, 881}

\bibitem[\protect\citeauthoryear{{White}, {Hernquist}  \& {Springel}}{{White}
  et~al.}{2001}]{2001ApJ...550L.129W}
{White} M.,  {Hernquist} L.,   {Springel} V.,  2001, \mn@doi [\apjl]
  {10.1086/319644}, \href
  {https://ui.adsabs.harvard.edu/abs/2001ApJ...550L.129W} {550, L129}

\makeatother
\end{thebibliography}



\appendix

\section{{Mock halo tests}}

\begin{figure}
\centering  
\includegraphics[width=0.48\textwidth]{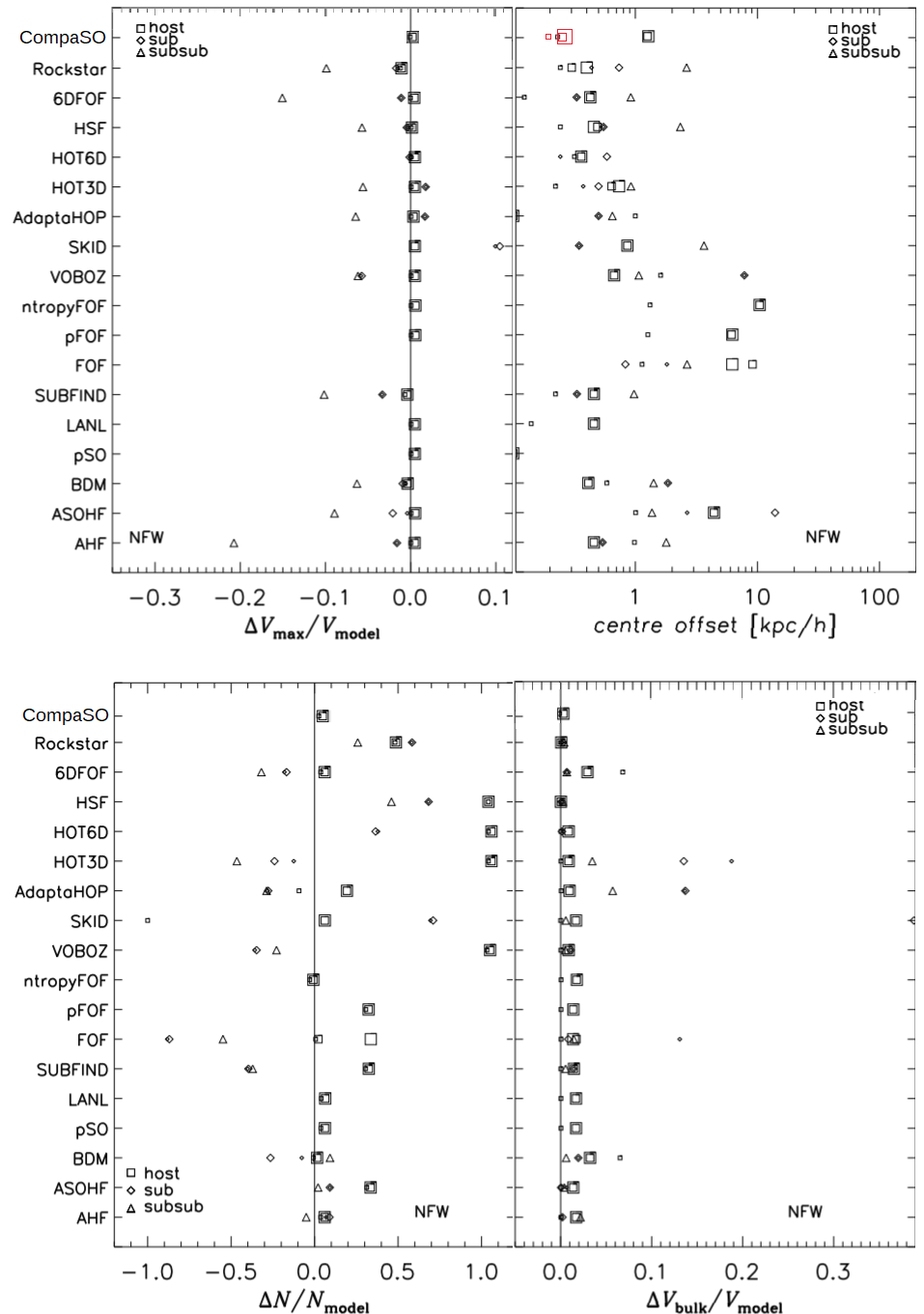} \\
\caption{Recovery of various halo properties by \textsc{CompaSO} and other halo finders: maximum circular velocity (upper left), offset of the halo centre (upper right), number of particles (lower left), bulk velocity (lower right). Displayed are the results for only the host haloes (squares), as \textsc{CompaSO} does not output subhalo properties. The size of the squares indicates the test sequence (i.e. larger symbols for haloes containing more sub(sub)haloes). Adapted from \citet{2011MNRAS.415.2293K}.}
\label{fig:mocks}
\end{figure}

{
All of the tests presented in the main body of this work are conducted using cosmological boxes. A useful test, therefore, is to compare the performance of \textsc{CompaSO} with that of other halo finders in a controlled environment. \citet{2011MNRAS.415.2293K} offers a comparison of the halo properties retrieved by a number of different halo finders, spanning a wide range of techniques such as friends-of-friends (FOF), spherical-overdensity (SO) and phase-space based algorithms. We perform the mock halo tests provided in \citet{2011MNRAS.415.2293K} on \textsc{CompaSO}. We note, however, that only a subset of these tests are relevant for \textsc{CompaSO} (in particular, those measuring halo properties), since it does not output any subhalo information apart from statistics with respect to the centre of the largest subhalo, which are only used to define the halo properties more robustly.
}

{
In Fig.~\ref{fig:mocks}, we show \textsc{CompaSO} alongside the other halo finders explored in \citet{2011MNRAS.415.2293K} for three different scenarios involving NFW mock haloes: a single-halo system, a halo-subhalo system, and a halo-subhalo-subsubhalo system. We only report the host halo properties for each of these and indicate them with a square (small, medium, and large, respectively). The properties explored are maximum circular velocity ($V_{\rm max}$), offset of the halo centre, number of particles ($N$), bulk velocity ($V_{\rm bulk}$), expressed as fractional differences, $\Delta X/X_{\rm model}$, with respect to the input value, $X_{\rm model}$. For more information, see \citet{2011MNRAS.415.2293K}. We see that \textsc{CompaSO} recovers all four properties very well. In particular, its measurement of $V_{\rm max}$ and $N$ is among the best, outperforming nearly all phase-space finders. The recommended \textsc{CompaSO} proxy for the bulk velocity is \texttt{v\_L2com}, the mean velocity of all particles in the largest subhalo. Its value depends on the subhalo density threshold, \texttt{SODensityL2} (see Table \ref{tab:names}), with higher values leading to a more accurate measure of the input bulk velocity. Finally, the recommended halo centre, defined as \texttt{x\_L2com}, i.e. the mean position of all particles in the largest subhalo, is denoted with black squares and exhibits a similar dependence on the subhalo density threshold threshold with higher threshold values yielding less of an offset. We see that using the default threshold value yields similar results to the bulk of the halo finders. Using the particle centre of the largest L2 halo, shown in red squares, performs better, in agreement with our finding in Fig.~\ref{fig:rad_mass}.
}

\section{Parameter choices}
In this appendix, we explore several of the parameter
choices made in the \textsc{CompaSO} halo-finding
algorithm. The procedure for obtaining 
the \textsc{CompaSO} halo catalogue is detailed in 
Section~\ref{sec:meth.algo}.
There are several free parameters that need to be
fixed when performing halo-finding using \textsc{CompaSO}. Here we discuss some of these choices.

\subsection{Linking length, eligibility radius, density threshold}
\label{app:choices}
The parameters that have the most direct impact on 
the halo statistics such as the halo mass function
and the two-point clustering are: 
$\texttt{FoFLinkingLength}$, $\texttt{L0DensityThreshold}$,
and \texttt{SO\_alpha\_eligible}.
(for definitions, see Table~\ref{tab:names}).
First, we vary the modified FoF linking length of the L0
groups since it is possible that the L1 \textsc{CompaSO}
haloes are confined by the L0 edges rather than their
SO density thresholds. Similarly, we decrease the
threshold on the kernel density, as the higher value
that we have set might be exluding particles that would
otherwise have joined L1 haloes. Finally, we study
the effect of the eligibility radius, which determines
the radius from the center of a given halo at which
particles are eligible to form new halo nuclei.
Here we perform a few tests by varying these parameters
in 5 small simulation boxes: 
$\texttt{GF\_N512\_L150-fiducial}$,
$\texttt{GF\_N512\_L150-alpha0.7}$,
$\texttt{GF\_N512\_L150-alpha0.9}$
$\texttt{GF\_N512\_L150-b0.30}$
$\texttt{GF\_N512\_L150-thresh40}$, 
each with box-size $L_{\rm box} = 150 \ {\rm Mpc}/h$
and $N_{\rm part} = 512^3$ particles. We note that
the resolution of these boxes matches very well
that of the \texttt{AbacusSummit\_highbase\_c000\_ph100} box studied in the
rest of the paper.
In Table~\ref{tab:small_app}, we specify for 
each of the small boxes what values we adopt
for the parameters that we vary. We note that
the subsequent analysis is performed on the $z = 0$
snapshot.

\begin{table*}
\begin{center}
\begin{tabular}{c c c c c c} 
 \hline\hline
 \backslashbox{Parameters}{\texttt{GF\_N512\_L150}} & \texttt{fiducial} & \texttt{alpha0.7} & \texttt{alpha0.9} & \texttt{b0.30} & \texttt{thresh40} \\ [0.5ex]
 \hline
\texttt{BoxSize} & 150 Mpc$/h$ & 150 Mpc$/h$ & 150 Mpc$/h$ & 150 Mpc$/h$ & 150 Mpc$/h$ \\
\texttt{NP} & $512^3$ & $512^3$ & $512^3$ & $512^3$ & $512^3$ \\
\texttt{ParticleMassHMsun} & $2.189 \times 10^9 \ M_\odot/h$ & $2.189 \times 10^9 \ M_\odot/h$ & $2.189 \times 10^9 \ M_\odot/h$ & $2.189 \times 10^9 \ M_\odot/h$ & $2.189 \times 10^9 \ M_\odot/h$ \\
\texttt{FoFLinkingLength} & 0.25 & 0.25 & 0.25 & 0.30 & 0.25 \\
\texttt{SO\_alpha\_eligible} & 0.8 & 0.7 & 0.9 & 0.8 & 0.8 \\
\texttt{L0DensityThreshold} & 60 & 60 & 60 & 60 & 40 \\
\hline \hline
\end{tabular}
\end{center}
\caption{Specifications of the small boxes (of the series \texttt{GF\_N512\_L150}) used to test the \textsc{CompaSO} parameter choices. See Table~\ref{tab:names} for a more detailed description of the variables names. Here we vary the parameters that have the most direct impact on halo statistics such as the halo mass function and the two-point correlation function.}
\label{tab:small_app}
\end{table*}

\begin{figure}
\centering  
\includegraphics[width=0.48\textwidth]{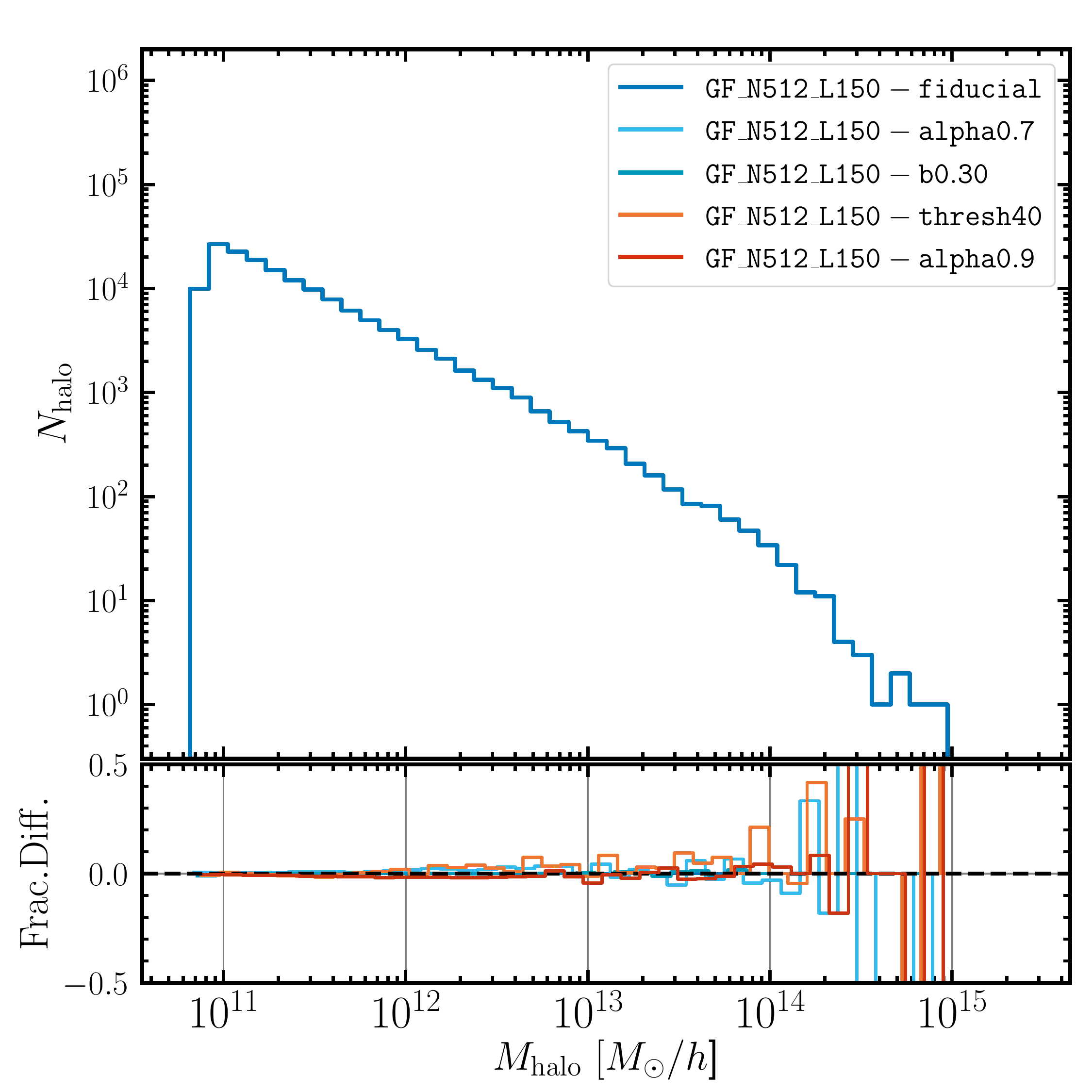} \\
\includegraphics[width=0.48\textwidth]{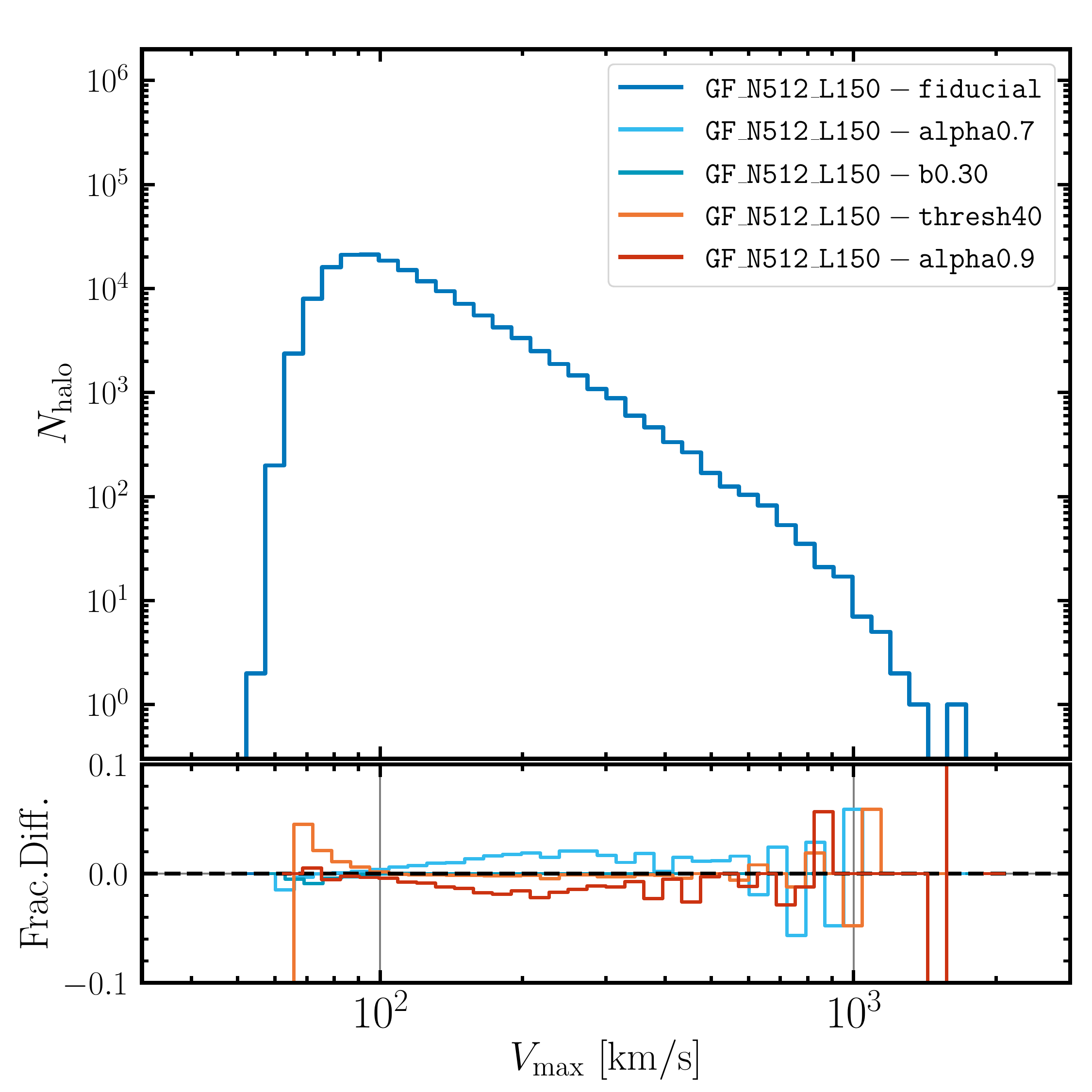}
\caption{Halo mass functions of the \textsc{CompaSO}-defined groups. The \textit{upper panel} shows the traditional halo mass function computed as the number of haloes as a function of halo mass for both finders, while the \textit{lower panel} shows the number of haloes as a function of the maximum circular velocity $V_{\rm max}$. We see that the differences between the test boxes are quite small on all scales except for the high-mass regime in the top panel, where we have very few objects. In the bottom panel, we see that the most substantial effect on the halo function results from varying the eligibility radius parameter.}
\label{fig:hmf_app}
\end{figure}

In Fig.~\ref{fig:hmf_app}, we show the halo mass functions
for the 5 test boxes computed using the total halo mass
($M_{\rm halo}$) as well as the maximum circular velocity
($V_{\rm max}$) analogously to Fig.~\ref{fig:hmf}. In the
top panel, showing the number of haloes as a function of
halo mass, we see that there are negligible differences
between the boxes in the small-mass regime, i.e. the 
fractional difference with respect to the fiducial
case is less than $10\%$
for haloes of mass $M_{\rm halo} \lesssim 10^{14} \ {\rm Mpc}/h$. 
The difference is more pronounced for larger haloes
$M_{\rm halo} \gtrsim 10^{14} \ {\rm Mpc}/h$, 
but noticing a trend in that regime is also more difficult since
we have much fewer examples of such haloes.
In the lower panel, we display
the number of haloes as a function of maximum circular
velocity $V_{\rm max}$. We notice that on all scales, 
the fractional difference with respect to the fiducial
sample is less than 5\%, suggesting that the 
halo-finding parameters have a rather small effect on the
resulting halo mass function. This is reassuring to
see as it shows that the \textsc{CompaSO} haloes are
unlikely to be affected by algorithmic flaws such as
hitting the edge of the L0 group when defining L1
haloes within (tested through \texttt{GF\_N512\_L150\-b0.30})
and having too high kernel density threshold that
excludes many particles from joining L0 groups
(tested through \texttt{GF\_N512\_L150-thresh40}).
The most substantial effect on the halo mass function
comes from varying the eligibility radius. Intuitively,
setting a higher value of $R_{\rm L1,elig}$ reduces the
number of substructures defined as distinct haloes on
the outskirts of larger haloes, which goes in the
same direction as the \textsc{ROCKSTAR} catalogues (see
Fig.~\ref{fig:hmf} in which we present the fractional
difference with respect to \textsc{ROCKSTAR}).
The opposite holds when choosing a lower value of
the eligibility radius. The default value for the
\textsc{AbacusSummit} boxes has been chosen as $R_{\rm L1,elig}=
0.8 \ R_{\rm L1}$ since at the boundaries of large
haloes the density profile drops significantly (roughly
by a factor of 5) and thus any structure that satisfies
the eligibility criteria at that radius (see 
Section~\ref{sec:meth.algo}) is likely to be a
not-yet-merged smaller halo in orbit around a larger object.
Yes smaller values of the eligibility radius produce
wedge-like gouges on the surface of large haloes.

\begin{figure*}
\centering  
\includegraphics[width=1.\textwidth]{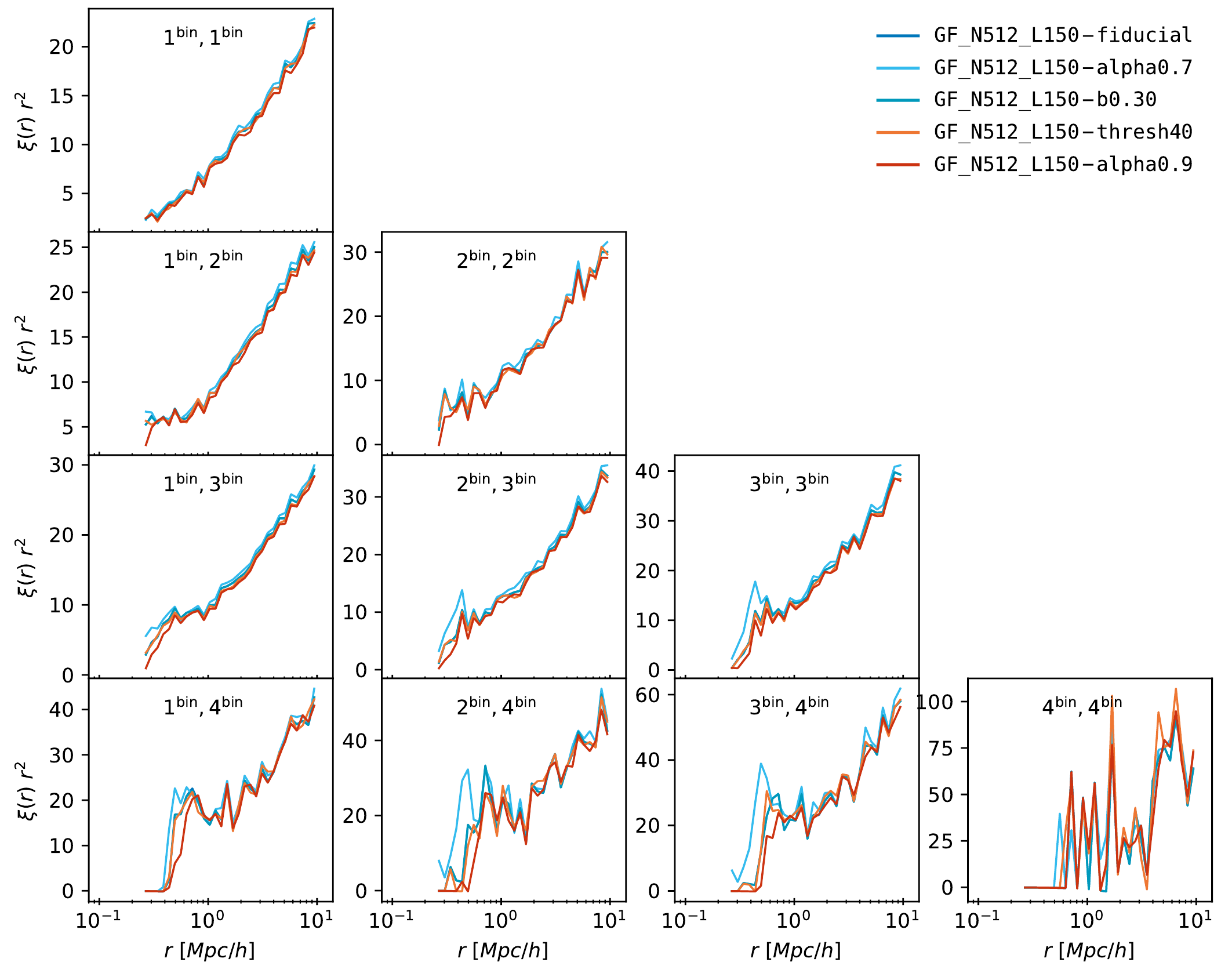}
\caption{Two-point correlation functions
of the \textsc{CompaSO} haloes in several smaller
test boxes. The four mass bins considered in this analysis are:
$1^{\rm st} \ {\rm bin}$: $\log M = 12.0 \pm 0.2$,
$2^{\rm nd} \ {\rm bin}$: $\log M = 12.5 \pm 0.2$, 
$3^{\rm rd} \ {\rm bin}$: $\log M = 13.0 \pm 0.2$,
$4^{\rm th} \ {\rm bin}$: $\log M = 13.5 \pm 0.4$,
measured in $M_\odot/h$, and we have computed cross-correlations
between each pair. We see that the curves look
very similar for all parameter choices, with the largest
difference being observed when varying the eligibility
radius $R_{\rm L1, elig}$. A large value of $R_{\rm L1, elig}$ reduces the bump in the cross-correlations,
whereas a small value produces an excess of small haloes
on the outskirts of other objects which enhances
the observed peaks.}
\label{fig:corr_app}
\end{figure*}

In Fig.~\ref{fig:corr_app}, we study the cross-correlations
between haloes belonging to the following four mass bins:
$1^{\rm st} \ {\rm bin}$: $\log M = 12.0 \pm 0.2$,
$2^{\rm nd} \ {\rm bin}$: $\log M = 12.5 \pm 0.2$, 
$3^{\rm rd} \ {\rm bin}$: $\log M = 13.0 \pm 0.2$,
$4^{\rm th} \ {\rm bin}$: $\log M = 13.5 \pm 0.4$,
where the mass is measured in $M_\odot/h$. Similarly
to the halo mass function plots, we find that most
of the halo-finding parameters have little effect
on the correlation functions, which is a reassuring
result. The largest difference among all cases is
seen again when varying the eligibility radius,
$R_{\rm L1, elig}$. A large value of the eligibility radius
implies more conservative identification of haloes on
the outskirts of large objects, which suppresses the
peak in the cross-correlations and again leads to
more similar behavior to the \textsc{ROCKSTAR} haloes,
whereas a smaller value of $R_{\rm L1, elig}$ increases
the bump in the cross-correlations, suggesting an
even larger excess of small objects lurking around
large haloes. As argued above, the choice of the fiducial
value is motivated by our desire to identify haloes in dense
environments where objects are in close orbits but
not yet merged. To test the ``realness'' of these haloes,
one would need to study their merger trees. This
analysis is summarised in S. Bose et al. (2020) in
prep., and some basic findings are presented in Section \ref{sec:clean.persist}.

\subsection{Smoothing scale of the weighting kernel}
\label{app:smoothing}
In this section, we discuss the choice of smoothing scale of the weighting kernel. If we pick a larger smoothing scale, we would get a less noisy
density estimate,
but only high-density particles would be eligible to be part of L0 (i.e. modified FoF) groups, and
we could miss lower-mass haloes. Therefore, we aim to selecting a smoothing scale that is generous to the smallest haloes and does not affect the largest ones. We note that despite using a smoothing kernel, the FoF group finding (and the density cut) will still leave orphaned particles near the boundary. Nevertheless, a smoothing kernel of the type adopted in \Abacus{Abacus} is better than FoF at including those boundary particles, as we will show below.

The smoothed density field is related to the unsmoothed one through the smoothing function $W(r;b_{\rm kernel}) = 1-r^2/b_{\rm kernel}^2$, so adopting the shorthand notation $b \equiv b_{\rm kernel}$, we get:
\begin{equation}
  \delta_{b}(\boldsymbol{r})
  =
  \frac{3}{4\pi R^3}\int d^3r' \,
  W\left(|\boldsymbol{r} - \boldsymbol{r}'|; b \right)
  \delta(\boldsymbol{r}'),
\end{equation}
where $\delta(\boldsymbol{r})$ is a 3-dimensional density contrast and $b$ is the smoothing scale.

First, we consider the expected particle count $\bar N$ and variance for a uniform background (i.e. $\delta(\boldsymbol{r}) = \Delta = {\rm const.}$):
\begin{equation}
  \bar N  = \int d^3r \Delta \left( 1-r^2/b^2 \right) = \frac{2}{5} \frac{4 \pi}{3} b^3 \Delta,
\end{equation}
\begin{equation}
  {\rm Var}[N] = \int d^3r \Delta \left( 1-r^2/b^2 \right)^2 = \frac{4}{7} \bar N,
\end{equation}
where we have assumed the particle mass to be equal to 1.
We would generally like to be sensitive to density contrasts around $\Delta = 60$. If $b = 0.4$, then $N = 6.4 \pm 1.4$,
which implies S/N $\approx 4.4$. Keeping that in mind, let us now study the convolution of the weighting kernel and SIS profile.

\begin{figure}
\centering  
\includegraphics[width=0.48\textwidth]{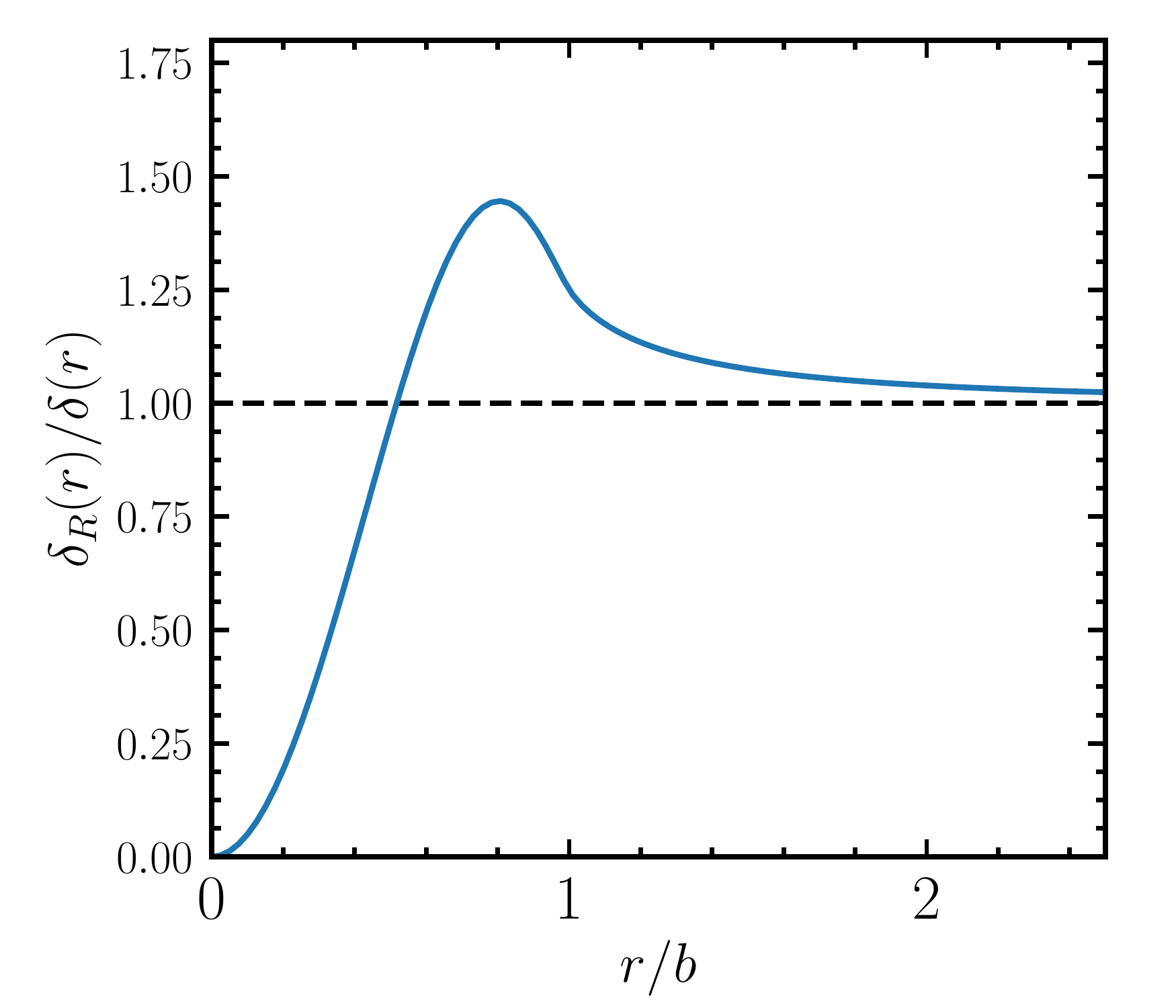}
\caption{Ratio of the convolved SIS density profile $\rho(r) = A/r^2$ with the \Abacus{Abacus} weighting kernel $W(r;b) = 1-r^2/b^2$ to the SIS density profile.}
\label{fig:SIS}
\end{figure}

These convolutions are often more easily computed in Fourier space. The Fourier counterpart of the smoothing function is given by
\begin{equation}
  \tilde{W}(kb) \equiv
  \int d^3r \, e^{-i\boldsymbol{k}\cdot\boldsymbol{r}}
  W\left(|\boldsymbol{r}|; b\right),
\end{equation}
which is real for a spherical smoothing function.  
The smoothing kernel adopted in the \Abacus{Abacus} simulations is a rescaled version of the Epanechnikov kernel \citep{Epanechnikov:1969:NPE}
\begin{equation}
    W(r; b) = 1 -{r^2}/{b^2},
\end{equation}
and its Fourier counterpart is
\begin{equation}
    \tilde{W}(kb) = 6 \frac{4\pi b^3}{3} \frac{j_2(kb)}{(kb)^2}.
\end{equation}
To choose the smoothing scale, we study the convolution of the kernel density estimator with a spherical isothermal sphere (SIS) profile:
\begin{equation}
    \delta(r) = \frac{200}{3} \left(\frac{R_{\rm 200m}}{r}\right)^2 \equiv \frac{A}{r^2},
\end{equation}
where we have expressed the profile using the halo radius containing 200 times the mean density, $R_{\rm 200m}$. The Fourier transform of this profile is:
\begin{equation}
    \tilde{\delta}(k) = \frac{2\pi^2 A}{k} .
\end{equation}
We perform the convolution as follows:
\begin{equation}
  \delta_b(\boldsymbol{r})
  = \frac{3}{4\pi b^3} \int \frac{d^3k}{(2 \pi)^3} \tilde{W}( k b) \  \tilde{\delta}(k) \ e^{-i \boldsymbol{k} \cdot \boldsymbol{r}} .
\end{equation}
Performing the angular integrals and substituting the Fourier transforms of $\tilde{W}(kb)$ and $\tilde{\delta}(k)$, we arrive at:
\begin{equation}
  \delta_b(\boldsymbol{r})
  = \int_0^\infty \frac{k^2 dk}{(2 \pi)^2} \ \frac{3}{2} \ \frac{j_2(kb)}{(kb)^2} \ \frac{2\pi^2 A}{k} \ 2 j_0(kb).
\end{equation}
We can now apply relation 6.574.3 (after expressing the spherical Bessel functions, $j_\nu$, as normal Bessel functions, $J_\nu$) from \citet{gradshteyn2007} to obtain the final result for the smoothed density field:
\begin{equation}
    \delta_b(\boldsymbol{r}) = 
\begin{cases}
    \frac{A}{r^2} \ F(1, 1/2; 7/2; \frac{b^2}{r^2}),& \text{if } r > b\\
    5 \ \frac{A}{r^2} \  \frac{r^2}{b^2} \ F(1, -3/2; 3/2; \frac{r^2}{b^2}),   & \text{if } r < b .
\end{cases}
\end{equation}
In Fig. \ref{fig:SIS}, we show the ratio between the smoothed field and the SIS profile. This ratio peaks around $\sim$1.45 for $r \approx 0.8 b$, and
at $r = b$, it is 1.25. It exceeds 1 when $r \gtrsim 0.53 b$, and asymptotically approaches 1 for $r \gg b$.

Getting a sense of this behavior is very useful. If we would like to be sensitive to lower-mass haloes and have the density remain unchanged for higher-mass haloes, then we can pick a smoothing scale similar to the virial radius, $R_{\rm 200 m}$, of these lower-mass haloes. That way, the smoothed density will tend to over-include particles for the smaller haloes and return to ``normal'' for the bigger ones. If we have a SIS with $N_{\rm 200m} = 40$, that implies $R_{\rm 200 m} \approx 0.36$, measured in units of the mean interparticle spacing. So using $b = 0.4$ is quite well matched to our smaller haloes of $\sim$40 particles, for which S/N is around 4. Combining this with an FoF linking length of $l_{\rm FoF} = 0.25$, which corresponds to $\Delta = 41$, would hence render the L0 group-finding fairly complete.

We can also repeat this exercise for steeper profiles, $\rho(r) = A/r^{x}$ with $x > 2$, and would find that in those cases our smoothing kernel is even more inclusive for lower-mass haloes that have $b \approx R_{\rm 200 m}$.


\bsp	
\label{lastpage}
\end{document}